\documentclass[twocolumn]{aastex61}
\received{\today}
\revised{\today}
\accepted{XXXX}

\usepackage{graphicx}
\usepackage[normalem]{ulem}            
\usepackage{txfonts}
\usepackage{amssymb}
\usepackage{enumerate}
\usepackage{subfigure}
\newcommand{\BE}{\begin{equation}}
\newcommand{\EE}{\end{equation}}
\newcommand{\BA}{\begin{equation}}
\newcommand{\EA}{\end{equation}}

\renewcommand{\vec}[1]{{\mathbf #1}}

\newcommand{\diverg}{ {\bf \nabla} \cdot}

\newcommand{\avec}{ \vec A}
\newcommand{\avecp}{ \vec A_p}
\newcommand{\avecj}{ \vec A_{\mathrm{j}}}

\newcommand{\bb}{\vec B}
\newcommand{\bbp}{\vec B_p}

\newcommand{\vv}{ \vec v}

\newcommand{\llnr}[1]{{\bf \color{magenta}{[]}} \color{black}} 

\shorttitle{Time variations of the non-potential and the volume threading magnetic helicities}
\shortauthors{Linan et al.}

\begin{document}

\title{Time variations of the non-potential and volume-threading magnetic helicities}

\correspondingauthor{Luis Linan}
\email{luis.linan@obspm.fr}

 \author{
 	L. Linan
	}	
  \affil{LESIA, Observatoire de Paris, Universit\'e PSL, CNRS, Sorbonne Universit\'e, Univ. Paris Diderot, Sorbonne Paris Cit\'e, 5 place Jules Janssen, 92195 Meudon, France         }
 
 \author[0000-0002-2900-0608]{
 	\'E. Pariat
	}		
  \affil{LESIA, Observatoire de Paris, Universit\'e PSL, CNRS, Sorbonne Universit\'e, Univ. Paris Diderot, Sorbonne Paris Cit\'e, 5 place Jules Janssen, 92195 Meudon, France         }

 \author{
	K. Moraitis
	}
		
  \affil{LESIA, Observatoire de Paris, Universit\'e PSL, CNRS, Sorbonne Universit\'e, Univ. Paris Diderot, Sorbonne Paris Cit\'e, 5 place Jules Janssen, 92195 Meudon, France }
 
 \author{
	G. Valori}

  \affil{Mullard Space Science Laboratory, University College London, Holmbury St. Mary, Dorking, Surrey, RH5 6NT, UK
         }
         \author{
	J. Leake
	}

  \affil{NASA Goddard Space Flight Center, 8800 purplebelt Rd, purplebelt MD, 22071, USA
         }



\begin{abstract}
Relative magnetic helicity is a gauge invariant quantity suitable for the study of the magnetic helicity content of heliospheric plasmas. Relative magnetic helicity can be decomposed uniquely into two gauge invariant quantities, the magnetic helicity of the non-potential component of the field, and a complementary volume-threading helicity. Recent analysis of numerical experiments simulating the generation of solar eruptions have shown that the ratio of the non-potential helicity to the total relative helicity is a clear marker of the eruptivity of the magnetic system, and that the high value of that quantity could be a sufficient condition for the onset of the instability generating the eruptions. The present study introduces the first analytical examination of the time variations of these non-potential and volume-threading helicities. The validity of the analytical formulas derived are confirmed with analysis of three-dimensional (3D) magnetohydrodynamics (MHD) simulations of solar coronal dynamics. Both the analytical investigation, and the numerical application show that, unlike magnetic helicity, the non-potential and the volume-threading helicities are not conserved quantities, even in the ideal MHD regime. A term corresponding to the transformation between the non-potential and volume-threading helicities frequently dominates their dynamics. This finding has an important consequence for their estimation in the solar corona: unlike with relative helicity, their volume coronal evolution cannot be ascertained by the flux of these quantities through the volume's boundaries. Only techniques extrapolating the 3D coronal field will enable both the proper study of the non-potential and volume-threading helicities, and the observational analysis of helicity-based solar-eruptivity proxies.

\end{abstract}




%

\section{Introduction} \label{sec:Introduction}

Magnetic helicity was originally introduced by \citet{Elsasser56} as a volume integral related to the the three-dimensional (3D) distribution of the magnetic field. \citet{Moffatt69} provided a physical interpretation of this integral, showing that it was intimately related with the Gauss linking number, hence that magnetic helicity quantitatively describes the level of entanglement of magnetic field lines in a magnetised plasma. Magnetic helicity is of particular interest within the ideal magnetohydrodynamic (MHD) paradigm, as it is a strictly conserved quantity \citep{Woltjer58}; creation or dissipation of helicity are forbidden and helicity can only be transported. From observations of the dynamics of plasmas in tokamak experiments, \citet{Taylor74} conjectured that even in non-ideal MHD, the dissipation of magnetic helicity should be relatively weak. \citet{Pariat15b} recently presented numerical evidence that magnetic helicity is indeed very well conserved, even when strong non-ideal effects such as those associated with solar eruptions develop. Thanks to this conservation property, physical quantities based on magnetic helicity are increasingly studied in natural plasmas where the MHD paradigm applies, e.g. in solar/stellar interiors and atmosphere, as well as solar/stellar winds \citep{Valori16,Miesch16,Brandenburg17}. 
 
In the solar context, the conservation of magnetic helicity provides a natural explanation for the existence of ejecta transporting away excess magnetic helicity that cannot indefinitely accumulate in the solar atmosphere; Coronal Mass Ejections (CMEs), and magnetic clouds and their underlying twisted magnetic structures \citep[e.g.][]{Burlaga95,Demoulin08}, appear to be the necessary consequence of magnetic helicity conservation \citep{Rust94,Low96}.
Magnetic helicity and its conservation is a topic of study when trying to link solar eruptions, CMEs and their interplanetary counterparts \citep{Dasso03,Dasso05a,Luoni05,Mandrini05,Dasso09,Hu14,Demoulin16,Patsourakos16,Patsourakos17}.
Magnetic helicity conservation is also invoked as an essential element that impacts the dynamics of magnetic reconnection \citep[e.g.][]{Linton01,Linton02,DelSordo10}, solar/stellar dynamos \citep[e.g.][]{Brandenburg05,Candelaresi12}, the formation of solar filaments \citep[e.g.][]{Antiochos13,Knizhnik15,ZhaoL15}, and the generation of solar eruptions \citep[e.g.][]{Kusano04,Longcope07b,Priest16}. Because of this hypothesis, important efforts to estimate the magnetic helicity in the solar coronal have been carried out over the last decades \citep{Demoulin07,Demoulin09,Valori16}.

Because of the physical requirement of gauge transformation invariance, Elsasser's magnetic helicity can generally not be used to study natural plasmas. \citet{BergerField84} and \citet{Finn85} introduced a gauge invariant quantity related to magnetic helicity than can be practically used with natural plasmas: the relative magnetic helicity. The direct estimation of relative helicity requires the knowledge of the full distribution in the 3D volume studied, while state-of-the-art solar observations only provide measurements on a 2D surface, the solar photosphere. Estimation of relative helicity by volume integration is thus model dependent and requires 3D extrapolation of the magnetic field. A few methods estimating relative magnetic helicity \citep{Rudenko11,Thalmann11,Valori12,YangS13} by such volume-integration have been recently developed and benchmarked \citep[cf. review of][]{Valori16}, enabling their use to study observed solar active regions \citep{Valori13,Moraitis14,GuoY17,Polito17,Temmer17,James18}. 

Because of the inherent observational difficulty to measure helicity from volume-integration, alternative methods have been developed \citep[e.g][]{Demoulin06,Longcope07a,Kazachenko09,Kazachenko10,Kazachenko12,GuoY10,GuoY13b,GuoY17,Georgoulis12} which rely on an implicit model of the solar coronal field. However, the historically most commonly used method to evaluate magnetic helicity relies on the calculation of the flux of helicity through the solar photosphere and the time integration of this flux to obtain the helicity that accumulates into the corona. This approach was originally developed by \citet{Chae01} and has received further improvements \citep{Pariat05,Chae07,LiuY12,LiuY13,Dalmasse14,Dalmasse18}. This method does not make any specific assumption about the coronal magnetic field. It however heavily relies on the helicity conservation principle since it assumes that the time-accumulated boundary flux of helicity is a good approximation of the volume helicity, with the coronal helicity dissipation being null, and its ejection through eruptions negligible, at least during the formation phase of active regions.

Relative helicity is not the only magnetic-helicity-based quantity which has been studied. \citet{Berger85} used a formulation of magnetic helicity and relative magnetic helicity where the magnetic field is decomposed into poloidal and toroidal components. Consistent with this approach, \citet{Low06} introduced the primitive helicity in a two-flux description in which the field is represented by Euler potentials. These examples of the linear decomposition of a magnetic field into the sum of untwisted fields, respects the requirement of gauge independence and easily allows the establishment of the helicity transport equation in Lagrangian variables \citep{Webb10,Webb11}.

\citet{Berger03} also showed that magnetic helicity can be decomposed into two gauge-invariant quantities, the current-carrying magnetic helicity and the volume-threading helicity. To our knowledge, \citet{Moraitis14} were the first to estimate and follow these quantities. In the numerical simulations of the formation of an active region that they analysed, the current-carrying helicity exhibited large fluctuations around the onset of eruptions. Recently, \citet{Pariat17} analysed the properties of these magnetic helicities in seven parametric simulations of flux emergence \citep{Leake13b,Leake14a}, during which solar-like active regions are formed, both non-eruptive and eruptive. \citet{Pariat17} observed that the ratio of the current-carrying helicity to the relative helicity was an excellent marker of the eruptive state of the system: only the simulations that would eventually erupt presented high values of that ratio, and only at times before the eruption. Non-eruptive simulations, as well as the eruptive simulations after the eruption, displayed low values of the ratio. \citet{Zuccarello18} further studied this ratio on different numerical experiments \citep{Zuccarello15}. In four line-tied boundary-driven numerical simulations of solar coronal eruptions (and a non-eruptive control case) for which the eruption/instability time was precisely estimated they showed that the eruptions were taking place for the same value of the helicity ratio, within the helicity measurement precision. Other physical quantities such as magnetic energies did not present the same behaviour. They concluded that the eruption process could be related to a threshold in the helicity ratio, and that this quantity is not only related to the eruptivity of the system but may be directly associated to the eruption driver. 

From these recent promising results stems the need to better understand the contribution entering into the relative helicity decomposition. This is the goal of the present study that aims to provide an analytical formulation for the time variation of both the current-carrying helicity and the volume-threading helicity. Only the study of their time variations allows us to ascertain if these quantities are independently conserved in ideal MHD, in the same way as was done for magnetic helicity. 

Additionally, if relative helicity can be studied and tracked from its flux through the photosphere, it is mostly thanks to its conservation property. Can the same be applied to the terms of the relative helicity decomposition? Can the flux of the current-carrying helicity be solely used to study its accumulation in the solar corona? Or are 3D extrapolation/modelling of the solar corona necessary steps to analyse the decomposition of magnetic helicity in solar active regions? 

In the present manuscript, we first present the time variation of the terms in the helicity decomposition (see Section \ref{sec:helicity}). Using data from three different simulations of solar-like phenomena (active region formation via flux emergence with or without eruption, and a boundary-driven solar jet), we verify our analytical derivations and study carefully the dynamics of the the current-carrying and volume-threading helicities in the different phases of the simulations: magnetic energy build-up and impulsive energy release (see Section \ref{sec:num_test}). In the conclusion (see Section \ref{sec:Conclusion}), we discuss the impact of the non-conservation of these helicities for their estimations in solar observations.

\section{Magnetic Helicity } \label{sec:helicity}
\subsection{Relative magnetic helicity} \label{sec:relative_helicity}
A scalar description of the geometrical properties of magnetic field lines is provided by magnetic helicity, which is defined as follows :
\begin{equation} \label{Eq:ClassicH}
H_{m}= \int_{V}^{} \textbf{A}\cdot\textbf{B} \, \mathrm{d}V= \int_{V}^{} \textbf{A}\cdot\nabla\times\textbf{A} \, \mathrm{d}V \, ,
\end{equation}
with $\textbf{A}$ the vector potential of the studied magnetic field, $\textbf{B}$, which is prescribed in the fixed volume $V$ bounded by the surface $S$. This definition is however unpractical for the study of most natural plasmas. Indeed, under the gauge transformation, 
$\textbf{A}'\longrightarrow\textbf{A}+\nabla\psi$, with $\psi$ an arbitrary function, $H_{m}$ is invariant if and only if $V$ is a magnetically bounded volume, 
i.e. if $\textbf{B}\cdot \textbf{n}|_{\mathrm{S}}=0$, \textbf{n} being the outward-pointing unit vector normal on $S$). This condition is not satisfied when considering the solar corona, as important  magnetic fluxes are threading through the solar photosphere. This led to the introduction of the concept of relative magnetic helicity by \citet{BergerField84}, in which a gauge invariant magnetic helicity is computed introducing a reference field. The most commonly used reference magnetic field is the potential field, $\textbf{B}_{\mathrm{p}}$, the unique current-free field having the same distribution of flux $\textbf{B}$ through the surface $S$. It satisfies:
\begin{equation}
\left\{
\begin{array}{l}
\nabla\times\textbf{B}_{\mathrm{p}}=0 \\
\textbf{n}\cdot(\textbf{B}-\textbf{B}_{\mathrm{p}})|_{\mathrm{S}}=0.
\end{array} \right. 
\end{equation}
The potential field can thus be defined from a scalar function $\phi$ such as $\nabla\phi=\textbf{B}_{\mathrm{p}}$. The scalar potential is computed as the solution of the Laplace equation :
\begin{equation} \label{eq:laplace}
\left\{
\begin{array}{l} \Delta\phi=0 \\
\frac{\partial\phi}{\partial n}|_{\mathrm{S}} = (\textbf{n}\cdot\textbf{B})|_{\mathrm{S}}.
\end{array} \right. 
\end{equation}

The definition of relative magnetic helicity that we use throughout the paper is the one given by \citet{Finn85}:
\begin{equation}
H_{\mathrm{v}}= \int_{V}^{} (\textbf{A}+\textbf{A}_{\mathrm{p}})\cdot(\textbf{B}-\textbf{B}_{\mathrm{p}}) \, \mathrm{d}V. \label{eq:h}
\end{equation}
This choice enables $H_{\mathrm{v}}$ to be independently invariant to the gauge transformations of both $\textbf{A}$ and $\textbf{A}_p$. Relative magnetic helicity can be divided in two terms \citep{Berger03} :
\begin{eqnarray}
 H_{\mathrm{v}}&=&H_{\mathrm{j}}+H_{\mathrm{pj}}\\
 H_{\mathrm{j}}&=&\int_{V}^{} (\textbf{A}-\textbf{A}_{\mathrm{p}})\cdot(\textbf{B}-\textbf{B}_{\mathrm{p}}) \, \mathrm{d}V \label{eq:hj}\\ 
H_{\mathrm{pj}}&=&2\int_{V}^{} \textbf{A}_{\mathrm{p}}\cdot(\textbf{B}-\textbf{B}_{\mathrm{p}}) \, \mathrm{d}V. \label{eq:hpj}
\end{eqnarray}

Considering the non-potential magnetic field, $\textbf{B}_{\mathrm{j}}=\textbf{B}-\textbf{B}_{\mathrm{p}}$, the vector $\textbf{A}_{\mathrm{j}}$ defined as $\textbf{A}_{\mathrm{j}}=\textbf{A}-\textbf{A}_{\mathrm{p}}$, is a vector potential of $\bb_{\mathrm{j}}$ since it verifies:
\begin{eqnarray}
\nabla\times\textbf{A}_{\mathrm{j}}&=&\nabla\times\textbf{A}-\nabla\times\textbf{A}_{\mathrm{p}} \nonumber \\ 
&=&\textbf{B}-\textbf{B}_{\mathrm{p}} = \textbf{B}_{\mathrm{j}}.  \\ \nonumber
\end{eqnarray}
Thus, $H_{\mathrm{j}}$ and $H_{\mathrm{pj}}$ can be expressed as :
\begin{eqnarray}
 H_{\mathrm{j}}&=&\int_{V}^{} \textbf{A}_{\mathrm{j}}\cdot \textbf{B}_{\mathrm{j}} \, \mathrm{d}V\\ 
H_{\mathrm{pj}}&=&2\int_{V}^{} \textbf{A}_{\mathrm{p}}\cdot \textbf{B}_{\mathrm{j}} \, \mathrm{d}V.
\end{eqnarray}
The term $H_{\mathrm{j}}$ thus corresponds to the classical magnetic helicity (Equation \ref{eq:hj}) of the non-potential magnetic field $\bb_{\mathrm{j}}$, for which $S$ is a flux surface by construction, while $H_{\mathrm{pj}}$ can be associated to a volume-threading helicity between $\textbf{B}_{\mathrm{p}}$ and $\bb_{\mathrm{j}}$. Since the decomposition of $\bb$ into $\bb_p$ and $\bb_{\mathrm{j}}$ is unique, the decomposition of $H_v$ in $H_{\mathrm{j}}$ and $H_{\mathrm{pj}}$ is also unique. This decomposition is further relevant since both $H_{\mathrm{j}}$ and $H_{\mathrm{pj}}$ are gauge invariant.

Studying the time evolution of the relative magnetic helicity, \citet{Pariat15b} established the following equation :
\begin{equation} \label{eq:dhdt_diss}
\frac{\text{d}H_{\mathrm{v}}}{\text{d}t}=\left.\frac{\text{d}H_{\mathrm{v}}}{\text{d}t}\right|_{\mathrm{Diss}}+\left.\frac{\text{d}H_{\mathrm{v}}}{\text{d}t}\right|_{\mathrm{Bp,\, var}}+F_{\mathrm{Vn}}+F_{\mathrm{Bn}}+F_{\mathrm{AAp}}+F_{\mathrm{\phi}}
\end{equation}
with :
\begin{eqnarray}
\left.\frac{\text{d}H_{\mathrm{v}}}{\text{d}t}\right|_{\mathrm{Diss}}&=&-2\int_{V}^{} (\textbf{R}\cdot\textbf{B}) \, \mathrm{d}V \\
 \left.\frac{\text{d}H_{\mathrm{v}}}{\text{d}t}\right|_{\mathrm{Bp,\, var}}&=&2\int_{V}^{} \frac{\partial \phi }{\partial t}\nabla\cdot\textbf{A}_{\mathrm{p}} \,\mathrm{d}V \\
 F_{\mathrm{Vn}}&=&-2\int_{\mathrm{S}}^{} (\textbf{B}\cdot\textbf{A})\textbf{v}\cdot\mathrm{d}S \\
 F_{\mathrm{Bn}}&=&2\int_{\mathrm{S}}^{} (\textbf{v}\cdot\textbf{A})\textbf{B}\cdot\mathrm{d}S \\
 F_{\mathrm{AAp}}&=&\int_{\mathrm{S}}^{} (\textbf{A}-\textbf{A}_{\mathrm{p}})\times\frac{\partial }{\partial t}(\textbf{A}+\textbf{A}_{\mathrm{p}}) \cdot\mathrm{d}S \\
 F_{\mathrm{\phi}}&=&-2\int_{\mathrm{S}}^{} \frac{\partial \phi }{\partial t}\textbf{A}_{\mathrm{p}}\cdot \mathrm{d}S
\end{eqnarray}
with $\textbf{R}$ the non-ideal MHD contribution to the electric field, such that $\textbf{E}=-\vv \times \bb + \textbf{R}$. Equation (\ref{eq:dhdt_diss}) assumes that the dynamics follow ideal MHD at the boundary of the domain, i.e. $\left.\textbf{R}\right|_S=0$. Without that hypothesis one would have a non-ideal term such as:
\begin{equation} \label{eq:dhdt}
\frac{\text{d}H_{\mathrm{v}}}{\text{d}t}=\left.\frac{\text{d}H_{\mathrm{v}}}{\text{d}t}\right|_{\mathrm{Non-ideal}}+\left.\frac{\text{d}H_{\mathrm{v}}}{\text{d}t}\right|_{\mathrm{Bp,\, var}}+F_{\mathrm{Vn}}+F_{\mathrm{Bn}}+F_{\mathrm{AAp}}+F_{\mathrm{\phi}}
\end{equation}
with
\begin{eqnarray}
\left.\frac{\text{d}H_{\mathrm{v}}}{\text{d}t}\right|_{\mathrm{Non-ideal}}&=&-2\int_{V}^{} (\nabla\times\textbf{R})\cdot\textbf{A} \, \mathrm{d}V \\
&=&\left.\frac{\text{d}H_{\mathrm{v}}}{\text{d}t}\right|_{\mathrm{Diss}}+F_{\mathrm{Non-ideal}}
\end{eqnarray}
and
\begin{equation}
F_{\mathrm{Non-ideal}}=-2\int_{\mathrm{S}}^{} (\textbf{R}\times\textbf{A}) \cdot \mathrm{d}S.
\end{equation}
This surface term is usually neglected in observation, but it can have its importance in specific simulations where it must then be explicitly calculated. 
We also note that if the non-ideal term of $\textbf{E}$ derives from a scalar potential, i.e. if there is a function $\Theta$ with $\textbf{R}=\nabla\Theta$, one would have :
\begin{equation}
 \left.\frac{\text{d}H_{\mathrm{v}}}{\text{d}t}\right|_{\mathrm{Non-ideal}}=0.
\end{equation}

The helicity time variation (Equation \ref{eq:dhdt}) contains both volume and flux contributions and cannot, in general, be expressed as a function of boundary values alone. In an active solar-like case, \citet{Pariat15b} showed that the dissipation term $\left.\text{d}H_{\mathrm{v}}/\text{d}t\right|_{\mathrm{Diss}}$ is very small, even though strong non-ideal effects are present. They also stated that imposing the Coulomb gauge to $\textbf{A}_{\mathrm{p}}$ makes $\left.\text{d}H_{\mathrm{v}}/\text{d}t\right|_{\mathrm{Bp,\, var}}$ null and hence relative magnetic helicity is a conserved quantity in ideal MHD, i.e. its variations in a volume are only due to flux transfers through the boundaries.

\subsection{Time variations of the non-potential magnetic helicity} \label{sec:dhjdt}

Following \citet{Pariat15b}, we aim to determine the time variation of the current-carrying helicity, $H_{\mathrm{j}}$. Assuming that the volume $V$ is fixed, 
we differentiate $H_{\mathrm{j}}$ in time:

\begin{equation}
\frac{\text{d}H_{\mathrm{j}}}{\text{d}t}=\int_{V}^{} \frac{\partial }{\partial t}\textbf{A}_{\mathrm{j}}\cdot(\textbf{B}-\textbf{B}_{\mathrm{p}})\,\mathrm{d}V \\
+\int_{V}^{} \textbf{A}_{\mathrm{j}}\cdot\frac{\partial }{\partial t}(\textbf{B}-\textbf{B}_{\mathrm{p}}) \, \mathrm{d}V 
\end{equation}
 Using the Gauss divergence theorem for the first term and after a combination with the second integral, we find :
\begin{eqnarray} \label{eq:22}
\frac{\text{d}H_{\mathrm{j}}}{\text{d}t}=&2&\int_{V}^{} \textbf{A}_{\mathrm{j}}\cdot\frac{\partial \textbf{B} }{\partial t}\, \mathrm{d}V 
-2\int_{V}^{}\textbf{A}_{\mathrm{j}}\cdot\frac{\partial \textbf{B}_{\mathrm{p}} }{\partial t} \, \mathrm{d}V \nonumber \\
&+&\int_{\mathrm{S}}^{} \textbf{A}_{\mathrm{j}}\times\frac{\partial }{\partial t}\textbf{A}_{\mathrm{j}} \cdot\mathrm{d}S
\end{eqnarray}
We choose to keep this form for the flux term and to treat the volume terms separately. We note that only the sum of these terms is properly defined physically, i.e. is gauge invariant: the three terms are not independently gauge invariant. Using the scalar potential $\phi$ of $\textbf{B}_{\mathrm{p}}$, and the Gauss divergence theorem, we can decompose the second term of Equation (\ref{eq:22}):
\begin{eqnarray}
-2\int_{V}^{}\textbf{A}_{\mathrm{j}}\cdot\frac{\partial \textbf{B}_{\mathrm{p}} }{\partial t} \, \mathrm{d}V=&-&2\int_{\mathrm{S}}^{} \frac{\partial \phi }{\partial t}\textbf{A}_{\mathrm{j}}\cdot \mathrm{d}S \nonumber \\
&+&2\int_{V}^{} \frac{\partial \phi }{\partial t}\nabla\cdot\textbf{A}_{\mathrm{j}} \,\mathrm{d}V 
\end{eqnarray}

We then use the Faraday law : $\partial \textbf{B}/\partial t = - \nabla \times \textbf{E}$. The first volume term of Equation (\ref{eq:22}) can be written as : 
\begin{eqnarray}
2\int_{V}^{} \textbf{A}_{\mathrm{j}}\cdot\frac{\partial \textbf{B} }{\partial t}\, \mathrm{d}V=&2&\int_{\mathrm{S}}^{} \textbf{A}_{\mathrm{j}}\cdot \nabla \times (\vv \times \bb - \textbf{R}) \, \mathrm{d}V \nonumber\\
=&2&\int_{V}^{} (\nabla \times (\vv \times \bb))\cdot\textbf{A}_{\mathrm{j}} \, \mathrm{d}V \nonumber \\
-&2&\int_{V}^{} (\nabla \times \textbf{R})\cdot\textbf{A}_{\mathrm{j}} \, \mathrm{d}V 
\end{eqnarray}
As a last step in the decomposition, we use the Gauss divergence theorem on the ideal term with : 
\begin{eqnarray}
2\int_{V}^{} (\nabla \times (\vv \times \bb))\cdot\textbf{A}_{\mathrm{j}} \, \mathrm{d}V
=&-&2\int_{V}^{} ((\textbf{v}\times\textbf{B})\cdot\textbf{B}_{\mathrm{p}}) \, \mathrm{d}V \nonumber \\ 
&-&2\int_{\mathrm{S}}^{} (\textbf{B}\cdot\textbf{A}_{\mathrm{j}})\textbf{v}\cdot\mathrm{d}S \nonumber \\
&+&2\int_{\mathrm{S}}^{} (\textbf{v}\cdot\textbf{A}_{\mathrm{j}})\textbf{B}\cdot\mathrm{d}S 
\end{eqnarray}

Finally, the variation of the magnetic helicity of the non-potential magnetic field can be decomposed as :
\begin{eqnarray} \label{eq:dhjdt}
\frac{\text{d}H_{\mathrm{j}}}{\text{d}t}&=&\left.\frac{\text{d}H_{\mathrm{j}}}{\text{d}t}\right|_{\mathrm{Non-ideal}}+\left.\frac{\text{d}H_{\mathrm{j}}}{\text{d}t}\right|_{\mathrm{Bp,\, var}}+\left.\frac{\text{d}H_{\mathrm{j}}}{\text{d}t}\right|_{\mathrm{Transf}} \nonumber\\
&+&F_{\mathrm{Vn,\, Aj}}+F_{\mathrm{Bn,\, Aj}}+F_{\mathrm{Aj,\, Aj}}+F_{\phi,\, Aj} 
\end{eqnarray}
with
\begin{eqnarray}
 \left.\frac{\text{d}H_{\mathrm{j}}}{\text{d}t}\right|_{\mathrm{Non-ideal}}&=&- 2\int_{V}^{} \nabla \times \textbf{R}\cdot\textbf{A}_{\mathrm{j}} \, \mathrm{d}V \label{eq:NoId_Aj}\\
 \left.\frac{\text{d}H_{\mathrm{j}}}{\text{d}t}\right|_{\mathrm{Transf}}&=&-2\int_{V}^{} (\textbf{v}\times\textbf{B})\cdot\textbf{B}_{\mathrm{p}} \, \mathrm{d}V \label{eq:Ftransf_Aj}\\ 
 \left.\frac{\text{d}H_{\mathrm{j}}}{\text{d}t}\right|_{\mathrm{Bp,\, var}}&=&2\int_{V}^{} \frac{\partial \phi }{\partial t}\nabla\cdot\textbf{A}_{\mathrm{j}} \,\mathrm{d}V \label{eq:Fvar_Aj}\\
 F_{\mathrm{Vn,\, Aj}}&=&-2\int_{\mathrm{S}}^{} (\textbf{B}\cdot\textbf{A}_{\mathrm{j}})\textbf{v}\cdot\mathrm{d}S \label{eq:FVn_Aj}\\
 F_{\mathrm{Bn,\, Aj}}&=&2\int_{\mathrm{S}}^{} (\textbf{v}\cdot\textbf{A}_{\mathrm{j}})\textbf{B}\cdot\mathrm{d}S \label{eq:FBn_Aj}\\
 F_{\mathrm{Aj,\, Aj}}&=&\int_{\mathrm{S}}^{} \textbf{A}_{\mathrm{j}}\times\frac{\partial }{\partial t}\textbf{A}_{\mathrm{j}} \cdot\mathrm{d}S \label{eq:FAj_Aj}\\
 F_{\phi,\, Aj}&=&-2\int_{\mathrm{S}}^{} \frac{\partial \phi }{\partial t}\textbf{A}_{\mathrm{j}}\cdot \mathrm{d}S \label{eq:Fphi_Aj}
\end{eqnarray}
The decomposition obtained possesses similarities with the time variation of the relative helicity. Apart from $\left. \text{d}H_{\mathrm{j}}/\text{d}t\right|_{\mathrm{Transf}}$, all the terms that appear in the time variation of $H_{\mathrm{j}}$ (Equation \ref{eq:dhjdt}) have their equivalent in the decomposition of $\text{d}H/\text{d}t$ (Equation \ref{eq:dhdt}). 
 We find a flux $F_{Vn,\,Aj}$ related to the normal component of the velocity $v_n$, and $F_{\mathrm{Bn,\, Aj}}$ related to the normal component of the field, $B_n$. The volume term $\left.\text{d}H_{\mathrm{j}}/\text{d}t\right|_{\mathrm{Bp,\, var}}$ related with the time variation of the magnetic field $\textbf{B}_{\mathrm{p}}$ also appears. The difference with the terms of the time variation of the relative helicity is the dependence on $\avecj$ instead of $\avec$. 

Unlike $H_v$, even in ideal MHD the time variation of $H_{\mathrm{j}}$ contains both volume and flux contributions. The term $\left.\text{d}H_{\mathrm{j}}/\text{d}t\right|_{\mathrm{Transf}}$ in Equation (\ref{eq:Ftransf_Aj}) is generally not null in ideal MHD (see Section \ref{sec:transf} for more discussion about this term). Theoretically, $H_{\mathrm{j}}$ is not a conserved quantity of ideal MHD, unlike the classical magnetic helicity $H_{m}$, and the relative magnetic helicity $H_v$ written in specific gauge conditions.

The majority of terms depend on the difference $\avecj$ between the two vector potentials $\avec$ and $\avecp$. Therefore by imposing specific relations between these, it is possible to eliminate some of the contributions to the time variation of $H_{\mathrm{j}}$ (see Section \ref{sec:cond}). 

Moreover, since the individual terms are not gauge-invariant, only their sum has true physical relevance. The intensity of the flux terms depends on the gauge selected.

\subsection{Time variation of the volume-threading helicity} \label{sec:dhpjdt}
Now considering $H_{\mathrm{pj}}$, following similar steps as for $H_{\mathrm{j}}$, we can derive the general equation of its time variation:
\begin{eqnarray} \label{eq:dhpjdt}
\frac{\text{d}H_{\mathrm{pj}}}{\text{d}t}&=&\left.\frac{\text{d}H_{\mathrm{pj}}}{\text{d}t}\right|_{\mathrm{Non-ideal}}+\left.\frac{\text{d}H_{\mathrm{pj}}}{\text{d}t}\right|_{\mathrm{Bp,\, var}} + \left.\frac{\text{d}H_{\mathrm{pj}}}{\text{d}t}\right|_{\mathrm{Transf}} \nonumber\\
&+&F_{\mathrm{Vn,\, Ap}}+F_{\mathrm{Bn,\, Ap}}+F_{\mathrm{Aj,\, Ap}}+F_{\mathrm{\phi,\, Ap}} 
\end{eqnarray}
with
\begin{eqnarray}
\left.\frac{\text{d}H_{\mathrm{pj}}}{\text{d}t}\right|_{\mathrm{Non-ideal}}&=& - 2\int_{V}^{} \nabla \times \textbf{R}\cdot\textbf{A}_{\mathrm{p}} \, \mathrm{d}V \label{eq:NoId_Ap} \\
 \left.\frac{\text{d}H_{\mathrm{pj}}}{\text{d}t}\right|_{\mathrm{Transf}}&=&2\int_{V}^{} (\textbf{v}\times\textbf{B})\cdot\textbf{B}_{\mathrm{p}} \, \mathrm{d}V \label{eq:Ftransf_Ap} \\
 \left.\frac{\text{d}H_{\mathrm{pj}}}{\text{d}t}\right|_{\mathrm{Bp,\, var}}&=&2\int_{V}^{} \frac{\partial \phi }{\partial t}\nabla\cdot(\avecp-\avecj) \,\mathrm{d}V \label{eq:Fvar_Ap} \\
 F_{\mathrm{Vn,\, Ap}}&=&-2\int_{\mathrm{S}}^{} (\textbf{B}\cdot\textbf{A}_{\mathrm{p}})\textbf{v}\cdot\mathrm{d}S \label{eq:FVn_Ap}\\
 F_{\mathrm{Bn,\, Ap}}&=&2\int_{\mathrm{S}}^{} (\textbf{v}\cdot\textbf{A}_{\mathrm{p}})\textbf{B}\cdot\mathrm{d}S \label{eq:FBn_Ap}\\
 F_{\mathrm{Aj,\, Ap}}&=&2\int_{\mathrm{S}}^{} \textbf{A}_{\mathrm{j}}\times\frac{\partial }{\partial t}\textbf{A}_{\mathrm{p}} \cdot\mathrm{d}S \label{eq:FAj_Ap}\\
 F_{\mathrm{\phi,\, Ap}}&=&-2\int_{\mathrm{S}}^{} \frac{\partial \phi }{\partial t}(\avecp-\avecj)\cdot \mathrm{d}S \label{eq:Fphi_Ap}
\end{eqnarray}

As with $H_{\mathrm{j}}$, the time variation of $H_{\mathrm{pj}}$ cannot be expressed solely through boundary fluxes, and thus $H_{\mathrm{pj}}$ is not a conserved quantity even in ideal MHD when $\textbf{R}=0$, unlike $H_{\mathrm{v}}$, and this is due to the transfer term of Equation (\ref{eq:Ftransf_Ap}).

\subsection{Helicity exchange} \label{sec:transf}

\subsubsection{Helicity exchange between $H_{\mathrm{j}}$ and $H_{\mathrm{pj}}$}
As expected, by summing the time variations of the non-potential and the volume-threading magnetic helicities (Equations \ref{eq:dhjdt} and \ref{eq:dhpjdt}) we obtain the time variation of the relative magnetic helicity (Equation \ref{eq:dhdt}):
\begin{equation}
\frac{\text{d}H_{\mathrm{v}}}{\text{d}t}=\frac{\text{d}H_{\mathrm{j}}}{\text{d}t}+\frac{\text{d}H_{\mathrm{pj}}}{\text{d}t}.
\end{equation}
Each term in Equation (\ref{eq:dhdt}) has indeed its counterpart in the decomposition of $H_{\mathrm{j}}$ and $H_{\mathrm{pj}}$, e.g. the sum of $F_{\mathrm{Vn,\, Ap}}$ and $F_{\mathrm{Vn,\, Aj}}$ gives $F_{\mathrm{Vn}}$. 

However, the time variations of $H_{\mathrm{pj}}$ and $H_{\mathrm{j}}$  each possess a volume contribution that is not present in $\text{d}H_{\mathrm{v}}/\text{d}t$. These terms, $\left.\text{d}H_{\mathrm{x}}/\text{d}t\right|_{\mathrm{Transf}}$ with $x$ being either $j$ or $pj$, allow the transfer of helicity between $H_{\mathrm{pj}}$ and $H_{\mathrm{j}}$. They correspond to oppositely signed quantities, i.e.:
\begin{equation}
\left.\frac{\text{d}H_{\mathrm{j}}}{\text{d}t}\right|_{\mathrm{Transf}}=- \left.\frac{\text{d}H_{\mathrm{pj}}}{\text{d}t}\right|_{\mathrm{Transf}}.
\end{equation}
The helicity transfer term is a volume quantity which allows the
transformation of one form of helicity into the other. This transformation occurs within the full domain of study. Our analysis of several numerical experiments (see Section \ref{sec:num_test}) shows that this quantity actually reaches high values compared to the other terms and can dominate the evolution of both $H_{\mathrm{j}}$ or $H_{\mathrm{pj}}$.

It is thus essential to have a precise understanding of the term $\left.\text{d}H_{j}/\text{d}t\right|_{Transf}$. A study at this term permits us to quantify the exchange between the helicities $H_{\mathrm{j}}$ and $H_{pj}$ in the volume, an exchange that does not affect the relative helicity, $H_{v}$. It is particularly worth noting that $\left.\text{d}H_{j}/\text{d}t\right|_{Transf}$ is a gauge invariant quantity. It indeed only depends on $\vv$, $\bb$, and $\bbp$ and does not have any vector potential contribution.

There are several possible decompositions for $\left.\text{d}H_{\mathrm{j}}/\text{d}t\right|_{\mathrm{Transf}}$. However, despite several different attempts, we could not find a way to express this term solely as a flux contribution, i.e. as an integral on the boundary. For example, we can write : 
\begin{eqnarray}
-2\int_{V}^{} (\textbf{v}\times\textbf{B})\cdot\textbf{B}_{\mathrm{p}} \, \mathrm{d}V=-&2&\int_{V}^{} ((\textbf{v}\times\textbf{B})\cdot\nabla\phi) \, \mathrm{d}V \nonumber \\
=-&2&\int_{\mathrm{S}}^{} \phi(\textbf{v}\times\textbf{B})\cdot\mathrm{d}S \nonumber \\
+&2&\int_{V}^{} \phi\nabla\cdot(\textbf{v}\times\textbf{B})\mathrm{d}V 
\end{eqnarray}
A volume integral contribution however remains. We have numerically tested this decomposition (in a similar manner as performed in Section \ref{sec:num_test}) and concluded that it presents no advantage over its $(\textbf{v}\times\textbf{B})\cdot\textbf{B}_{\mathrm{p}}$ form.

\subsubsection{Helicity exchange with the surrounding environment} \label{sec:surr_envi}

In the specific case of the resistive MHD, the non-ideal contribution to the electric field can be explicitly written as : $\textbf{R}=\eta \nabla \times \textbf{B}$ with $\eta$ corresponding to the magnetic resistivity. Using the Gauss-divergenge theorem, the non-ideal term in Equation (\ref{eq:dhjdt}) can be decomposed into a surface term and a dissipation term :
\begin{eqnarray} 
\left.\frac{\text{d}H_{\mathrm{j}}}{\text{d}t}\right|_{\mathrm{Non-ideal}}&=&-2\int_{V}^{} \nabla \times (\eta\nabla\times\textbf{B})\cdot\textbf{A}_{\mathrm{j}} \, \mathrm{d}V\\
&=&\left.\frac{\text{d}H_{\mathrm{j}}}{\text{d}t}\right|_{\mathrm{Diss}}+F_{\mathrm{Non-ideal,\, Aj}} \label{eq:NonIdealDec_Hj}
\end{eqnarray}
with
\begin{eqnarray}
\left.\frac{\text{d}H_{\mathrm{j}}}{\text{d}t}\right|_{\mathrm{Diss}}&=&-2\int_{V}^{} \eta(\nabla\times\textbf{B})\cdot\textbf{B}_{\mathrm{j}} \, \mathrm{d}V \label{eq:Diss_Aj}\\
F_{\mathrm{Non-ideal,\, Aj}}&=&-2\int_{\mathrm{S}}^{} \eta(\nabla\times\textbf{B})\times\textbf{A}_{\mathrm{j}} \cdot \mathrm{d}S. \label{eq:SNoId_Aj}
\end{eqnarray}
The dissipation term, $\text{d}H_{\mathrm{j}}/\text{d}t|_{\mathrm{Diss}}$, as well as the transfer term, $\text{d}H_{\mathrm{j}}/\text{d}t|_{\mathrm{Transf}}$, is gauge invariant. By defining $\text{d}H_{\mathrm{j}}/\text{d}t|_{\mathrm{Own}}$ such as:
\begin{eqnarray} \label{eq:Own_Hj}
 \left.\frac{\text{d}H_{\mathrm{j}}}{\text{d}t}\right|_{\mathrm{Own}}=&&\left.\frac{\text{d}H_{\mathrm{j}}}{\text{d}t}\right|_{\mathrm{Bp,\, var}}+F_{\mathrm{Non-ideal,\, Aj}} \nonumber \\
 &+&F_{\phi,\, Aj}+F_{\mathrm{Vn,\, Aj}}+F_{\mathrm{Bn,\, Aj}}+F_{\mathrm{Aj,\, Aj}}
 \end{eqnarray}
we obtain an equation for the time variation of $H_{\mathrm{j}}$ which is formed solely of gauge invariant terms: 
\begin{equation} \label{eq:dhjdt_gaugeinv}
\frac{\text{d}H_{\mathrm{j}}}{\text{d}t}=\left.\frac{\text{d}H_{\mathrm{j}}}{\text{d}t}\right|_{\mathrm{Own}}+\left.\frac{\text{d}H_{\mathrm{j}}}{\text{d}t}\right|_{\mathrm{Diss}}+\left.\frac{\text{d}H_{\mathrm{j}}}{\text{d}t}\right|_{\mathrm{Transf}}.
\end{equation}
Similarly, we can construct a time variation of $H_{\mathrm{pj}}$ with gauge invariant terms only:
\begin{eqnarray} \label{eq:dhpjdt_gaugeinv}
\frac{\text{d}H_{\mathrm{pj}}}{\text{d}t}=\left.\frac{\text{d}H_{\mathrm{pj}}}{\text{d}t}\right|_{\mathrm{Own}}+\left.\frac{\text{d}H_{\mathrm{pj}}}{\text{d}t}\right|_{\mathrm{Diss}}-\left.\frac{\text{d}H_{\mathrm{j}}}{\text{d}t}\right|_{\mathrm{Transf}} 
\end{eqnarray}
with
\begin{eqnarray}
 \left.\frac{\text{d}H_{\mathrm{pj}}}{\text{d}t}\right|_{\mathrm{Diss}}&=&-2\int_{V}^{} \eta(\nabla\times\textbf{B})\cdot\textbf{B}_{\mathrm{p}} \, \mathrm{d}V \label{eq:Diss_Ap} \\
 \left.\frac{\text{d}H_{\mathrm{pj}}}{\text{d}t}\right|_{\mathrm{Own}}&=&\left.\frac{\text{d}H_{\mathrm{pj}}}{\text{d}t}\right|_{\mathrm{Bp,\, var}}+F_{\mathrm{Non-ideal,\, Ap}} \nonumber\\
&+&F_{\mathrm{Vn,\, Ap}}+F_{\mathrm{Bn,\, Ap}}+F_{\mathrm{Aj,\, Ap}}+F_{\mathrm{\phi,\, Ap}} \label{eq:Own_Hpj}
\end{eqnarray} 
and
\begin{equation}
F_{\mathrm{Non-ideal,\, Aj}}=-2\int_{\mathrm{S}}^{} \eta(\nabla\times\textbf{B})\times\textbf{A}_{\mathrm{p}} \cdot \mathrm{d}S. \label{eq:SNoId_Ap}
\end{equation}

The "Own" terms correspond to the proper helicity variation of either
$H_{\mathrm{j}}$ or $H_{\mathrm{pj}}$. They do not strictly speaking correspond to a
flux through the boundary since a volume contribution is also present
for both $H_{\mathrm{j}}$ and $H_{\mathrm{pj}}$ : $\left.\text{d}H_{\mathrm{x}}/\text{d}t\right|_{\mathrm{Bp,\, var}}$. These volume
contributions are however gauge dependent. A particular choice of
gauges (the Coulomb gauge for $\textbf{A}$ and $\textbf{A}_{\mathrm{p}}$) can nonetheless make
them null. Written with this choice of
gauges,$\left.\text{d}H_{\mathrm{j}}/\text{d}t\right|_{\mathrm{Own}}$ and
$\left.\text{d}H_{\mathrm{pj}}/\text{d}t\right|_{\mathrm{Own}}$ then only appear as
pure boundary flux contributions, that would correspond to transfer
of helicities between the studied domain and its surrounding environment. Like $\text{d}H_{\mathrm{j}}/\text{d}t|_{\mathrm{Transf}}$, both "Own" terms are independently gauge invariant quantities. Equations (\ref{eq:dhjdt_gaugeinv})-(\ref{eq:dhpjdt_gaugeinv}) are thus only involving independently gauge invariant terms. Their analysis is thus of particular interest, as will be shown in our study of the helicity evolution in different numerical experiments (cf. Section \ref{sec:num_test}).

\subsection{Specific gauge conditions} \label{sec:cond}
While the variation of $H_{\mathrm{j}}$ and $H_{\mathrm{pj}}$ can be generally described by Equations (\ref{eq:hj})-(\ref{eq:hpj}) for any gauge, the choice of some specific additional constraints on the gauge allows us to simplify the expression of their time variations and possibly their computation.

A first possible additional constraint is to use the Coulomb gauge for the vector potential of the potential field, $\avecp$, i.e. :
\begin{equation}
\nabla\cdot\textbf{A}_{\mathrm{p}}=0 \label{eq:coulomb_ap}
\end{equation}
In this gauge, the volume contributions related to the variation of the potential field become : 
\begin{eqnarray}
 \left.\frac{\text{d}H_{\mathrm{pj}}}{\text{d}t}\right|_{Bp,\, var,\, cond\, (\ref{eq:coulomb_ap}) }&=&-\left.\frac{\text{d}H_{\mathrm{j}}}{\text{d}t}\right|_{Bp,\, var,\, cond\, (\ref{eq:coulomb_ap}) }\\
 &=&-2\int_{V}^{} \frac{\partial \phi }{\partial t}\nabla\cdot\textbf{A} \,\mathrm{d}V 
\end{eqnarray}

While condition (\ref{eq:coulomb_ap}) leads to a cancellation of $\text{d}H_{\mathrm{v}}/\text{d}t|_{\mathrm{Bp,\, var}}$ for the relative helicity \citep[e.g.][]{Pariat15b}, this it is not the case for the evolution of its components $H_{\mathrm{j}}$ and $H_{\mathrm{pj}}$.

Another possible additional constraint that can be imposed is to link the vector potentials $\avec$ and $\avecp$ on the boundary, i.e.:
\begin{equation}
\avec|_{\mathrm{S}}- \avecp|_{\mathrm{S}}=\textbf{A}_{\mathrm{j}}|_{\mathrm{S}}=0 \label{eq:aj_null}
\end{equation}
This condition ensures the nullity of $F_{AAp,\, Ap}$ and all the fluxes of the form $F_{\alpha,Aj}$ (with $\alpha\in\{Vn, Bn, Aj, \phi, Non-ideal\}$) in Equation (\ref{eq:Own_Hj}). Under such a condition, the time variation of the non-potential magnetic helicity can be described only as a volume variation, consisting of the sum of the transfer term, $\left.\text{d}H_{\mathrm{j}}/\text{d}t\right|_{\mathrm{Transf}}$, and the term $\left.\text{d}H_{\mathrm{j}}/\text{d}t\right|_{\mathrm{Bp,\, var}}$. There is no contribution of helicity due to any fluxes through the boundary. 

Another possible constraint is to eliminate the normal component of $\avecp$ on the boundary: \begin{equation}
\textbf{A}_{\mathrm{p}}\cdot\textbf{n}|_{\mathrm{S}}=0 \label{eq:Ap_bond}
\end{equation}
This choice combined with the previous one Equation (\ref{eq:aj_null}) leads to the elimination of the term $F_{\mathrm{\phi,\, Ap}}$.

Combining these conditions and assuming that the evolution follows the ideal MHD evolution, i.e. supposing
\begin{equation} \label{eq:condition_combination}
\left\{
\begin{array}{l} 
\textbf{R} =0 \\
\nabla\cdot\textbf{A}_{\mathrm{p}}=0 \\
\textbf{A}_{\mathrm{p}}\cdot\textbf{n}|_{\mathrm{S}}=0 \\
\avec|_{\mathrm{S}}- \avecp |_{\mathrm{S}}=0 
\end{array} \right. 
\end{equation}
as is frequently assumed when studying the flux of relative magnetic helicity \citep[e.g.][]{Demoulin09}, we obtain the following form of the time variations of $H_{\mathrm{j}}$ and $H_{\mathrm{pj}}$ :
\begin{equation}
 \left.\frac{\text{d}H_{\mathrm{j}}}{\text{d}t}\right|_{cond.\, (\ref{eq:condition_combination})}=2\int_{V}^{} \frac{\partial \phi }{\partial t}\nabla\cdot\textbf{A} \,\mathrm{d}V - 2\int_{V}^{} (\textbf{v}\times\textbf{B})\cdot\textbf{B}_{\mathrm{p}} \, \mathrm{d}V
\end{equation}
and
\begin{eqnarray}
 \left.\frac{\text{d}H_{\mathrm{pj}}}{\text{d}t}\right|_{cond.\, (\ref{eq:condition_combination})}= &2&\int_{V}^{} (\textbf{v}\times\textbf{B})\cdot\textbf{B}_{\mathrm{p}} \, \mathrm{d}V -2\int_{V}^{} \frac{\partial \phi }{\partial t}\nabla\cdot\textbf{A} \,\mathrm{d}V \nonumber \\ 
 +&2&\int_{\mathrm{S}}^{} ((\vv \cdot\textbf{A}_{\mathrm{p}})\textbf{B}-(\textbf{B}\cdot\textbf{A}_{\mathrm{p}})\textbf{v})\cdot\mathrm{d}S  \end{eqnarray}

With these specific conditions $H_{\mathrm{j}}$ does not exchange with the outside environment (no surface fluxes). The flux of the total relative helicity, $H_{\mathrm{v}}$, is uniquely due to the flux of $H_{\mathrm{pj}}$ through the boundaries. $H_{\mathrm{pj}}$ can additionally undergo variations in the volume. These volume variations are actually helicity exchanges with $H_{\mathrm{j}}$ (cf. Equation (\ref{eq:Ftransf_Aj}) and Equation (\ref{eq:Ftransf_Ap})). The time evolution of $H_{\mathrm{j}}$ is indeed only related to volume conversion from $H_{\mathrm{pj}}$. 

While condition (\ref{eq:condition_combination}) appears particularly strong, it is \textit{de facto} a usual assumption in observational methods analyzing the flux of helicity through the solar photosphere. Our analysis highlights the internal exchange in the volume between the two helicities. It indicates that the traditional helicity flux method cannot be used to evaluate the current-carrying component, $H_{\mathrm{j}}$, in observations. This point will be further highlighted in our analysis of numerical simulations of the dynamics of the solar corona (cf. following Section) and in the conclusion (see Section \ref{sec:Conclusion}).

\section{Numerical tests} \label{sec:num_test}

In order to quantify the transfer of helicity and to numerically validate the time variations of the two helicities (cf. Equation (\ref{eq:dhjdt}) and Equation (\ref{eq:dhpjdt})), we analyze the magnetic field of three different numerical simulations produced by two different 3D MHD codes.

\subsection{Test cases} \label{sec:test}

\subsubsection{Jet simulation} \label{sec:jet}
\begin{figure*}[!h]
  \centering
			\includegraphics[scale=0.6]{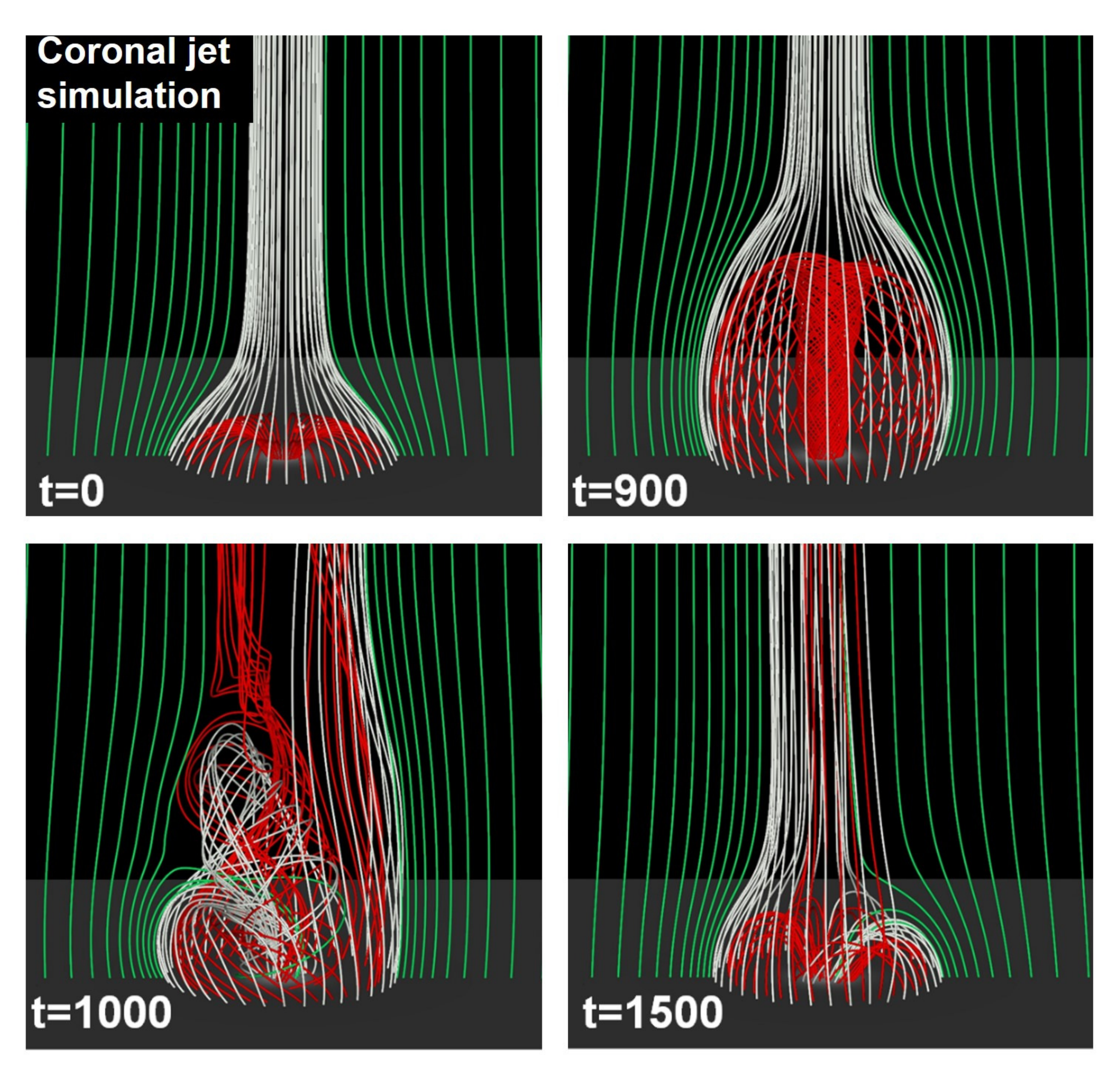}
			\caption{Snapshots of the magnetic field evolution during the generation of a coronal jet. The red field lines are initially closed. The green and white field lines are initially open. At t = 900 the system is in its pre-eruption stage. At t=1000 the system is erupting. Helicity is ejected upward along newly opened reconnected field lines. At t = 1500 the system is slowly relaxing to its final stage}
			\label{jet}
\end{figure*}

The first test case is a 3D MHD numerical simulation of the generation of a solar coronal
jet \citep{Pariat09a}. The initial magnetic field in this numerical experiment is formed by a 3D null point (cf. Figure \ref{jet}). The volume contains two different magnetic connectivity domains; open and closed. The simulation lasts between $t=0$ and $t=1600$ in the system's non-dimensional units. In the pre-eruptive phase $t\in [0,920]$, the energy and the helicity are accumulated by line-tied twisting motions of the central magnetic polarity. These motions preserve the distribution of the vertical magnetic field component, $B_z$, at the bottom boundary. Thanks to topological constraints, magnetic reconnection is inhibited during that phase \citep[see discussion in Section 2 of][]{Pariat09a}, and the dynamics of the system can thus be considered quasi-ideal MHD before $t\sim920$.

At around $t\sim920$, magnetic reconnection between closed and open field lines induce the formation of a jet. The period after $t\sim 920$, is designated as the non-ideal phase in opposition to the pre-jet phase. During this non-ideal MHD evolution, with intense magnetic reconnection, free magnetic energy is dissipated and released by the jet. The helicity is transferred outside through the domain boundaries by a non-linear Alfv\'enic wave constituting the jet \citep[cf.][for the physics of the driving mechanism of the jet]{Pariat16}. The jet lasts between $t\sim 920$ and 
$t\sim 1200$. After $t\sim 1200$, the jet plasma has left the domain and no high velocity upflows are present in the studied domain. The system slowly relaxes towards its initial potential state, thanks to low intensity reconnections. A low amplitude, large scale standing wave remains in the domain \citep{Pariat15b}. At the end of the simulation, very few twisted field lines remain.

Physical quantities are outputted with a cadence of $\Delta t=50 $ for $t\in [0,700]$ and $\Delta t=10 $ for $t\in [700,1600]$, the higher cadence allowing a better analysis of the non-ideal phase.
The output data grid corresponds to a sample of the simulation computation grid. The volume analysed in the present study is a 3D mesh of $129^3$ points, whose range is: $x\in [-6,6]$, $y\in [-6,6]$ and $z\in [0,12]$, hence corresponding to the central domain of the simulation of \citet{Pariat09a}. The data used for the present analysis corresponds to the full 3D velocity and magnetic vector fields. 

The magnetic field presents a finite level of non-solenoidality ($\diverg \bb \neq 0$), unavoidably induced by the discretized dataset, that limits the precision of the helicity computation \citep[cf. Section 7 of][]{Valori16}. 
To quantify the impact of this effect, following \citet{Valori16}, we determine the non-solenoidal energy $E_{ns}$ introduced by \citet{Valori13}. This energy corresponds to the sum of artifact additional contributions, due to finite non-solenoidality of $\bb$, to the Thompson decomposition of energy. As mentioned in \citet{Pariat15b}, the ratio $E_{ns}/E$ of the non-solenoidal energy to the total magnetic energy is lower than $0.1\%$: it highlights the excellent solenoidality of the magnetic field data. Since the non-solenoidality is the largest source of errors in helicity calculation \citet{Valori16}, such a small value supports the excellent precision on the helicity computation.

\subsubsection{Flux emergence simulations} \label{sec:emergence}

\begin{figure*}[!h]
  \centering
			\includegraphics[scale=0.4]{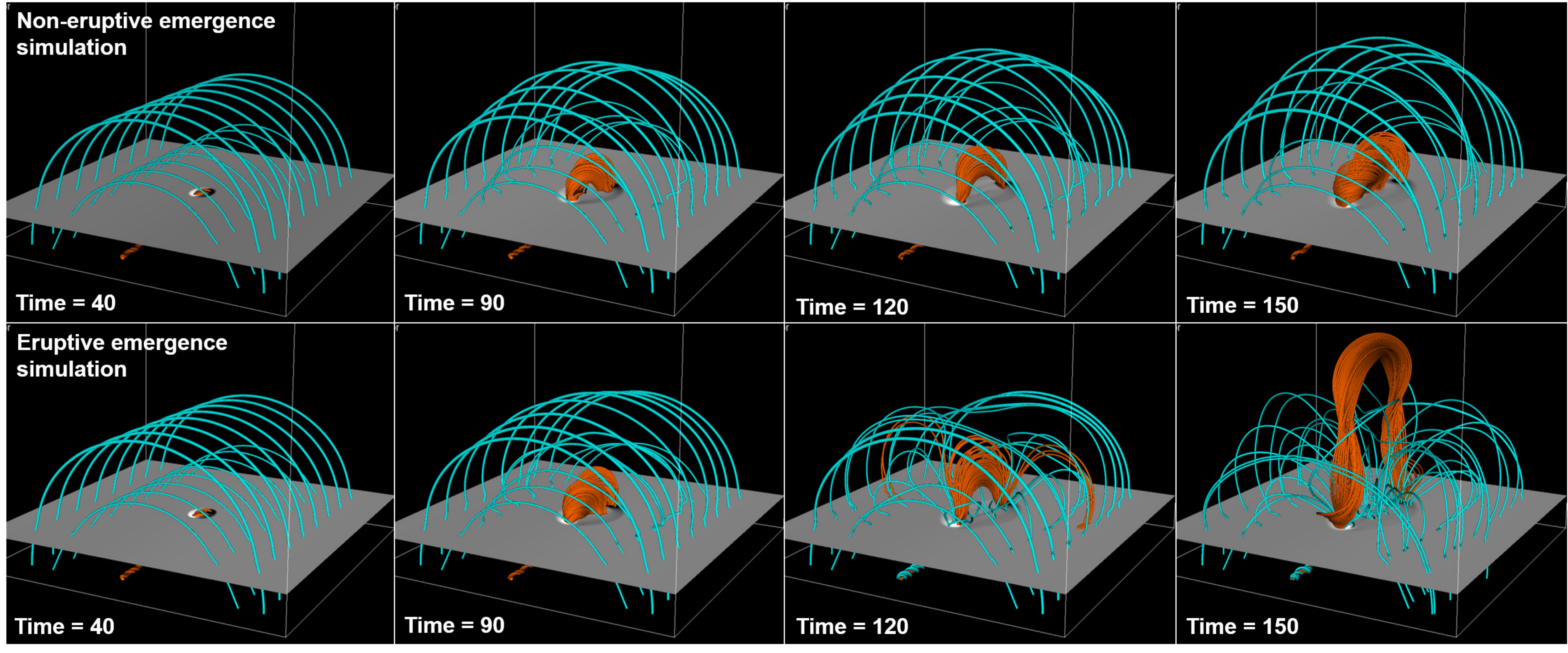}
			\caption{Snapshots comparing the evolution of the magnetic field in the eruptive (bottom row) and non-eruptive (top row) flux emergence simulations. The orange (respectively cyan) field lines initially belong to the emerging flux rope (respectively arcade). The grayscale 2D surface displays the distribution of the magnetic field at the simulated photospheric level. Only the coronal domain, above that boundary, is considered in the present study.}
			\label{eruption}
\end{figure*}

The next two tests employed in our study are based on the 3D visco-resistive MHD numerical experiments of \citet{Leake13b,Leake14a}. The two simulations considered here present the emergence of the same twisted magnetic flux rope from the upper convection zone into the stratified solar atmosphere (cf. Figure \ref{eruption}). In the coronal domain, a constant value for the resistivity is assumed, $\eta=0.01$. These simulations have also been analysed by \citet{Valori16}, \citet{Guennou17} and \citet{Pariat17}. 

For our present analysis, the flux rope emerges at the bottom boundary, the minimum temperature region emulating the solar photosphere, and forms a solar-like bipolar active region. The simulations last between $t=0$ and $t=200$ in the non-dimensional units of the system \citep[cf.][ for more details]{Leake13b}. The flux rope pierces through the photospheric boundary at $t=30$. Before that time, the flux rope is moving trough the convection zone, a domain not considered in our study. 

The same initial twisted magnetic flux rope emerges into a potential coronal field presenting an arcade structure of different intensity and orientation. In \citet{Leake13b} the magnetic arcade is parallel with the top magnetic field of the emerging flux rope. In this set-up, the emerging flux rope remains contained within the coronal domain and stabilises itself. No impulsive ejection of magnetic structure is observed. In the present paper, we analyse the particular simulation obtained with a medium intensity field arcade \citep[MD case in][]{Leake13b} , and we designate it as "non-eruptive emergence" throughout this manuscript \footnote{This simulation was noted "No Erupt MD" in \citet{Pariat17}.}. 

The second flux emergence simulation that we analyse has an opposite direction of the arcade field compared to the first simulation \citep[MD case in][]{Leake14a}. We refer to this simulation as "eruptive emergence" \footnote{It was noted as "Erupt MD" in \citet{Pariat17}.}. In this geometry, magnetic reconnections between the emerging flux rope and the surrounding arcade field are favoured. The emergence of the flux rope eventually induces an eruption: a secondary flux rope is formed, following the emergence, that eventually becomes unstable and is impulsively ejected upward toward the top boundary. The eruption develops between $t\sim120$ and $t\sim150$, a period during which the axis of the flux rope presents a high upward velocity before eventually leaving the numerical domain. During the post-eruptive phase, the system relaxes toward a stable configuration.
 
 The original simulations are performed on a 3D irregular Cartesian mesh. Only the coronal domain of the magnetic and velocity fields data are extracted and are remapped onto a 3D uniform Cartesian grid using trilinear interpolation. The analysed-volume range is: $x\in [-100,100]$, $y\in [-100,100]$ and $z\in [0.36,150]$ in the non-dimensional units of \citet{Leake13b,Leake14a}. While the domain analysed is the same as in \citet{Valori16,Pariat17}, the interpolation is performed on a grid with a higher resolution. The number of pixels is $1.5$ times larger in each direction compared to the previous helicity analyses. The grid is composed of $311$ mesh points along both the horizontal directions, $x$ and $y$, and $232$ along the vertical direction, $z$.

As with the jet simulation, the divergence of $\textbf{B}$ is not exactly null, and the energy decomposition presents a finite value of $E_{ns}$. The value of the ratio $E_{ns}/E$ is lower than $1.5\%$ once the flux rope starts to emerge. Thus the level of non-solenoidality remains low, ensuring a good reliability in our calculation of the helicity, as follows from the tests performed in \citet{Valori16}.

\subsection{Volume helicity estimation} \label{sec:helicity_estimation}

The output of the different simulations are datacubes, on a
 uniform cuboid Cartesian grid, of the magnetic field $\textbf{B}$,
 the plasma-velocity field, $\textbf{v}$ and of the plasma
 thermodynamical quantities (not used here). From the 3D magnetic
 field, the different magnetic helicities, $H_{\mathrm{v}}$, $H_{\mathrm{j}}$ and
$H_{\mathrm{pj}}$, can be directly computed at each time step. 

In the present study, we adopt the method of \citet{Valori12}: at each
time-step of the simulations, from $\textbf{B}(t)$ we compute its
respective potential magnetic field, $\textbf{B}_p(t)$, from its
scalar potential $\phi(t)$, which is obtained from a numerical solution of the Laplace Equation (\ref{eq:laplace}). 

Then, the respective potential vectors $\textbf{A}(t)$ and
$\textbf{A}_{\mathrm{p}}(t)$ can be computed with a few 1D integrations
\citep[cf. Equation (14) of][]{Valori12}. This method assumes that the potential vectors satisfy the DeVore gauge \citep{DeVore00}, in which
their vertical component is null at every instant and every point in the domain, i.e.:
\begin{equation}
A_{z}(x,y,z,t)=A_{p,z}(x,y,z,t)=0
\end{equation}
This condition does not uniquely define the vector
potential. Following \citet{Pariat15b, Pariat17},
respectively for the jet and the emerging flux simulations, 
our numerical computation of $\textbf{A}_{\mathrm{p}}$ enforces the Coulomb gauge ($\nabla \cdot \avecp = 0$). Following \citet{Pariat15b}, this gauge choice is referred to the DeVore-Coulomb gauge. 
Additionally, the 1D-integration starts at the top boundary of the domain (at height $z_{top}$), 
where we enforce the relation:
\begin{equation}
A(x,y,z=z_{top},t)_\perp=A_{\mathrm{p}}(x,y,z=z_{top},t)_\perp \label{eq:devore}
\end{equation}

Finally the helicities are obtained from this volume integration
method using Equations (\ref{eq:h}), (\ref{eq:hj}) and (\ref{eq:hpj}).
As mentioned earlier, the resulting helicity shows a residual gauge dependence due to the finite
level of solenoidality of $\textbf{B}$ unavoidably induced by the
discretized dataset. In our simulations, the errors introduced are
minimal and do not affect the helicitiy estimations, thanks to the low value of $E_{ns}$, (cf. Section \ref{sec:num_test}).

In Figure \ref{helicity}, we plot the time evolution of the three helicities of our study for
the different simulations. We note that for the non-eruptive
flux-emergence simulation, the relative helicity of the system is dominated by
the $H_{\mathrm{pj}}$ component. For the eruptive flux-emergence simulation,
the helicity is dominated by $H_{\mathrm{j}}$ before the onset of the eruption
(before $t\sim 120$), and then by $H_{\mathrm{pj}}$ once the system is relaxing
after the eruption. This property has been studied in
\citet{Pariat17} and \citet{Zuccarello18}, who showed that the ratio $H_{\mathrm{j}}/H_{\mathrm{v}}$ was a very
good marker of the eruptivity of these simulations. 

We observe for the
jet simulation that, while in the very beginning of the energizing,
the relative helicity is mostly constituted by $H_{\mathrm{pj}}$ (for $t<400$),
the system becomes dominated by $H_{\mathrm{j}}$, between $t\sim400$ and $t\sim
900$, up until the onset of the jet. After the jet formation, $H_{\mathrm{j}}$ decreases
substantially and the system is on average dominated by $H_{\mathrm{pj}}$. This
simulation thus seems to confirm the results of \citet{Pariat17} and \citet{Zuccarello18}: the
ratio $H_{\mathrm{j}}/H_{\mathrm{v}}$ appears closely related to the eruptivity in the
system: this ratio is low at the start of the energizing, where the
system is departing from a potential configuration, becomes high before
the onset of the eruptive behavior and drops to a low value after the
eruption, when the system relaxes to a stable state. 

\subsection{Time variation estimations} \label{sec:variation_estimation}%

One goal of the present study is to verify numerically our
analytical derivation of the time variations of the non-potential and volume threading helicities (Equations \ref{eq:hj} and \ref{eq:hpj}). From three successive outputs of the studied MHD
system, corresponding to three instants separated by
a time interval $\Delta t$, we directly compute their helicity
variation rate for the instant $t$ :
\begin{equation} \label{eq:Delta}
\frac{\Delta H_{\mathrm{x}}(t)}{\Delta t}=\frac{H_{\mathrm{x}}(t+\Delta t)-H_{\mathrm{x}}(t-\Delta t)}{2\Delta t}
\end{equation}
with $x$ standing either for $j$ or $pj$.

Along with the estimation of the helicities in the volume and their
time differentiation ($\Delta H_{\mathrm{j}}/\Delta t$ and $\Delta
H_{\mathrm{pj}}/\Delta t$, cf. Section \ref{sec:helicity_estimation} ), we also evaluate their instantaneous time variation, $\text{d}H_{\mathrm{j}}/\text{d}t$ and $\text{d}H_{\mathrm{pj}}/\text{d}t$ from
Equations (\ref{eq:dhjdt}) and (\ref{eq:dhpjdt}).

Our analysis is performed within the resistive MHD paradigm. Beside the resistive term, any other non-ideal contributions are not treated by the numerical solvers in the simulations. For our numerical computation of the helicity variations terms, the non-ideal effects are thus limited to the inclusion of the dissipation term in Equations (\ref{eq:Diss_Aj}), (\ref{eq:Diss_Ap}), (\ref{eq:SNoId_Aj}) and (\ref{eq:SNoId_Ap}). 

In the particular case of the jet simulation, the ideal MHD equations are solved, and the resistivity, $\eta$, is not specified. \citet{Pariat15b} demonstrated that the helicity dissipation is extremely low for that simulation. Consequently, in the following, for the jet simulation we assume that $\eta=0$.

The estimation of these latter quantities requires, in addition to the
3D magnetic field, $\textbf{B}$, the knowledge of the velocity field,
$\textbf{v}$, at the boundary of the domain. Moreover, using the magnetic field, the potential magnetic field and the vector potentials are obtained from the
volume estimation procedure (cf. Section \ref{sec:helicity_estimation}). The different terms that appear in the instantaneous time variation
(Equations \ref{eq:dhjdt} and \ref{eq:dhpjdt}) are estimated independently, even
though, as already noted in Section \ref{sec:transf}, most of them are not
independently gauge invariant. For the quantities that correspond to fluxes, the
determination of the surface integrals is calculated systematically as
the sum of the contributions from the six boundaries.

It should be noted that for the emergence simulations, for $t\in[25,30]$, when the top of the flux rope starts to pierce through the
photospheric-like layer, the bottom boundary of the dataset we
analyse, some quantities involve small values, which introduce numerical
errors in our helicity estimations. For example, this can be observed
in Figure \ref{helicity-3} as a small peak in the evolution of 
$H_{\mathrm{pj}}$ at $t=27$ for the eruptive emergence. This numerical
artefact leads to fluctuations in the estimation of most terms of the
instantaneous time variation, even though the helicities have very small values. To correct this problem, we thus set all helicity terms equal to zero for $t<27$.

\subsection{Numerical validation} \label{sec:num_validation}

In order to validate the analytical derivation of the time variation of the terms in the helicity decomposition (Equations \ref{eq:dhjdt} and \ref{eq:dhpjdt}), we will compare the estimation of the time derivative of the helicity rates obtained from the volume integration method ($\Delta H_{\mathrm{j}}/\Delta t$ and $\Delta H_{\mathrm{pj}}/\Delta t$ Equation (\ref{eq:Delta}) with the estimation of the instantaneous time variation ($dH_{\mathrm{j}}/dt$ and $dH_{\mathrm{pj}}/dt$, Equations (\ref{eq:dhjdt}) and (\ref{eq:dhpjdt})), as discussed in Section \ref{sec:variation_estimation}.

Since the numerical evaluation of both the instantaneous time variation of a quantity, and its time differentiation involve many operations, some numerical errors are always expected when comparing them. Studying the evolution of relative
magnetic helicity, \citet{Pariat15b} found that for the jet simulation,
during the quasi-ideal phase, before the onset of the jet, a relative error of
about $0.001$ exists between the instantaneous rate and the
differentiation of the volume quantity. Here, although using different routines, we obtain the same result for $H_{\mathrm{v}}$ as \citet{Pariat15b}. This previous test suggests that both results are numerically correct, and that these routines can be used for the calculation of the terms in our analytical time variations of $H_{\mathrm{j}}$ and $H_{\mathrm{pj}}$ (Equations \ref{eq:dhjdt} and \ref{eq:dhpjdt}). 
 
As noted in the previous section, since the simulations are \textit{de facto}
 solving the resistive MHD equation, the additional non-ideal contributions are not considered with this paradigm. In order to numerically validate our analytical formula and quantify
 the correspondence between the left and right hand terms of Equations (\ref{eq:dhjdt}) and (\ref{eq:dhpjdt}), we evaluate the differences:

\begin{eqnarray} \label{eq:Dn}
D_{\mathrm{n,x}}&=&\frac{\Delta H_{\mathrm{x}}}{\Delta t}-\frac{\text{d}H_{\mathrm{x}}}{\text{d}t}\end{eqnarray}
where $x$ corresponds to either $j$ or $pj$. 

\begin{figure*}[!hb]
  \subfigure{\label{error_noerupt-1} \includegraphics[width=9.4cm]{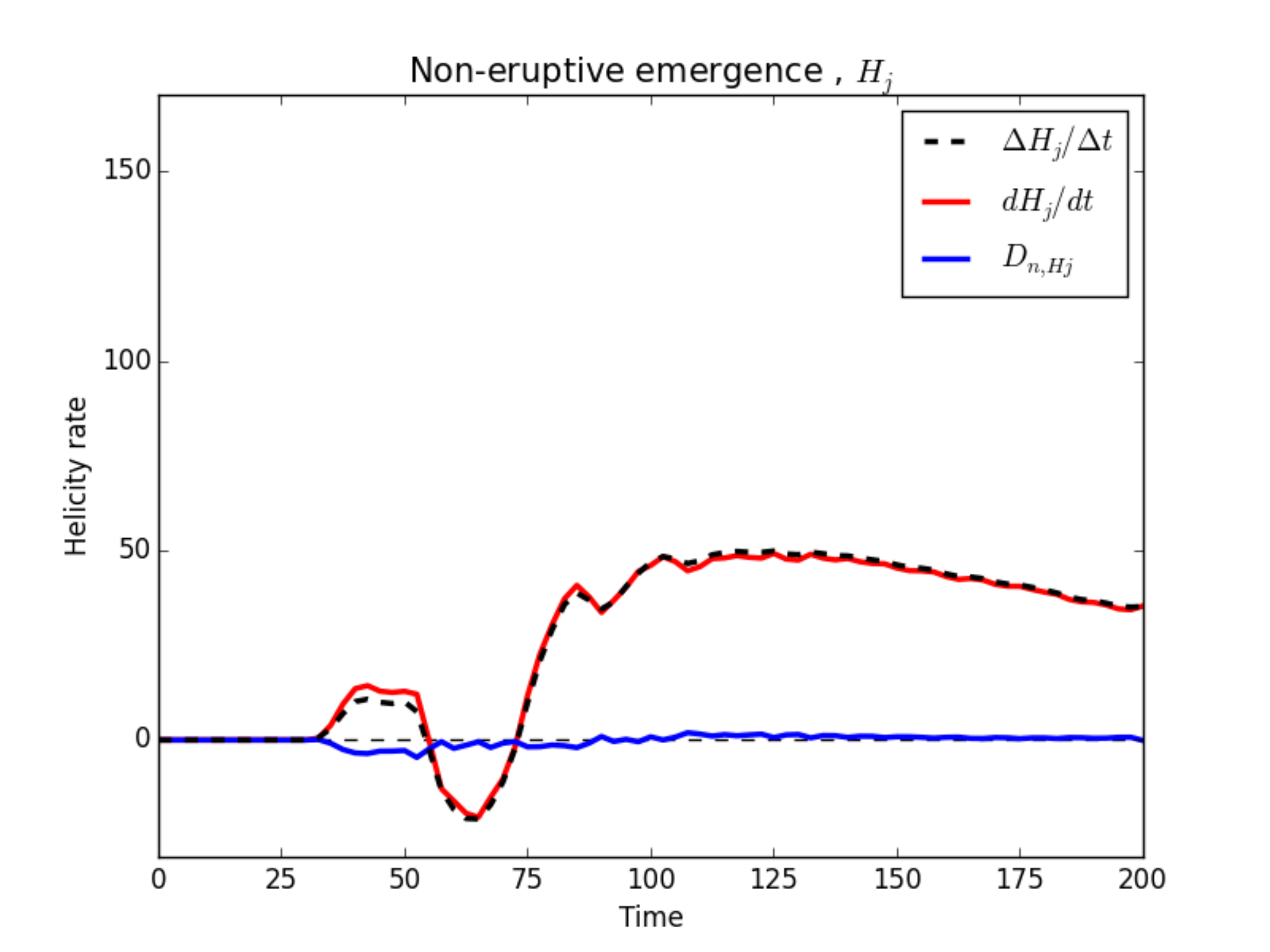}}
  \subfigure{\label{error_noerupt-2} \includegraphics[width=9.4cm]{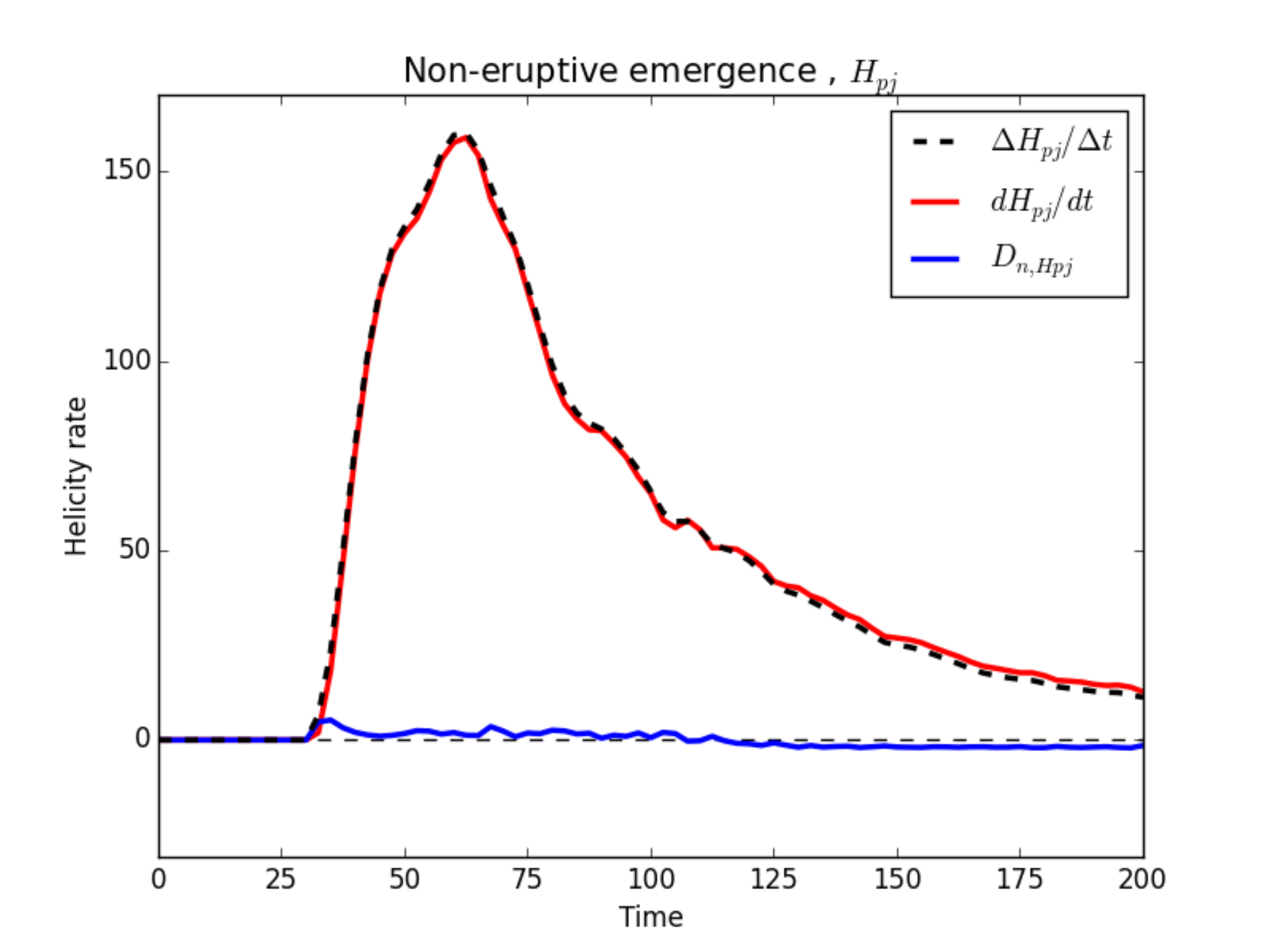}}
  \caption{Time evolution of the helicity variation rates, $\Delta H_{\mathrm{j}}/\Delta t$ and $\Delta H_{\mathrm{pj}}/\Delta t$ in (dashed black curves, Equation \ref{eq:Delta}), the instantaneous time variations $\text{d}H_{\mathrm{j}}/\text{d}t$ and $\text{d}H_{\mathrm{pj}}/\text{d}t$ (continuous red curves, Equations \ref{eq:dhjdt} and \ref{eq:dhpjdt}), and the differences of these quantities, $D_{\mathrm{n,j}}$ and $D_{\mathrm{n,pj}}$ (continuous blue curves, Equation \ref{eq:Dn}) for the non-eruptive flux-emergence. The left and right panels present the evolution of the non-potential helicity, $H_{\mathrm{j}}$, and volume-threading helicity, $H_{\mathrm{pj}}$, respectively}
  \label{error_noerupt}
\end{figure*}

\begin{figure*}[!h]
  \subfigure{\label{error_erupt-1} \includegraphics[width=9.4cm]{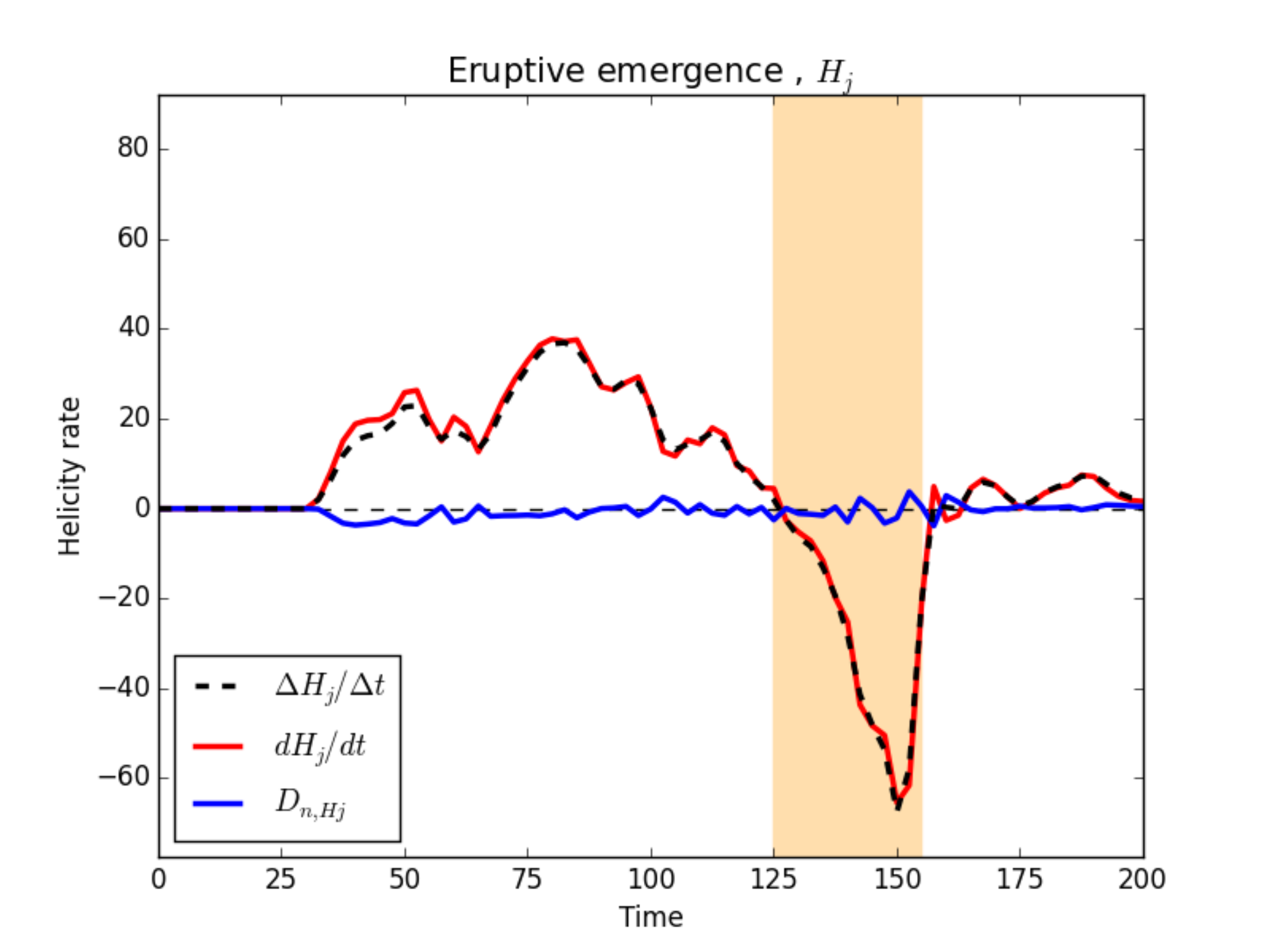}}
  \subfigure{\label{error_erupt-2} \includegraphics[width=9.4cm]{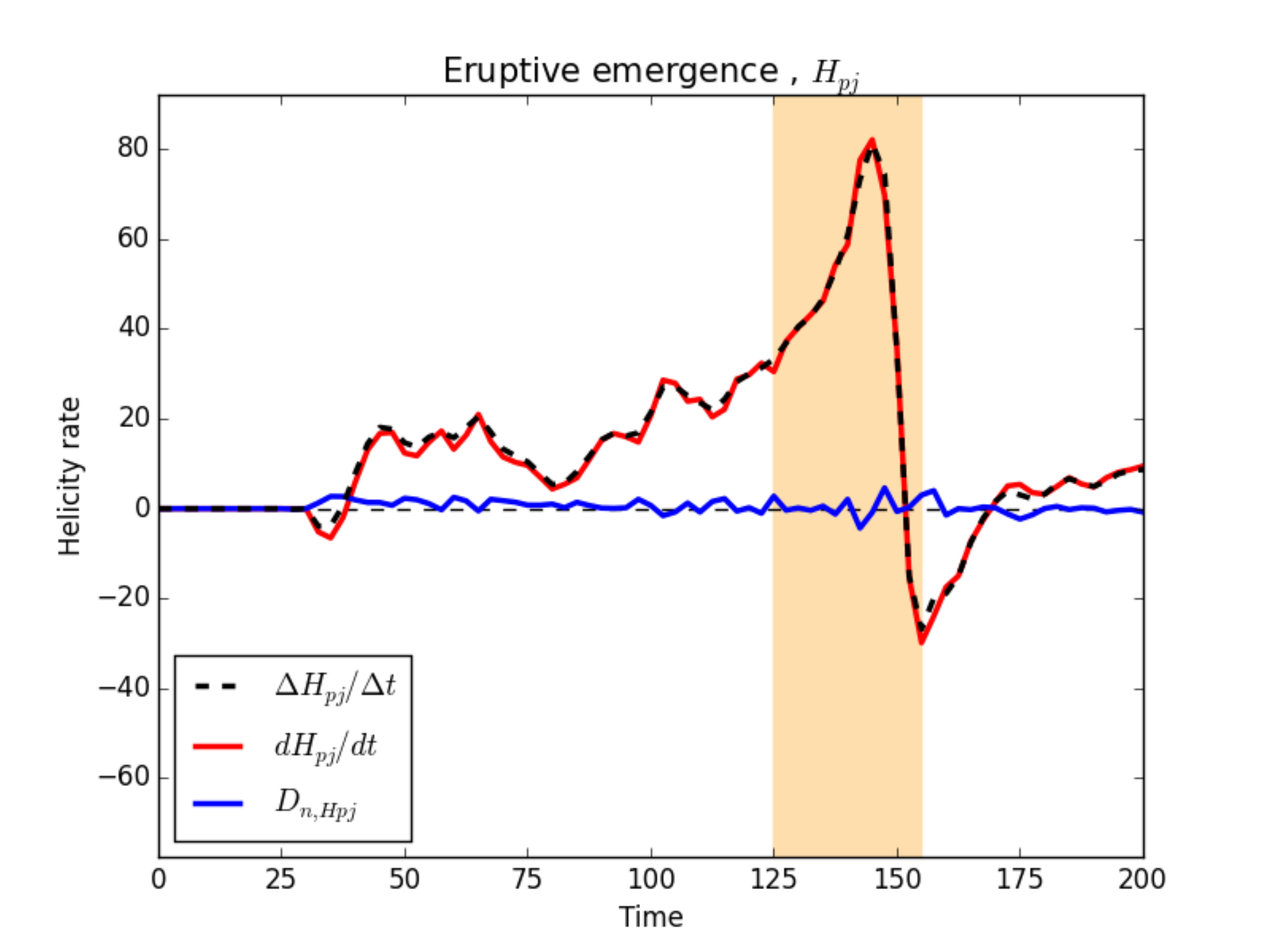}}
  \caption{Same as Figure \ref{error_noerupt} for the eruptive flux-emergence simulation. The yellow bands correspond to the eruptive phase.}
  \label{error_erupt}
\end{figure*}

\begin{figure*}[!h]
  \subfigure{\label{error_jet-1} \includegraphics[width=9.4cm]{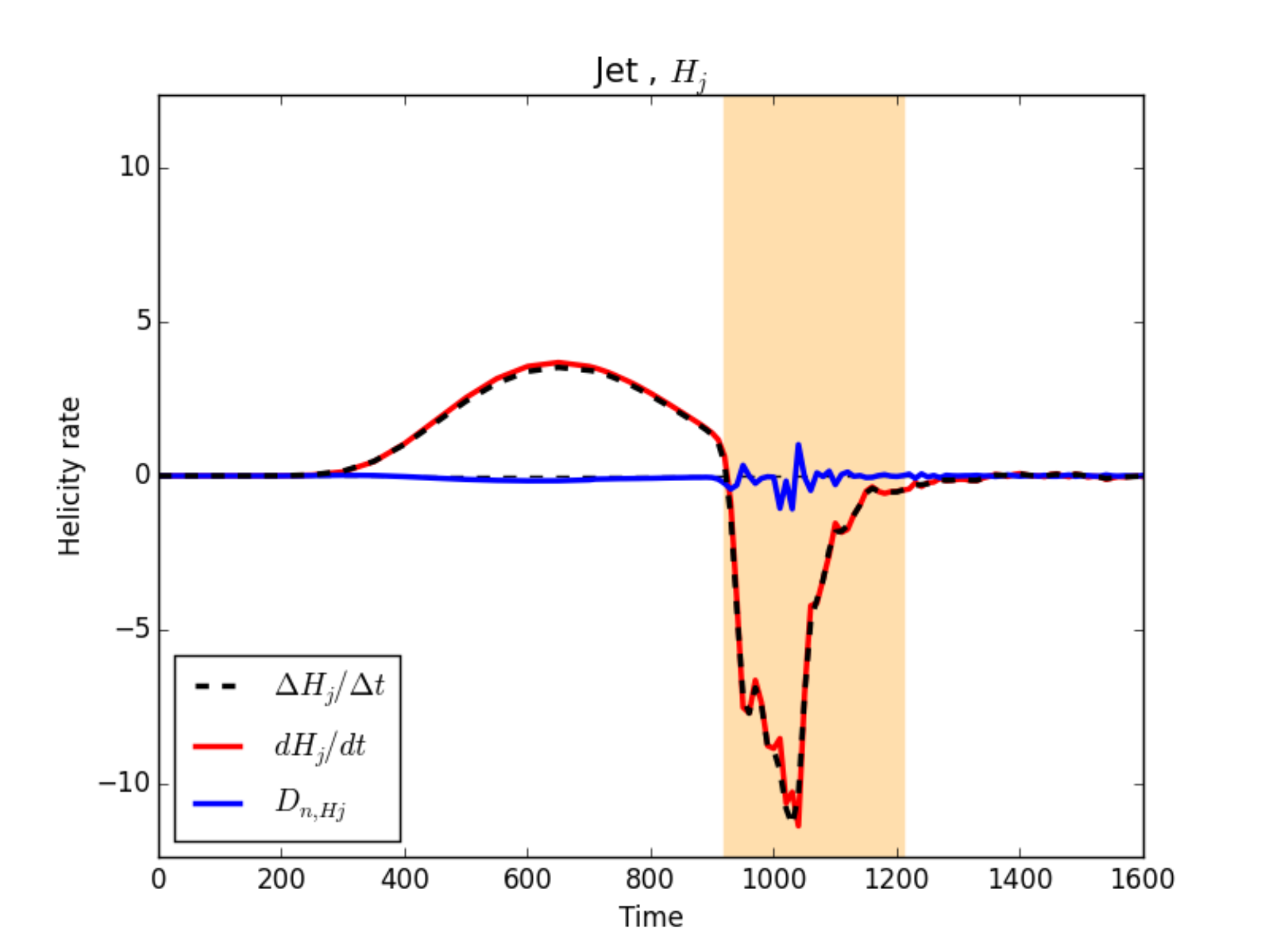}}
  \subfigure{\label{error_jet-2} \includegraphics[width=9.4cm]{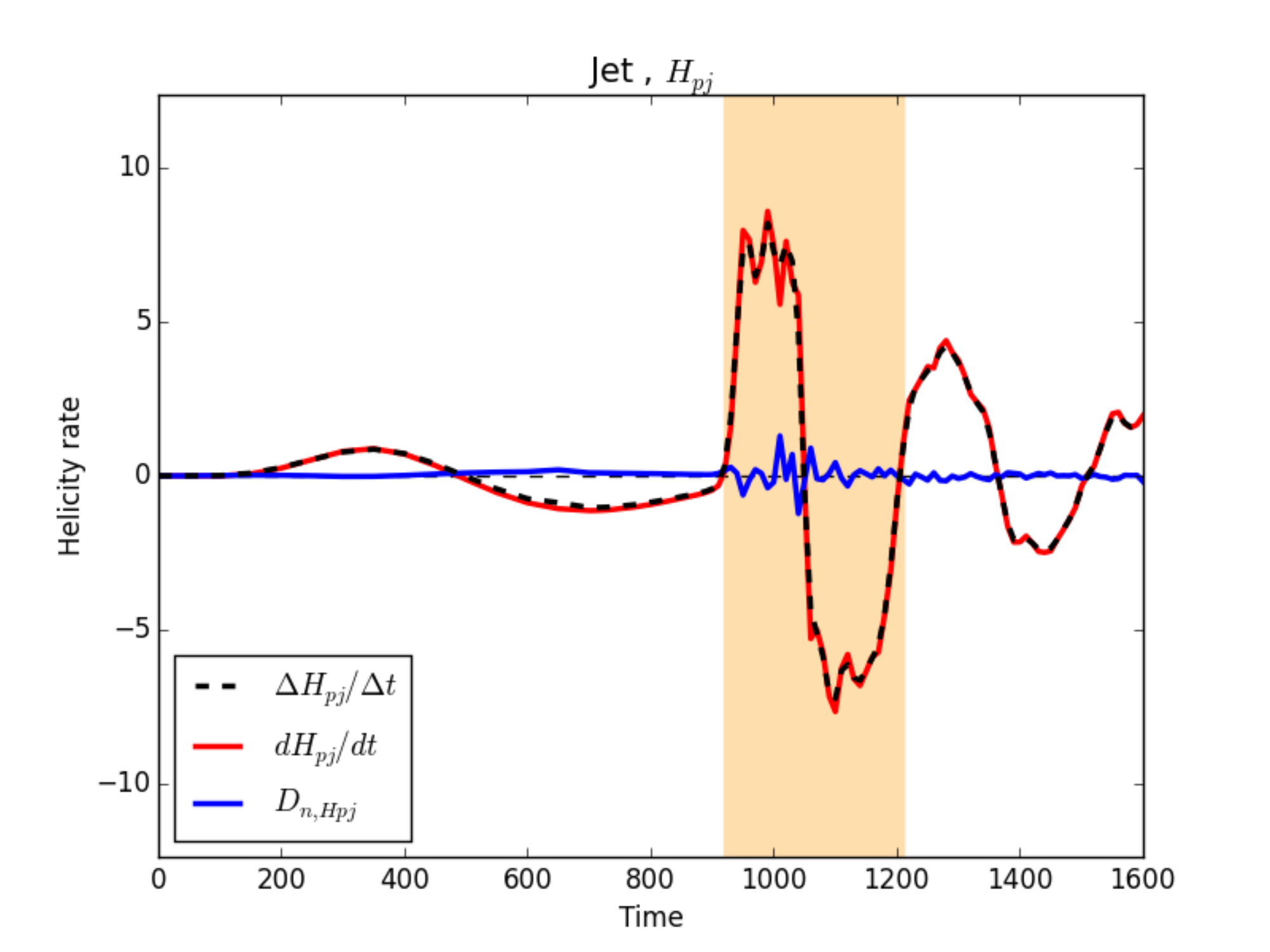}}
  \caption{Same as Figure \ref{error_noerupt} for the coronal jet simulation. The yellow bands correspond to the eruptive phase.}
  \label{error_jet}
\end{figure*}

Figures \ref{error_noerupt}, \ref{error_erupt} and \ref{error_jet} present the evolution of the time derivative of
the volume helicities, the instantaneous variation rate and
their difference. Figure \ref{error_jet} we note that during the quasi-ideal phase, before the
onset of the jet, $D_{\mathrm{n,x}}$ remains extremely small,
similarly to what has been obtained for the relative helicity
$H_{\mathrm{v}}$. The mean ratio of $|D_{\mathrm{n,j}}|$ to $|dH_{\mathrm{j}}/dt|$ is lower than $3\%$, and less than $7\%$ for $|D_{\mathrm{n,pj}}|$ to $|dH_{\mathrm{pj}}/dt|$
during that phase. Dissipation being null in ideal MHD, this very low
value is coherent with the theory and shows that numerical errors are
also very low. The low values of $|D_{\mathrm{n,x}}|$ confirms the validity of Equations (\ref{eq:dhjdt}) and (\ref{eq:dhpjdt}) about the helicity time variations, which constitute the central result of this study.
 
During the non-ideal phase of the jet simulation, strong non-ideal effects are present with intense magnetic reconnections. While the $D_{n}$ differences present larger values than
during the quasi-ideal phase, mostly through the form of fluctuating
peaks, their intensity is very weak in comparison with the values of $dH_{\mathrm{x}}/dt$. During the eruption phase the mean ratio of
$|D_{\mathrm{n,j}}|$ to $|dH_{\mathrm{j}}/dt|$ is also about
$12\%$, and $6\%$ for $|D_{\mathrm{n,pj}}|$ to $|dH_{\mathrm{pj}}/dt|$. Equations (\ref{eq:dhjdt}) and (\ref{eq:dhpjdt}) are thus valid to a
very high degree of accuracy.

As mentionned in Section \ref{sec:variation_estimation}, for the flux-emergence simulations, we are able to calculate non-ideal effects related to the resistivity. Thus, $D_{\mathrm{n,x}}$, presented in Figures \ref{error_noerupt} and \ref{error_erupt}, respectively for the, eruptive and the non-eruptive simulations. is almost entirely associated with numerical errors.

The difference between
$\Delta H_{\mathrm{x}}/\Delta t$ and $dH_{\mathrm{x}}/dt$ remains
relatively low with respect to the values reached by these
quantities. There is an exception however, during the very initial
phase of the flux emergence, both for the eruptive and non-eruptive
cases, during which the curves of $\Delta H_{\mathrm{j}}/\Delta t$ appear
distinct from the ones of $dH_{\mathrm{j}}/dt$. 
For $t\in[25,60]$, we note a high value of $D_{\mathrm{n,j}}$
for $H_{\mathrm{j}}$. This effect is more
particularly pronounced for the time variation of $H_{\mathrm{j}}$. During this
period the mean ratio of $|D_{\mathrm{n,j}}|$ to $|dH_{\mathrm{j}}/dt|$ is $14\%$ for both simulations.

We hypothesize that this value of $D_{\mathrm{n,x}}$ during that period is due
to a temporal under-sampling due to relatively low-cadence data available.
This phase indeed corresponds to the moment when
the flux rope emerges into  the coronal domain though the model photosphere, when the 
bottom boundary in our calculation thus exhibits relatively fast changes.

Neglecting this initial phase, one observes that the values of
$D_{\mathrm{n,x}}$ remain overall significantly smaller than the values of
$dH_{\mathrm{x}}/dt$. The mean ratios of
$|D_{\mathrm{n,j}}|$ to $|dH_{\mathrm{j}}/dt|$ and $|D_{\mathrm{n,pj}}|$ to $|dH_{\mathrm{pj}}/dt|$ are lower than $9\%$ for the eruptive emergence, and lower than $5\%$ for the non-eruptive simulation.

From the analysis of these three numerical simulations, we conclude that our estimation of
$\text{d}H_{\mathrm{x}}/\text{d}t$ enables a consistent
evaluation of the helicity, and that Equations (\ref{eq:dhjdt}) and (\ref{eq:dhpjdt}) for the time
variation of $H_{\mathrm{j}}$ and $H_{\mathrm{pj}}$ are satisfactorily numerically verified.

\subsection{Helicity fluxes} \label{sec:fluxes}
In Figures \ref{flux_hj} and \ref{flux_hpj}, we compare the different
terms which compose the instantaneous time variation of $H_{\mathrm{j}}$ (Equation \ref{eq:dhjdt}) and $H_{\mathrm{pj}}$ (Equation \ref{eq:dhpjdt}), respectively.
As described in Section \ref{sec:transf}, only the transfer term, the dissipation term and the sum of the fluxes, $\left.\text{d}H_{\mathrm{x}}/\text{d}t\right|_{\mathrm{Own}}$ 
are gauge invariant. The fluxes, $F_{\mathrm{\alpha,\, Ap}}$ and $F_{\mathrm{\alpha,\, Aj}}$ (with $\alpha\in\{Vn,Bn,Aj,\phi,Non-ideal\}$), are gauge dependent. Therefore, computed with different gauges, the curves in Figures
\ref{flux_hj} and \ref{flux_hpj} might be noticeably different, e.g. as shown in Appendix \ref{app:A}. As the individual terms nonetheless inform on the dynamics of the helicity variations within the adopted gauge, we briefly present their evolution here. 

In Figures \ref{flux_hj} and \ref{flux_hpj}, $\left.\text{d}H_{\mathrm{x}}/\text{d}t\right|_{\mathrm{Transf}}$, which is a volume contribution, is not small compared to the other fluxes. It confirms that $H_{\mathrm{pj}}$ and $H_{\mathrm{j}}$ are not conserved quantities. In other words their time variations can not be written as only the sum of surface contributions. We also see in Figure \ref{flux_hj}, that for the three simulations, the curves of $\Delta H_{\mathrm{j}}/\Delta t$ are frequently overlaid by $\left.\text{d}H_{\mathrm{j}}/\text{d}t\right|_{\mathrm{Transf}}$. The transfer term seems to be essential to understand the evolution of helicities. In particular for the jet simulation (cf. Figure \ref{flux_hj}, right panel) where the other terms are negligible compared to $\left.\text{d}H_{\mathrm{j}}/\text{d}t\right|_{\mathrm{Transf}}$. This term will thus be studied in more detail in the following section (cf. Section \ref{sec:num_transf}).

Unlike the jet simulation, the flux-emergence simulations do not possess a quasi ideal phase and a finite level of non-ideality is present as soon as the emerging flux rope enters in the coronal domain. This can be seen with non-null values of $\left.\text{d}H_{\mathrm{x}}/\text{d}t\right|_{\mathrm{Diss}}$ and $F_{\mathrm{Non-ideal}}$ presented in Figures \ref{flux_hj} and \ref{flux_hpj} during the whole simulation. As expected, the intensity of these terms is very low. In other words, non-ideal effects seem to have only a small impact on the temporal evolution of helicities.

A difference between the emergence simulations and the jet simulation is the mechanism by which magnetic energy and helicity accumulate in the system. In the jet simulation, helicity is inputted at the boundary by a purely rotational motion of a symmetric field distribution that leaves the potential field unchanged. Here symmetry and flow geometry accounts for lots of cancellation. Because of this, all the surface terms $F_{\mathrm{\alpha,\, Aj}}$ are almost null in Figure \ref{flux_hj}, right panel. Only $F_{\phi,\, Aj}$ displays a weak variation during the jet, indicating the sudden load change of the reference magnetic field $\textbf{B}_{\mathrm{p}}$ caused by the reconnections. 

On the contrary, for the emergence simulations, $A_{\mathrm{j}}$ changes during the passage of the flux rope through the bottom boundary. This leads to significant contributions of the different flux terms during the pre-eruptive phase, as displayed by the evolution of the terms $F_{\mathrm{Vn,\, Aj}}$ and $F_{\mathrm{Bn,\, Aj}}$. During the eruption, for the eruptive emergence simulation as with the jet case, the scalar potential $\phi$ changes abruptly, and therefore so are $\left.\text{d}H_{\mathrm{j}}/\text{d}t\right|_{\mathrm{Bp,\, var}}$ and $F_{\phi,\, Aj}$. It is a consequence of the boundary modification conditions for the Laplace equation due to the transit of eruption through the surface. 

The flux $F_{\mathrm{\alpha,\, Ap}}$ presented in Equation (\ref{eq:dhpjdt}) does not depend on $A_{\mathrm{j}}$ but on the vector potential $\textbf{A}_{\mathrm{p}}$. For the emergence simulations, $\textbf{A}_{\mathrm{p}}$ changes significantly in emergence phase, yielding the dominant role of $F_{\mathrm{Bn,\, Ap}}$ and $F_{\mathrm{Vn,\, Ap}}$. 

\subsection{Transfer between $H_{\mathrm{j}}$ and $H_{\mathrm{pj}}$} \label{sec:num_transf}
\begin{figure*}
  \subfigure{\label{invariant_noerupt -1} \includegraphics[width=9.4cm]{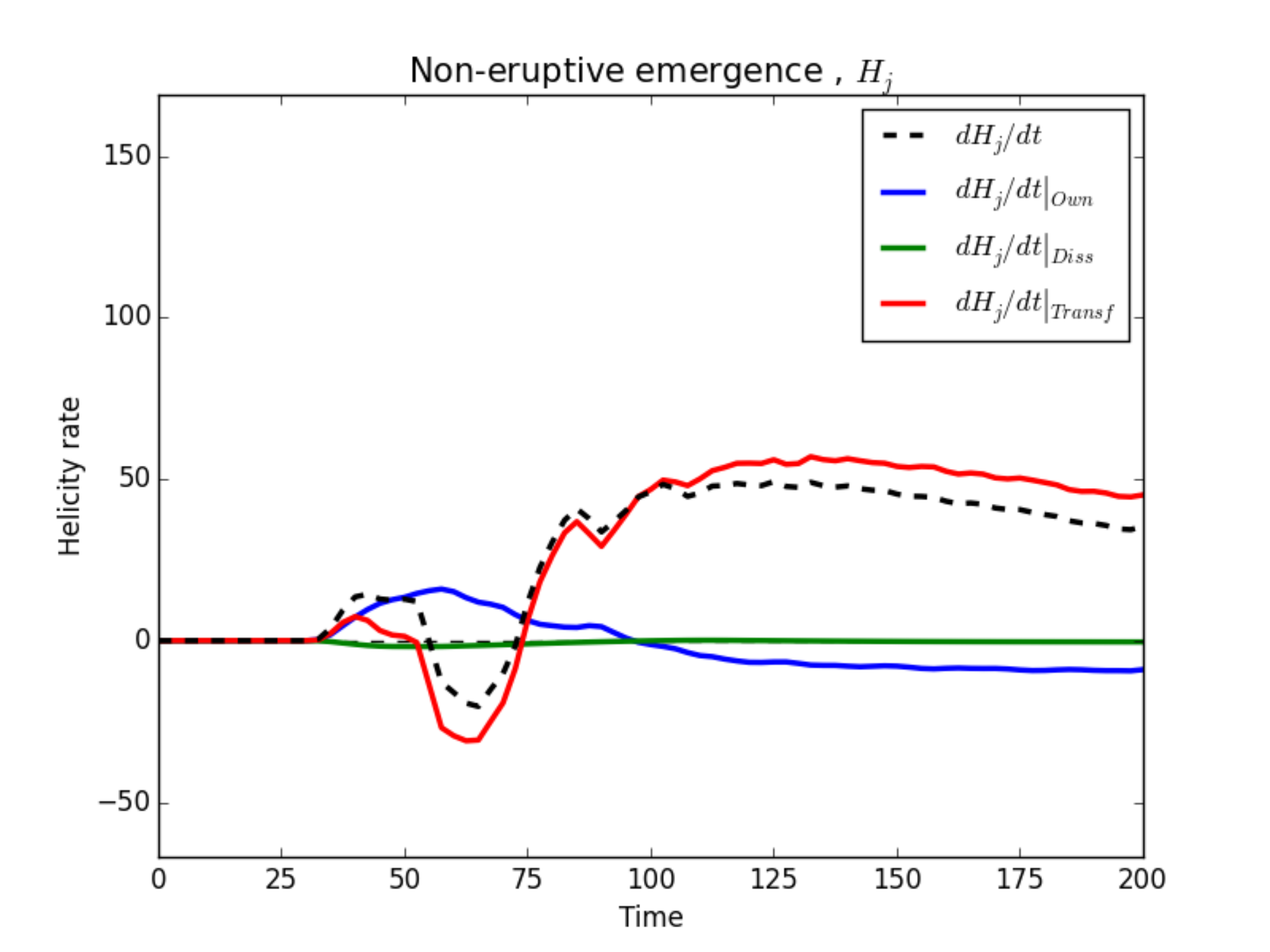}}
  \subfigure{\label{invariant_noerupt -2} \includegraphics[width=9.4cm]{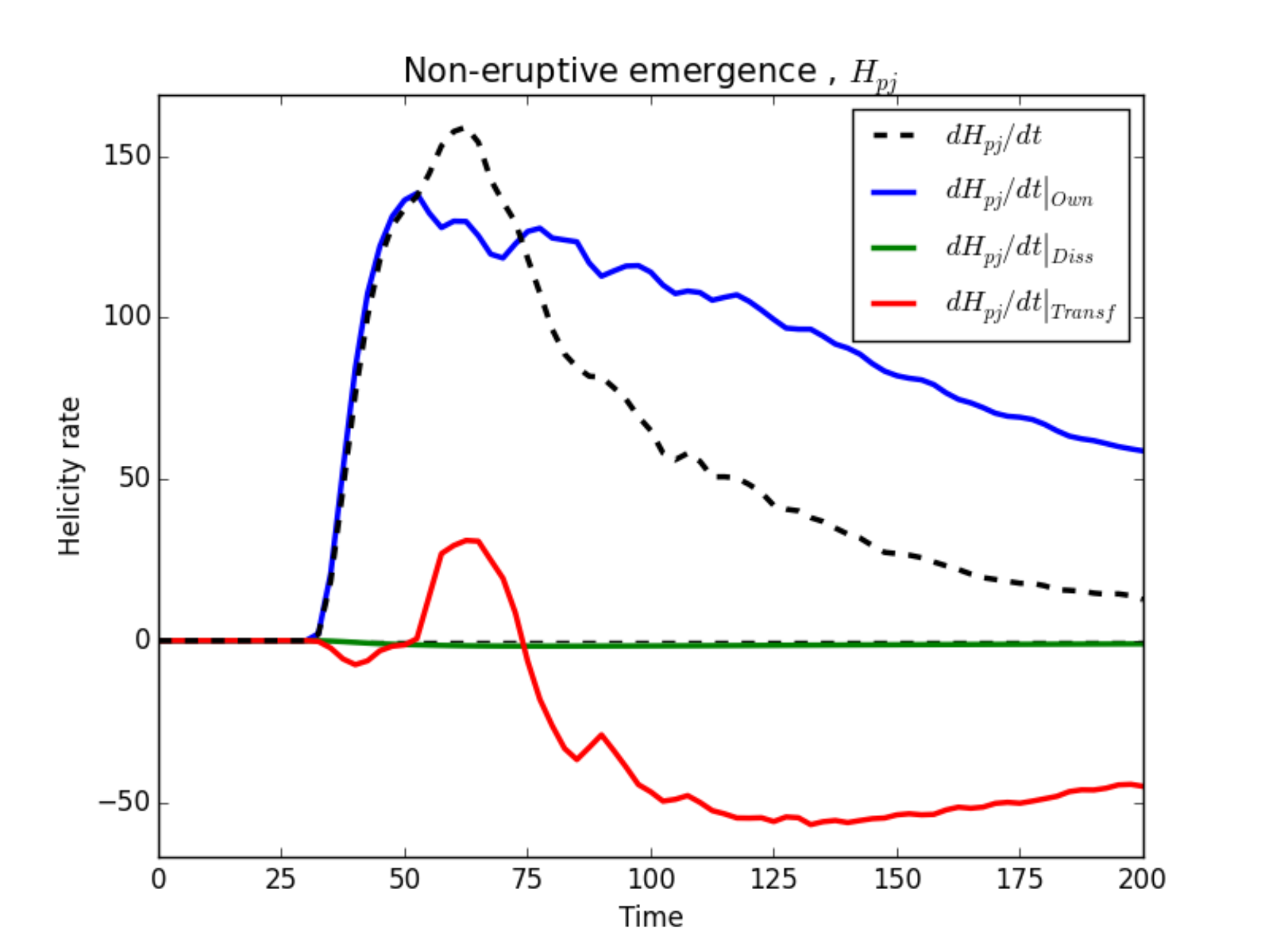}}
  \caption{Time evolution of the helicity variation rates, $\text{d}H_{\mathrm{j}}/\text{d}t$ and $\text{d}H_{\mathrm{pj}}/\text{d}t$ (dashed black curves, Equations \ref{eq:dhjdt} and \ref{eq:dhpjdt}), of the helicity transfer rate, $\left.\text{d}H_{\mathrm{j}}/\text{d}t\right|_{\mathrm{Transf}}$ and $\left.\text{d}H_{\mathrm{pj}}/\text{d}t\right|_{\mathrm{Transf}}$ (continuous red curves, Equations \ref{eq:Ftransf_Aj} and \ref{eq:Ftransf_Ap}), of the own rates, $\left.\text{d}H_{\mathrm{j}}/\text{d}t\right|_{\mathrm{Own}}$ and $\left.\text{d}H_{\mathrm{pj}}/\text{d}t\right|_{\mathrm{Own}}$ (continuous blue curves, Equation \ref{eq:Own_Hj} and \ref{eq:Own_Hpj}), the dissipation rates, $\left.\text{d}H_{\mathrm{j}}/\text{d}t\right|_{\mathrm{Diss}}$ and $\left.\text{d}H_{\mathrm{pj}}/\text{d}t\right|_{\mathrm{Diss}}$ (continuous green curves, Equations \ref{eq:Diss_Aj} and \ref{eq:Diss_Ap}) for the non-eruptive emergence simulation. The left and right panels present the evolution of the non-potential helicity, $H_{\mathrm{j}}$, and volume threading helicity, $H_{\mathrm{pj}}$, respectively. }
  \label{invariant_noerupt}
\end{figure*}

\begin{figure*}
  \subfigure{\label{invariant_erupt-1} \includegraphics[width=9.4cm]{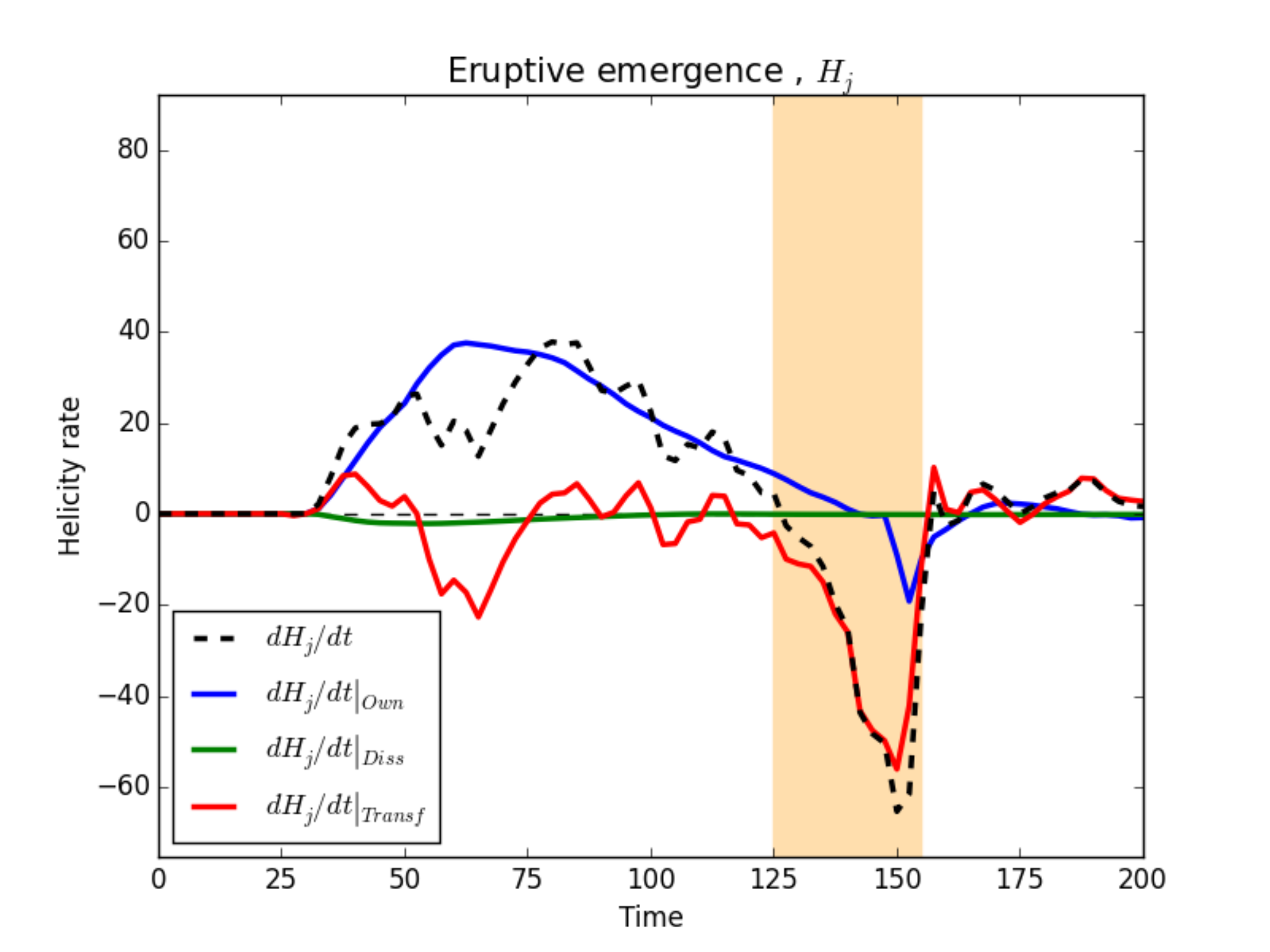}}
  \subfigure{\label{invariant_erupt-2} \includegraphics[width=9.4cm]{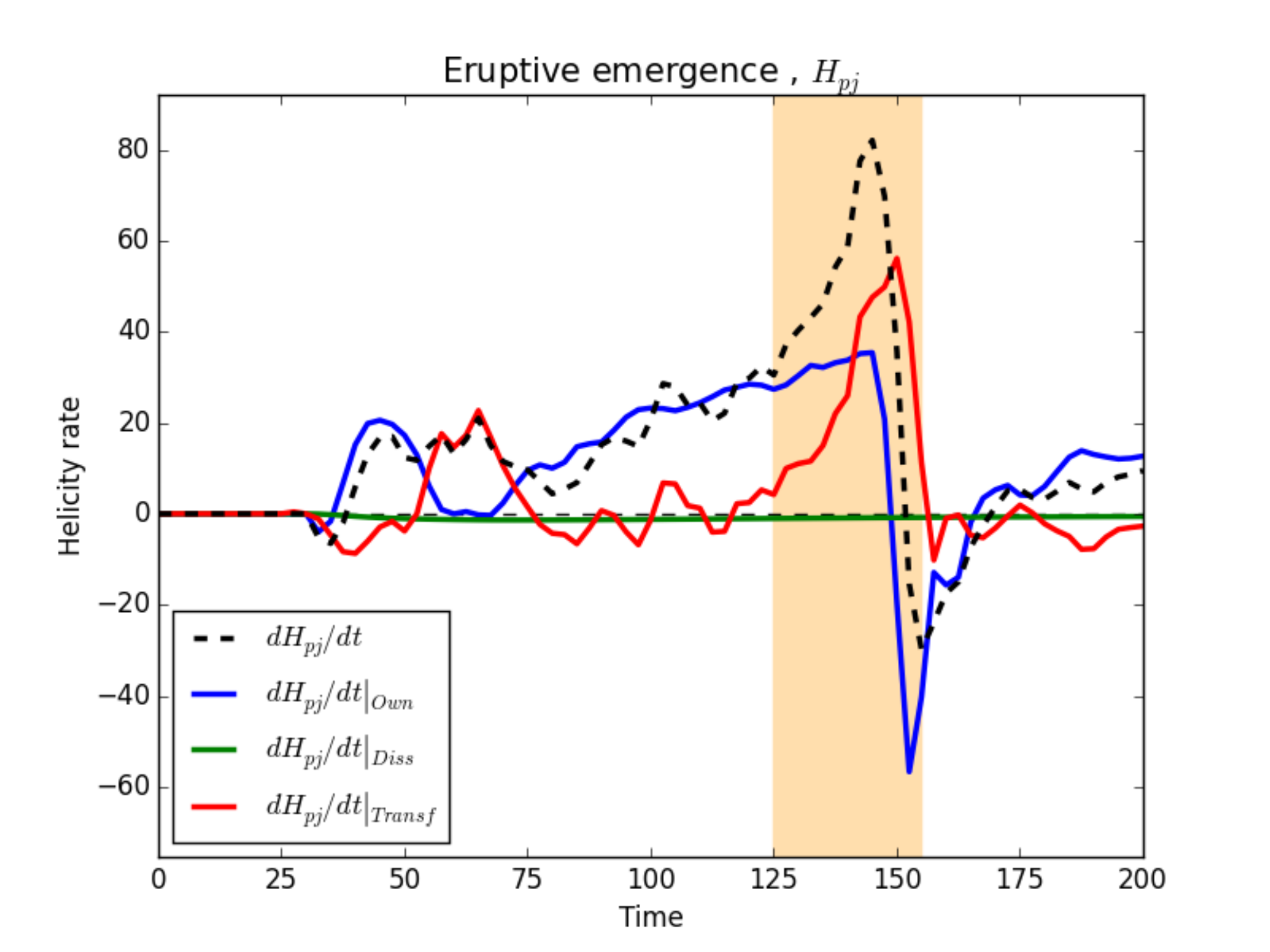}}
  \caption{Same as Figure \ref{invariant_noerupt} for the eruptive flux-emergence simulation. The yellow bands correspond to the eruptive phase.}
  \label{invariant_erupt}
\end{figure*}

Figures \ref{invariant_noerupt}, \ref{invariant_erupt}, and \ref{invariant_jet} show the time variation of $H_{j}$ and $H_{pj}$ in the gauge invariant form of Equations (\ref{eq:dhjdt_gaugeinv}) and (\ref{eq:dhpjdt_gaugeinv}) for the non-eruptive flux-emergence, the eruptive flux-emergence, and the jet simulations respectively. For the emergence simulations, the dissipation $\left.\text{d}H_{x}/\text{d}t\right|_{\mathrm{Diss}}$ which is explicitly calculated, is completely negligible compared to the two other terms. As mentioned in Section \ref{sec:fluxes}, non-ideal terms are not null but their intensity is very low. In the case of the relative magnetic helicity, a weak dissipation was sufficient to conclude that $H_{v}$  is very well conserved \citep{Pariat15b}. This is not the case for $H_{j}$ and $H_{pj}$ because of the transfer term.  Overall, in Figures \ref{invariant_noerupt}, \ref{invariant_erupt}, and \ref{invariant_jet} we observe that the volume transfer terms, $\left.\text{d}H_{x}/\text{d}t\right|_{Transf}$, tend to have important values and frequently dominate the helicity variations. This is the key result of the analysis of these numerical simulations. As discussed earlier (see Section \ref{sec:transf}), these terms are pure volume terms. The presence of the transfer thus confirm that $H_{j}$ and $H_{pj}$ are not conserved quantities, unlike relative magnetic helicity, which is very well conserved for these simulations. Additionally, we observe that the dynamics of the helicity decomposition is often dominated by the exchange between $H_{j}$ and $H_{pj}$, rather than their own fluxes through the volume boundaries. This is particularly true for $H_{j}$ for which the $\left.\text{d}H_{j}/\text{d}t\right|_{Own}$ term is usually unimportant in two out of three cases. This results have important consequences for the estimation of $H_{j}$ in the solar context, as it will be discussed in the conclusion (see Section \ref{sec:Conclusion}). 

\begin{figure*}[!h]
  \subfigure{ \includegraphics[width=6.2cm]{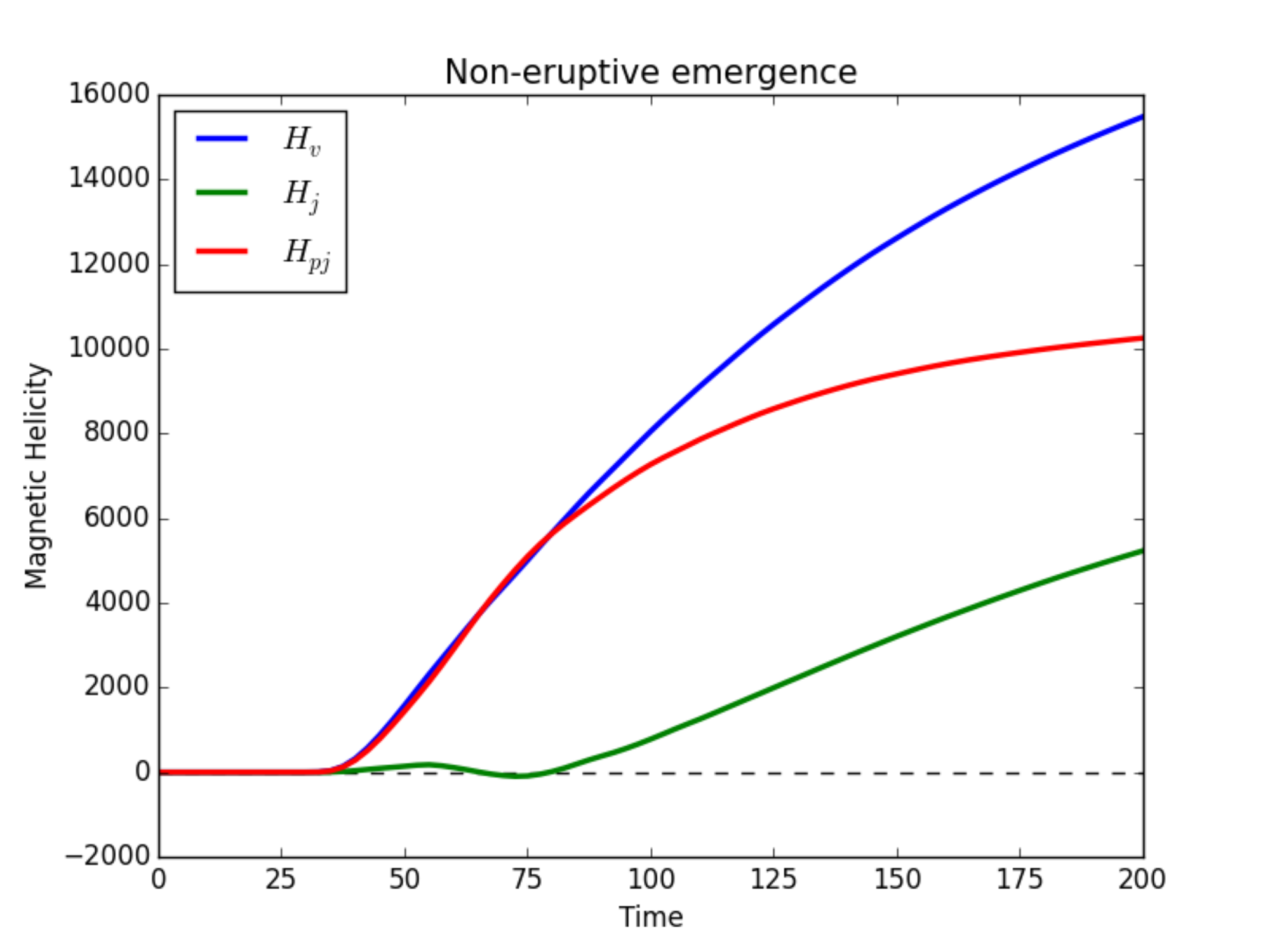}}
  \subfigure{\label{helicity-2} \includegraphics[width=6.2cm]{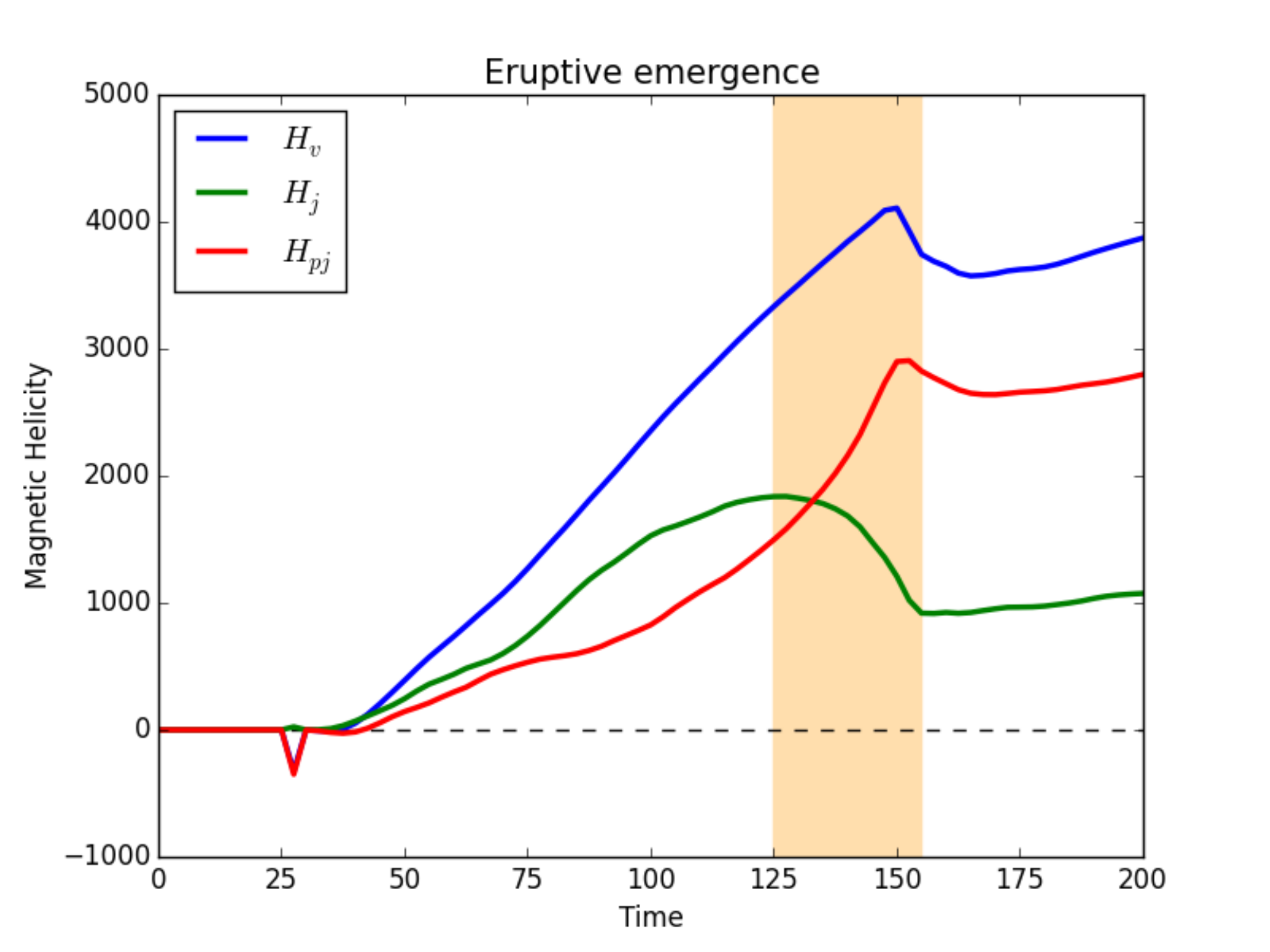}}
  \subfigure{\label{helicity-3} \includegraphics[width=6.2cm]{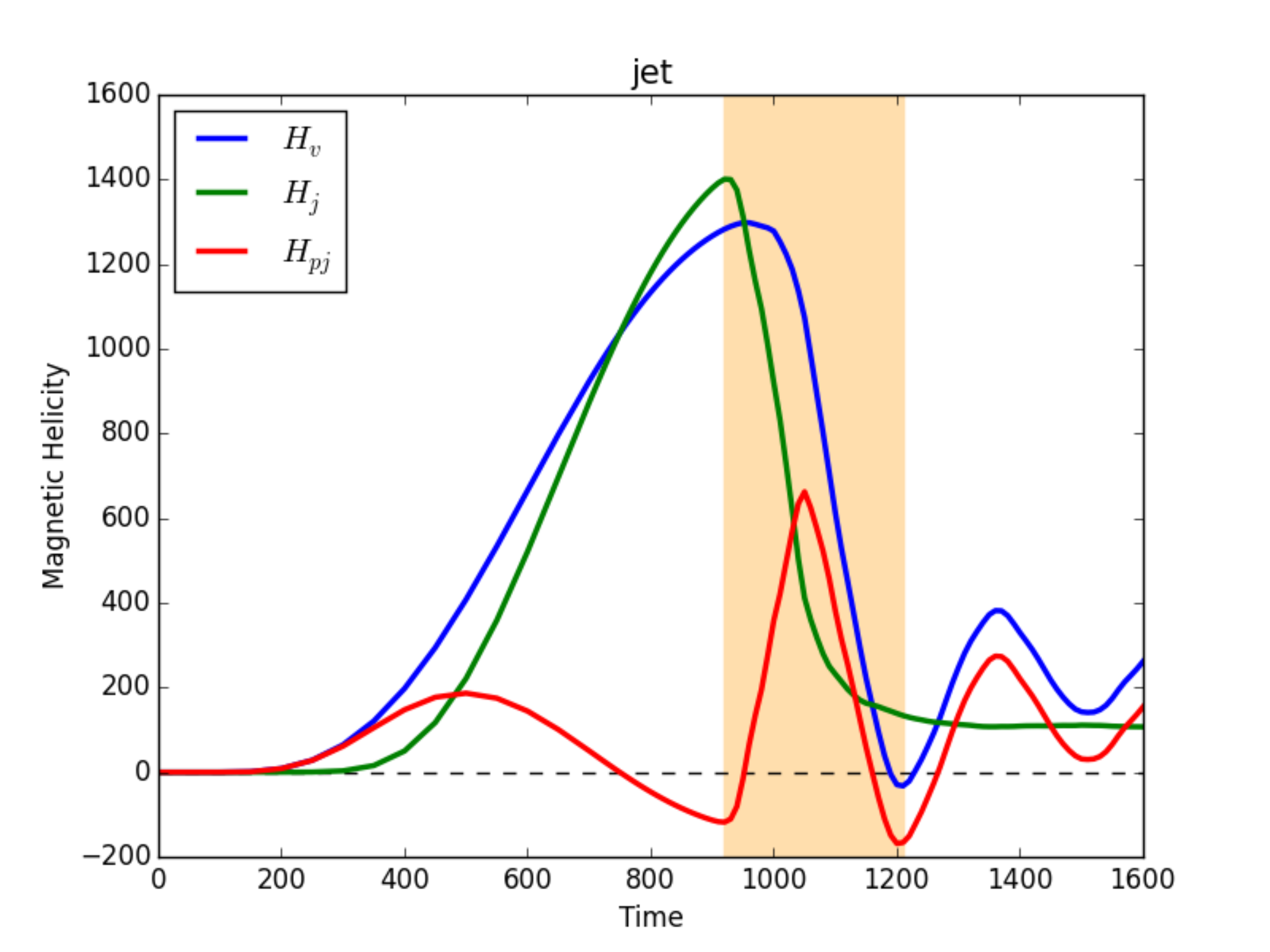}}
  \caption{Time evolution of the different magnetic helicities : the relative magnetic helicity (blue lines, Equation \ref{eq:h}), the volume threading helicity (red lines, Equation \ref{eq:hpj}) and the non-potential magnetic helicity (green lines, Equation \ref{eq:hj}), for the different simulations studied: from left to right, the non-eruptive emergence simulation, the eruptive emergence and the simulation of the generation of a solar coronal jet. For the eruptive emergence and the jet simulations the yellow vertical band corresponds to the eruptive phases. }
  \label{helicity}
\end{figure*}

Figures \ref{invariant_noerupt}-\ref{invariant_jet} along with Figure \ref{helicity} allow us to follow the dynamics of the exchange between $H_{\mathrm{j}}$ and $H_{\mathrm{pj}}$ in the volume, as well as their exchange with the environment, in the different simulations. The non-eruptive emergence case is the most straightforward to analyse. In the left panel of Figure \ref{helicity}, we distinguish two main behaviors: before $t\sim 80$, $H_{\mathrm{v}}$ is increasing along with $H_{\mathrm{pj}}$, and, after $t\sim 80$, $H_{\mathrm{j}}$ starts to grow. The intensity of $H_{\mathrm{pj}}$ is still increasing but is tending to a constant value. For $t\in[25,80]$, corresponding to the emergence of the flux rope, we note in Figure \ref{invariant_noerupt} that $\left.\text{d}H_{\mathrm{pj}}/\text{d}t\right|_{\mathrm{Own}}$ is the predominant flux (in particular due to $F_{\mathrm{Vn,\, Ap}}$ and $F_{\mathrm{Bn,\, Ap}}$ in this gauge, Figure \ref{flux_hpj}). During this period, $H_{\mathrm{j}}$ fluctuates weakly because of the exchange with $H_{\mathrm{pj}}$ (cf. left panel of Figure \ref{helicity}). Figure \ref{invariant_noerupt} (left panel) allows an additional understanding of Figure \ref{helicity} : after $t\sim 80$, $H_{\mathrm{j}}$ is increasing not because of its flux through the surface but only due to the transfer term, $\left.\text{d}H_{\mathrm{j}}/\text{d}t\right|_{\mathrm{Transf}}$. While $\left.\text{d}H_{\mathrm{pj}}/\text{d}t\right|_{\mathrm{Own}}$ (cf. Figure \ref{invariant_noerupt}, right panel) is still dominating the variation of $H_{\mathrm{pj}}$, a large portion of it is directly transferred to $H_{\mathrm{j}}$, thanks to the significant negative values of $\left.\text{d}H_{\mathrm{pj}}/\text{d}t\right|_{\mathrm{Transf}}$. These oppositely signed terms partly balance each other and explain why the rise of $H_{\mathrm{pj}}$ is slower than before $t\sim 80$ (cf. Figure \ref{helicity}, left panel). We also note that $\left.\text{d}H_{\mathrm{j}}/\text{d}t\right|_{\mathrm{Own}}$ is negligible during most of the simulation (left panel of Figure \ref{invariant_noerupt}): the low fluctuations of this term may be related to the helicity measurement error $D_{\mathrm{n,j}}$ (cf. Figure \ref{error_noerupt}) which are relatively important before $t\sim80$ (see Section \ref{sec:num_validation}).

The situation is more complicated for the eruptive emergence simulation. Indeed $\left.\text{d}H_{\mathrm{j}}/\text{d}t\right|_{\mathrm{Own}}$ presents a significant intensity during the whole simulation (cf. Figure \ref{invariant_erupt}). The own terms could be due to emergence flows but also to the pervasive reconnection between the emerging flux and the coronal arcade fields. This reconnection starts as soon as the flux emerges through the photosphere. Thus, $\left.\text{d}H_{\mathrm{x}}/\text{d}t\right|_{\mathrm{Own}}$ is more important than in the non-eruptive simulation, where there is less reconnection early on and where the "own" terms are likely due to the surface flows.

In Figure \ref{invariant_erupt}, after $t\sim 75$, the $\left.\text{d}H_{\mathrm{x}}/\text{d}t\right|_{\mathrm{Own}}$ terms are higher than the transfer term, $\left.\text{d}H_{\mathrm{j}}/\text{d}t\right|_{\mathrm{Transf}}$. The transfer term is weaker and hovers around zero during most of the pre-eruptive phase. Suddenly, after $t\sim 150$, $\left.\text{d}H_{\mathrm{j}}/\text{d}t\right|_{\mathrm{Transf}}$ becomes negative and dominant. While $H_{\mathrm{j}}$ decreases (cf. Figure \ref{helicity}, middle panel), $H_{\mathrm{pj}}$ increases quickly because of the conversion of $H_{\mathrm{j}}$. Consequently $H_{\mathrm{v}}$ still increases. However, after $t\sim 150$, $H_{\mathrm{pj}}$ decreases because of the ejection of the magnetic structure through the top boundary (and consequently a decrease of $H_{\mathrm{v}}$ is observed). We observe in Figure \ref{invariant_erupt}, as indicated by the dominant values of $\left.\text{d}H_{\mathrm{j}}/\text{d}t\right|_{\mathrm{Transf}}$ during the eruption phase, that the non-potential helicity $H_{\mathrm{j}}$ is not directly expelled through the system boundaries: during the eruption phase it is first transformed into $H_{\mathrm{pj}}$. This quantity is then ejected outside, around $t\sim 150$, as seen by a negative peak of $\left.\text{d}H_{\mathrm{pj}}/\text{d}t\right|_{\mathrm{Own}}$. Overall, Figure \ref{invariant_erupt} explains why the decrease of $H_{\mathrm{pj}}$ and $H_{\mathrm{v}}$ shows a delay with the decrease of $H_{\mathrm{j}}$ during the eruption (as seen in Figure \ref{helicity}, middle panel). 

In the jet simulation, the energy and helicity do not increase in the volume by the emergence of a magnetic structure, but by the boundary shearing motions. During the whole ideal phase, before $t\sim920$, $\left.\text{d}H_{\mathrm{j}}/\text{d}t\right|_{\mathrm{Own}}$ is negligible. The injection of helicity from the external domain is primarily provided by the flux of $H_{\mathrm{pj}}$ (thanks to $F_{\mathrm{Bn,\, Ap}}$ in the gauge used in this paper, cf. Figure \ref{flux_hpj}): only $H_{\mathrm{v}}$ and $H_{\mathrm{pj}}$ increase before $t\sim 350$ (cf. Figure \ref{helicity} right panel). 
While initially $\left.\text{d}H_{\mathrm{pj}}/\text{d}t\right|_{\mathrm{Own}}$ dominates 
$\left.\text{d}H_{\mathrm{j}}/\text{d}t\right|_{\mathrm{Transf}}$, very rapidly the situation changes: 
$\left.\text{d}H_{\mathrm{j}}/\text{d}t\right|_{\mathrm{Transf}}$ becomes the dominant term and $\text{d}H_{\mathrm{pj}}/\text{d}t$ 
becomes negative after $t\sim 500$ (cf. Figure \ref{invariant_jet}, right panel), $H_{\mathrm{pj}}$ decreases to the benefit of $H_{\mathrm{j}}$. 
In Figure \ref{helicity}, we see that $H_{\mathrm{v}}$ is increasing along with $H_{\mathrm{j}}$, and that $H_{\mathrm{pj}}$ 
decreases. Without the analysis of Figure \ref{invariant_jet}, one could imagine, when observing the right panel of Figure \ref{helicity}, that $H_{\mathrm{v}}$ is increasing directly due to the injection of $H_{\mathrm{j}}$. 
However the situation is more complex. Because of the injection of currents, the increase of $H_{\mathrm{v}}$ is still due to an injection of $H_{\mathrm{pj}}$ 
through the boundary, however a conversion from $H_{\mathrm{pj}}$ to $H_{\mathrm{j}}$ is occurring simultaneously at 
an even higher rate, as shown by the dominant values of $\left.\text{d}H_{\mathrm{j}}/\text{d}t\right|_{\mathrm{Transf}}$ (cf. Figure \ref{invariant_jet}). Hence, in Figure \ref{helicity} (right panel), occurs the decrease of $H_{\mathrm{pj}}$ and the increase of both $H_{\mathrm{j}}$ and $H_{\mathrm{v}}$.
At the moment of the jet, after $t\sim920$, we find again the same behavior as during the eruptive emergence simulation: a fast transfer of $H_{\mathrm{j}}$ to $H_{\mathrm{pj}}$ directly followed by ejection of $H_{\mathrm{pj}}$ through the boundaries, as indicated by the succession of a positive peak of $\left.\text{d}H_{\mathrm{pj}}/\text{d}t\right|_{\mathrm{Transf}}$, followed by a negative peak of $\left.\text{d}H_{\mathrm{pj}}/\text{d}t\right|_{\mathrm{Own}}$ (cf. Figure \ref{invariant_jet}, right panel). This conversion seems to be a marker of the beginning of eruptive activity.

Overall, the analysis of the evolution of the terms of the helicity decomposition in the three simulations shows that the flux $H_{\mathrm{j}}$ from the outside of the system, $\left.\text{d}H_{\mathrm{j}}/\text{d}t\right|_{\mathrm{Own}}$, is either positive or negligible compared to the transfer term. In that respect, while the increase of the non-potential helicity can be due to its own flux, the decrease of $H_{\mathrm{j}}$ seems to be mainly related with its conversion to $H_{\mathrm{pj}}$. Moreover this transfer (probably related to the ongoing magnetic reconnections) appears as the first phase of an eruption.

\begin{figure*}
  \subfigure{\label{primitive-1} \includegraphics[width=6.2cm]{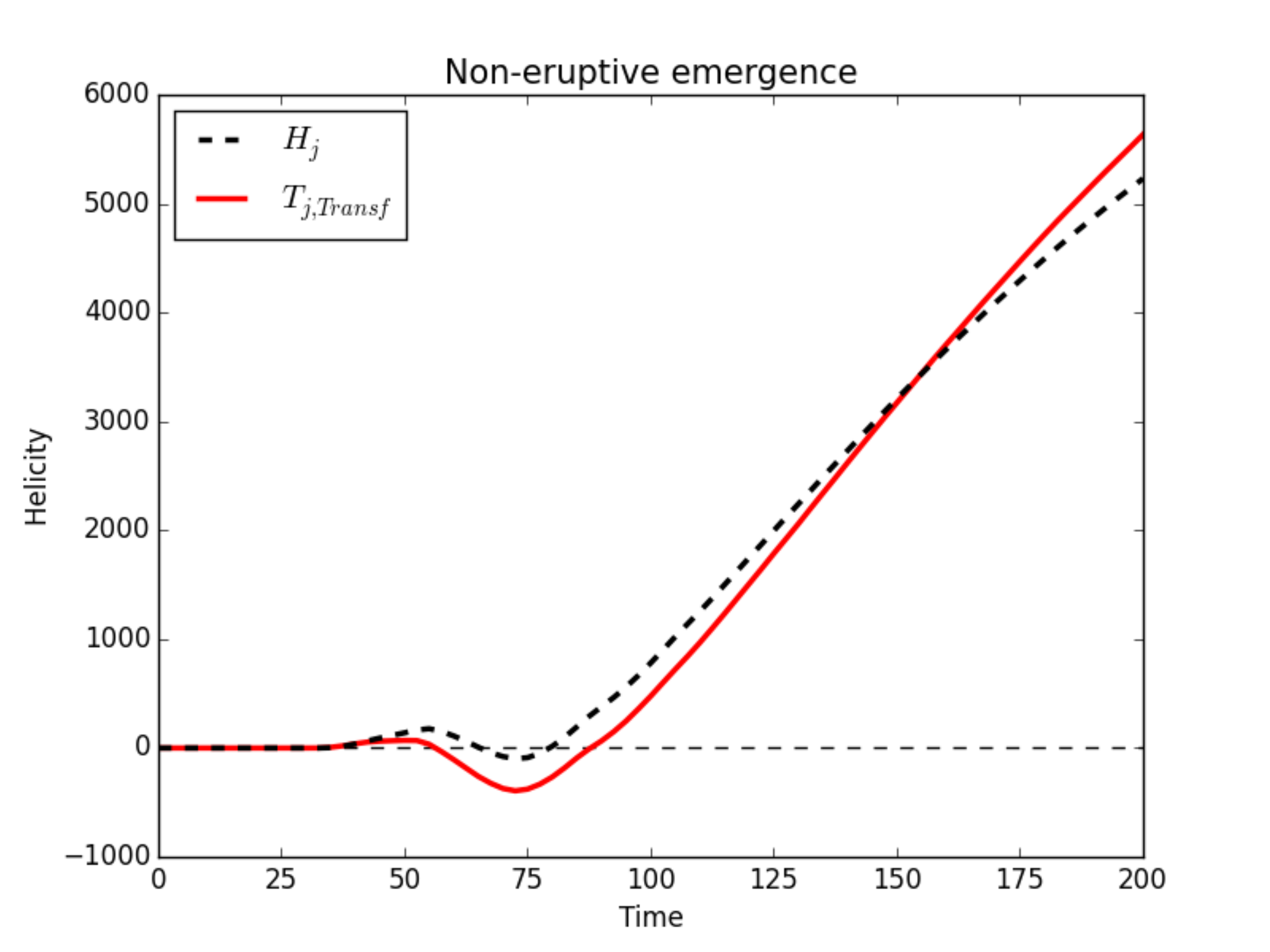}}
  \subfigure{\label{primitive-2} \includegraphics[width=6.2cm]{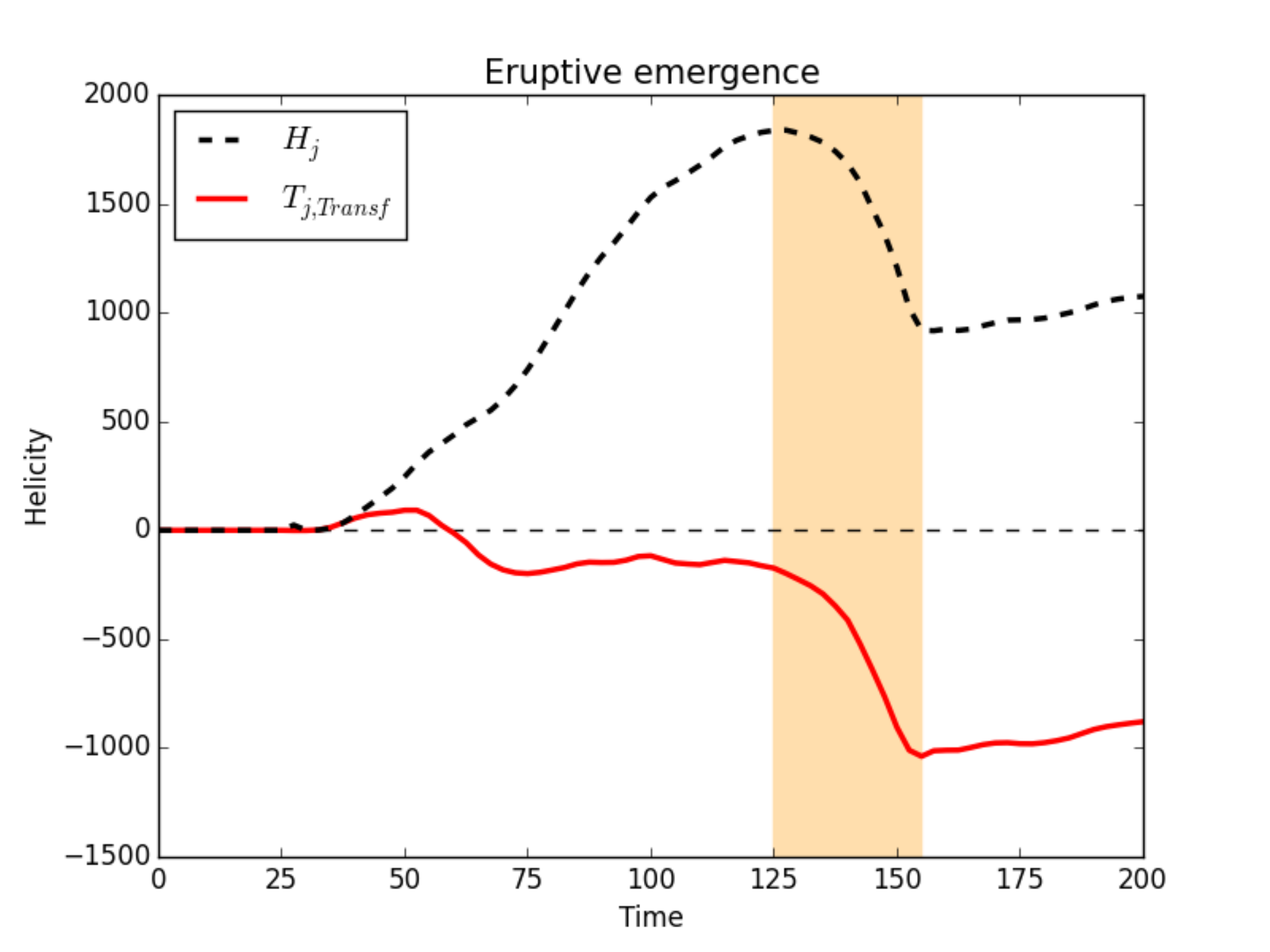}}
  \subfigure{\label{primitive-3} \includegraphics[width=6.2cm]{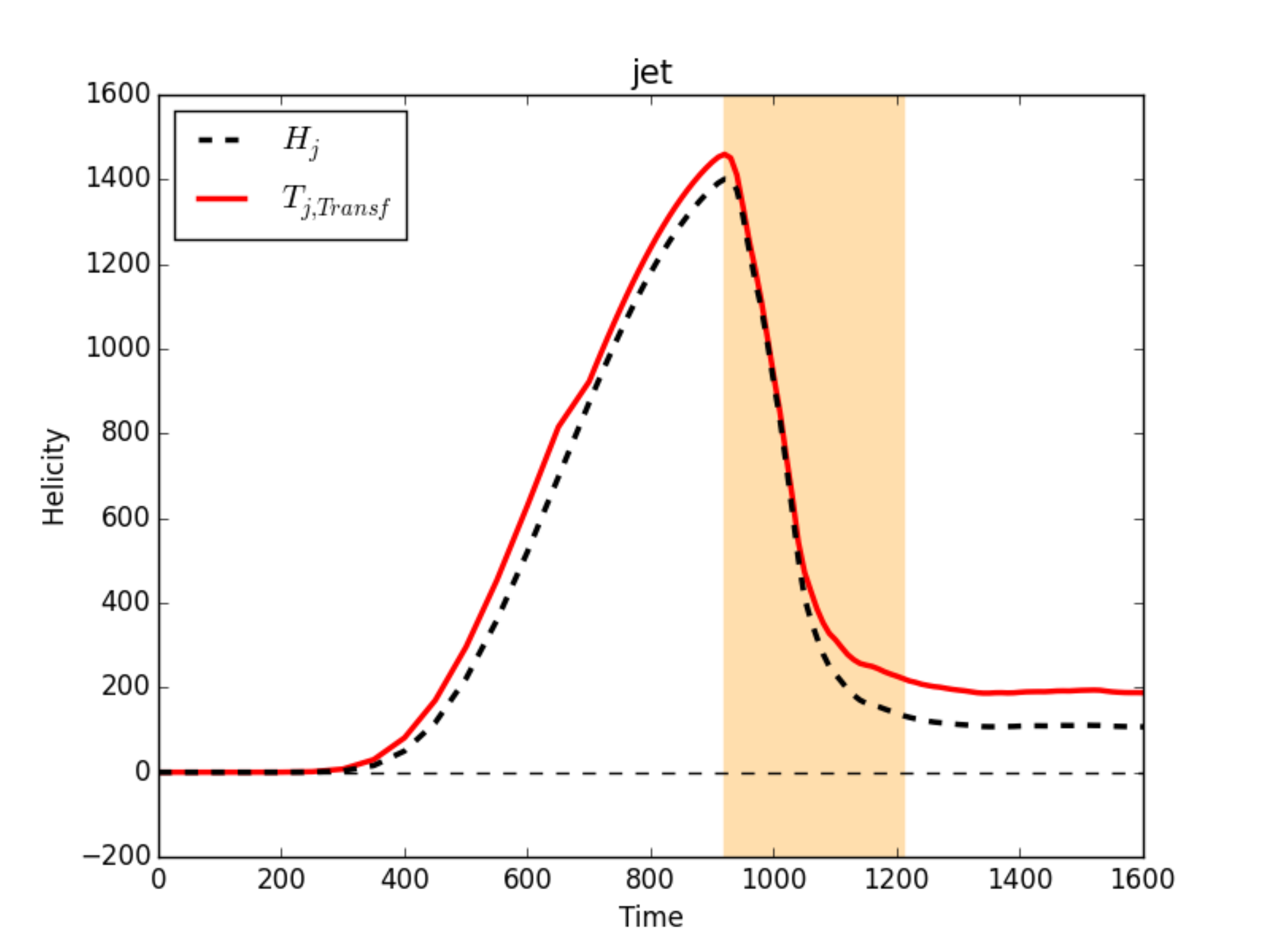}}
  \caption{
   Time evolution of the non-potential helicity $H_{\mathrm{j}}$ (dashed line, Equation \ref{eq:hj}) and the time integral of the transfer rate $T_{\mathrm{j, Transf}}$ (red lines, Equation \ref{eq:Tr}) for the non-eruptive emergence simulation (left panel), the eruptive emergence simulation (middle panel) and the simulation of the generation of a solar coronal jet (right panel). The yellow vertical bands are the same as in Figure \ref{helicity}.}
  \label{primitive}
\end{figure*}

As discussed before, $\left.\text{d}H_{\mathrm{j}}/\text{d}t\right|_{\mathrm{Transf}}$ frequently dominates the variation of $H_{\mathrm{j}}$, either during the non-/pre-eruptive or eruptive phases. To highlight its impact on the evolution of $H_{\mathrm{j}}$, Figure \ref{primitive} presents the time integral, $T_{j,Transf}$, of the transfer term : 
\begin{eqnarray}
T_{j,Transf}&=& \int_{}^{t} \left.\frac{\text{d}H_{\mathrm{j}}}{\text{d}t'}\right|_{\mathrm{Transf}} \, \mathrm{d}t' \nonumber \\
&=&-2 \int_{}^{t} \int_{V}^{} ((\textbf{v}\times\textbf{B})\cdot\textbf{B}_{\mathrm{p}}) \, \mathrm{d}V \, \mathrm{d}t'\label{eq:Tr}
\end{eqnarray}

As can be seen on the right panel of Figure \ref{primitive}, for the jet simulation, the behavior of $T_{j,Transf}$ and $H_{\mathrm{j}}$ are very similar: the curves overlap with only a small deviation toward the end of the simulation. The same is also true for the non-eruptive emergence simulation (cf. Figure \ref{primitive}, left panel). This result is the direct consequence of the relatively low value of $\left.\text{d}H_{\mathrm{j}}/\text{d}t\right|_{\mathrm{Own}}$ for these two simulations (cf. Figures \ref{invariant_noerupt} and \ref{invariant_jet}, left and right panels). The evolution of $H_{\mathrm{j}}$ in these two cases is directly related to its conversion from $H_{\mathrm{pj}}$. In the eruptive emergence simulation, Figure \ref{invariant_erupt}, the term $\left.\text{d}H_{\mathrm{j}}/\text{d}t\right|_{\mathrm{Own}}$ is not negligible compared to the transfer term. Consequently in the middle panel of Figure \ref{primitive}, $T_{j,Transf}$ does not overlap with $H_{\mathrm{j}}$. In other words, $H_{\mathrm{j}}$ can not be computed by only considering the integral of the transfer term which is what was expected if one only looks at the jet and non-eruptive simulations. However, $T_{j,Transf}$ also allows us to locate the eruption as can be done with $H_{\mathrm{j}}$. The instant $t\sim 125$, where $H_{\mathrm{j}}$ decreases exactly coincides with the main drop of the transfer term. This confirms that $H_{\mathrm{j}}$ losses are linked to a conversion toward $H_{\mathrm{pj}}$.

\section{Conclusion} \label{sec:Conclusion}

The present work focused on the analytical and numerical study of time variations of the terms in the decomposition of relative magnetic helicity, $H_v$, into the sum of the current-carrying helicity, $H_{\mathrm{j}}$, and the volume-threading helicity, $H_{\mathrm{pj}}$. After having introduced the different quantities, we analytically derived their time-derivatives (cf. Section \ref{sec:helicity}). We obtained the generalized equations for the time variation of $H_{\mathrm{j}}$ (cf. Equation \ref{eq:dhjdt}) and $H_{\mathrm{pj}}$ (cf. Equation \ref{eq:dhpjdt}). We considered special cases and gauge choices that can simplify the calculation of these formulae (cf. Section \ref{sec:cond}). 

The key outcome of our analytical derivation was to reveal a gauge-invariant quantity that controls the transfer of helicity between $H_{\mathrm{j}}$ and $H_{\mathrm{pj}}$, themselves gauge-invariant, ($\left.\text{d}H_{\mathrm{j}}/\text{d}t\right|_{\mathrm{Transf}}$ (cf. Equation \ref{eq:Ftransf_Aj}). Since this quantity is expressed with a volume term we deduce that neither $H_{\mathrm{j}}$ nor $H_{\mathrm{pj}}$ are conserved quantities in resistive or ideal MHD. While relative magnetic helicity can be build as an invariant quantity in ideal MHD, helicity can nonetheless be exchanged between $H_{\mathrm{j}}$ and $H_{\mathrm{pj}}$, and these latter quantities evolve even by ideal MHD motions. 


The time evolution of $H_v$, $H_{\mathrm{j}}$, $H_{\mathrm{pj}}$ and the terms entering in their time derivatives were then studied in three 3D MHD numerical simulations of solar coronal events (cf. Section \ref{sec:test}): the generation of a coronal jet \citep{Pariat09a}, the formation of a stable active region by flux-emergence \citep{Leake13b}, as well as the formation of an eruptive active region \citep{Leake14a}. These simulations present a sample of boundary forcing (line-tied vs flux-emergence) as well as dynamics, e.g. ideal evolution, magnetic reconnection and eruptions. 

The analysis of these numerical experiments allowed us first to confirm numerically the time variation equations of $H_{\mathrm{j}}$ and $H_{\mathrm{pj}}$ that were derived analytically (cf. Section \ref{sec:helicity_estimation}). This confirms, evidently in some cases, that $H_{\mathrm{j}}$ and $H_{\mathrm{pj}}$ are not individually invariants of ideal MHD. 

In particular we observed that in many cases, the transfer term $\left.\text{d}H_{\mathrm{j}}/\text{d}t\right|_{\mathrm{Transf}}$ was dominating the dynamics of $H_{\mathrm{j}}$ (cf. Section \ref{sec:num_transf}). We observed that the evolution of $H_{\mathrm{j}}$ is frequently uniquely controlled by the term $\left.\text{d}H_{\mathrm{j}}/\text{d}t\right|_{\mathrm{Transf}}$, which means that this quantity does not evolve due to boundary fluxes but builds-up through the transformation of $H_{\mathrm{pj}}$. This dynamics was observed both during the energy build-up phases of the evolution of the system, but as well during the eruption/ejection phases. During the energy injection phases, both for the jet and the stable emergence simulations, boundary fluxes first increase $H_{\mathrm{pj}}$, and part of it is then transformed into $H_{\mathrm{j}}$. During the eruption phases (in the jet and eruptive emergence simulations), $H_{\mathrm{j}}$ is first transformed into $H_{\mathrm{pj}}$ and it is the latter that is expelled from the domain by boundary fluxes. 

This finding has an important consequence on our ability to estimate $H_{\mathrm{j}}$ and $H_{\mathrm{pj}}$ in observed solar active regions. As discussed in the Introduction (see Section \ref{sec:Introduction}), the most common way to estimate relative magnetic helicity is by time integration of its flux through the solar photosphere. Since relative magnetic helicity is largely conserved, its photospheric flux dominates the evolution of the relative helicity in the coronal domain. From numerical simulations, it was shown that indeed the time integrated boundary fluxes closely match the amount of helicity in the system \citep{YangS13,Pariat15b,Pariat17}. \citet{Lim07} also confirmed from observational data, that the helicity flux accumulation indeed gives a proper estimation of the coronal helicity with a $\sim 15 \%$ relative error. Such flux-integration approach is however doomed to fail when applied to $H_{\mathrm{j}}$ and $H_{\mathrm{pj}}$ because of the volume transfer term their evolution is not dictated by their boundary fluxes. In the jet simulation for example, the time integration of the term $\left.\text{d}H_{\mathrm{j}}/\text{d}t\right|_{\mathrm{Own}}$ is order of magnitude smaller than the amount of $H_{\mathrm{j}}$ in the system.
The analysis of $H_{\mathrm{j}}$ and $H_{\mathrm{pj}}$ in observations thus has to rely on the volume-integration approach \citep{Valori16}. In this method, the 3D coronal magnetic field must be reconstructed from the 2D photospheric measurements, thanks to extrapolation techniques \citep{Wiegelmann12,Wiegelmann14}. Because of the inherent nature of $H_{\mathrm{j}}$, which describes non-potential fields, potential magnetic field reconstruction cannot estimate this quantity. Similarly, we believe that linear force-free extrapolation will only provide too crude an approximation of $H_{\mathrm{j}}$. The linear force-free approximation can be used to estimate relative helicity solely thanks to the boundary distribution of the normal component of the magnetic field \citep{Berger85,Lim07}. The impact of this approximation on the estimation of $H_{\mathrm{j}}$ and $H_{\mathrm{pj}}$ remains to be studied but it can be conjectured that, since the linear approximation effectively distributes current in the entire volume, it is inaccurate for systems, like the corona, where currents are spatially localized. These quantities being highly non-linear, non-force free extrapolations will probably be the unique way to properly approach an estimation of $H_{\mathrm{j}}$ and $H_{\mathrm{pj}}$, similarly to the study of \citet{James18} who provided the first estimation of $H_{\mathrm{j}}/H_v$ in observed data. 

\citet{Pariat17} and \citet{Zuccarello18} have shown that the helicity ratio $H_{\mathrm{j}}/H_v$ seems to be tightly related with the eruption process. This paper presents an additional evidence through the analysis of the coronal jet simulation of \citet{Pariat09a}. We show here that the jet is indeed triggered when the helicity ratio $H_{\mathrm{j}}/H_v$ attains a very high value. The value of this ratio drops significantly after the generation of the jet. If, as hinted by these studies, $H_{\mathrm{j}}$ is a key element for the eruptivity of solar active regions, the present study demonstrates the need for the development of quality 3D magnetic reconstructions of the solar magnetic field in order to measure $H_{\mathrm{j}}$ and $H_{\mathrm{pj}}$ in observed solar active regions. The Solar Orbiter mission and its PHI instrument, which will provide the first remote magnetic field observations complementing those from the Earth environment, may provide a unique opportunity to improve vector magnetic field measurements. 
Finally, the present work has highlighted yet another interesting behaviour of magnetic helicity. The non-potential helicity $H_{\mathrm{j}}$ evolution seems to be mostly driven by the volume transformation from $H_{\mathrm{pj}}$ rather than from its own flux. Further investigations on different magnetic field simulations are required to capture the dynamics of this quantity. More generally, magnetic helicity needs to be further understood through fundamental studies on its mathematical properties \citep{Oberti18}, on its physical interpretation \citep{Yeates13,Yeates14,Russell15,Aly18}, and on its proper measurement in the solar corona \citep{Dalmasse13,Dalmasse14,Dalmasse18,Valori16,GuoY17,Moraitis18}.


\begin{acknowledgements}
LL, EP, \& KM acknowledge the support of the French Agence Nationale pour la Recherche through the HELISOL project, contract n$^\circ$ ANR-15-CE31-0001. GV acknowledges the support of the Leverhume Trust, Research Projet Grant 2014-051. JEL is supported by NASA's LWS and HGI programs. The authors acknowledge access to the HPC resources of CINES under the allocations 2017--046331 made by GENCI (Grand Equipement National de Calcul Intensif).
\end{acknowledgements}

\appendix
\section{Helicities variation decomposition with a different gauge choice} \label{app:A}
In the core part of this study, in our application of the helicity estimation to numerical experiments (cf. Section \ref{sec:num_test}), we used the practical DeVore-Coulomb gauge for $\avec$ and $\avecp$ with the additional constraint that they have the same distribution at the top boundary (cf. condition (\ref{eq:devore})). This choice induces that all the fluxes $F_{\mathrm{\alpha,\, Aj}}$ (with $\alpha\in\{Vn,Bn,Aj,\phi\}$) in Equation (\ref{eq:dhjdt}) are null at the top boundary. To determine the impact of this choice, we perform the helicity calculations with a different condition:
\begin{equation}\label{eq:bot}
\textbf{A}(x,y,z=z_{bot},t)_\perp=\textbf{A}_{\mathrm{p}}(x,y,z=z_{bot},t)_\perp
\end{equation} 
i.e with $\textbf{A}$ and $\textbf{A}_{\mathrm{p}}$ having the same distribution at the bottom boundary, at $z=z_{bot}$ instead of the top boundary.

\begin{figure*}
  \includegraphics[width=18cm]{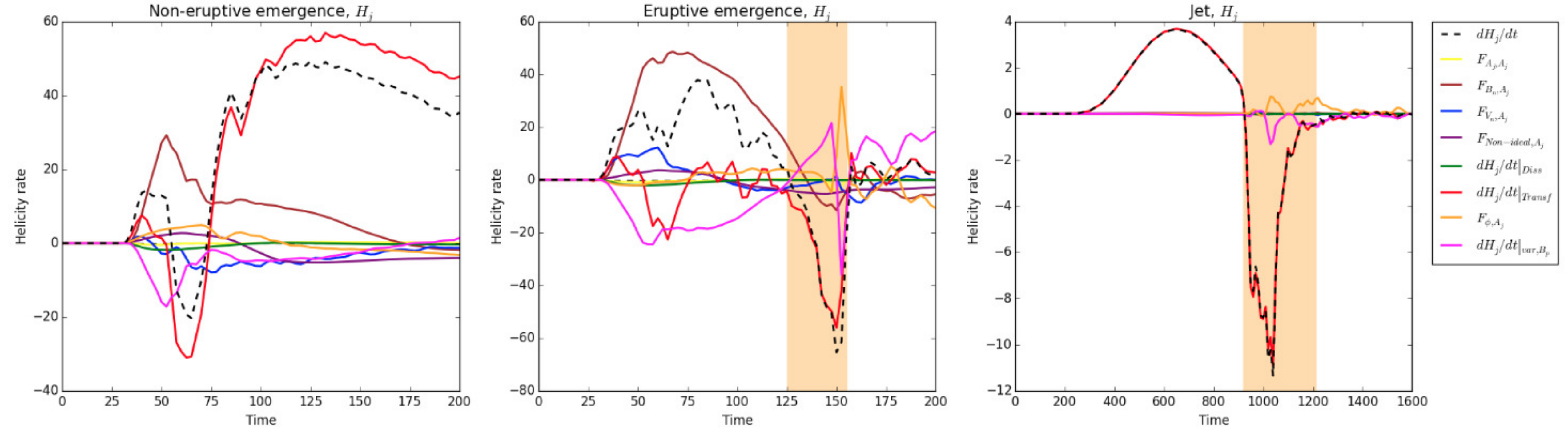}
   \caption{\label{flux_hj}
   Time evolution of the non-potential helicity variation rate (dashed black line; $\text{d}H_{\mathrm{j}}/\text{d}t$; Equation \ref{eq:dhjdt}) and the different terms constituting the instantaneous time variation of $H_{\mathrm{j}}$ (Equation \ref{eq:dhjdt}): $F_{\mathrm{Aj,\, Aj}}$ (yellow line, Equation \ref{eq:FAj_Aj}), $F_{\mathrm{Bn,\, Aj}}$ (brown line, Equation \ref{eq:FBn_Aj}) $F_{\mathrm{Vn,\, Aj}}$ (blue line, Equation \ref{eq:FVn_Aj}), $\left.\text{d}H_{\mathrm{j}}/\text{d}t\right|_{\mathrm{Transf}}$ (red line, Equation \ref{eq:Ftransf_Aj}), $F_{\mathrm{\phi,\, Aj}}$ (orange line, Equation \ref{eq:Fphi_Aj}), $\left.\text{d}H_{\mathrm{j}}/\text{d}t\right|_{\mathrm{Bp, var}}$ (magenta line, Equation \ref{eq:Fvar_Aj}), $F_{\mathrm{Non-ideal,\, Aj}}$ (purple line, Equation \ref{eq:SNoId_Ap}), $\left.\text{d}H_{\mathrm{j}}/\text{d}t\right|_{\mathrm{Diss}}$ (green line, Equation \ref{eq:Diss_Aj}). From left to right : the non-eruptive emergence simulation, the eruptive emergence and the simulation of the generation of a solar coronal jet. The yellow bands are the same as in Figure \ref{helicity}.}
\end{figure*}

\begin{figure*}
  \includegraphics[width=18cm]{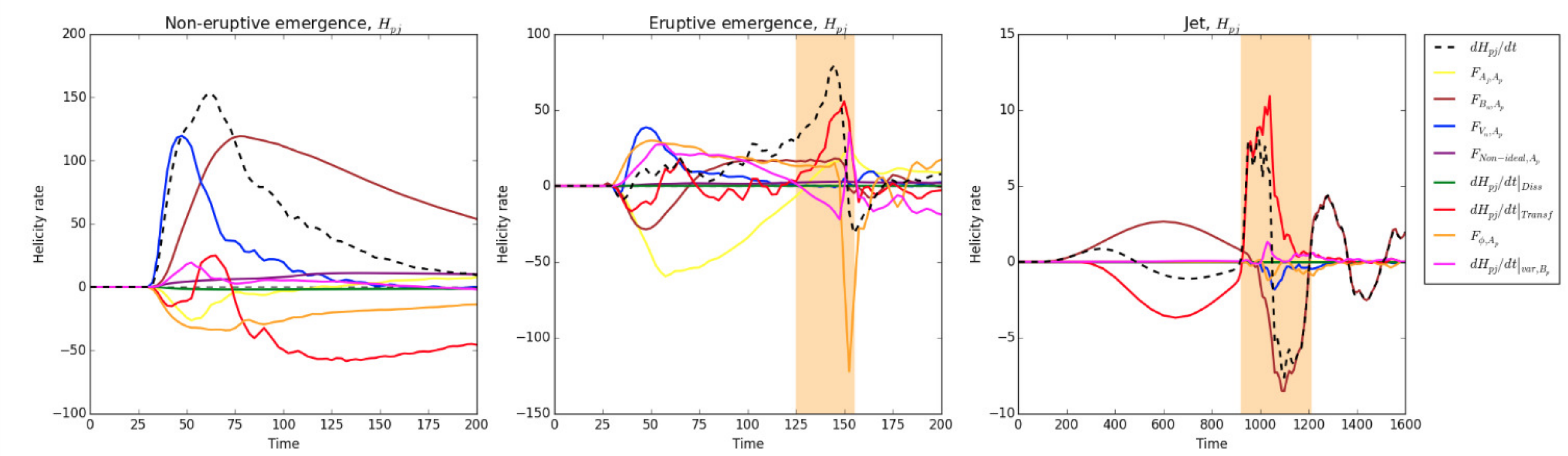}
  \caption{\label{flux_hpj}
   Time evolution of the volume -threading helicity variation rate (dashed black line; $\text{d}H_{\mathrm{pj}}/\text{d}t$; Equation \ref{eq:dhpjdt}) and the different terms constituting the instantaneous time variation of $H_{\mathrm{pj}}$ (Equation \ref{eq:dhpjdt}): $F_{\mathrm{Aj,\, Ap}}$ (yellow line, Equation \ref{eq:FAj_Ap}), $F_{\mathrm{Bn,\, Ap}}$ (brown line, Equation \ref{eq:FBn_Ap}) $F_{\mathrm{Vn,\, Ap}}$ (blue line, Equation \ref{eq:FVn_Ap}), $\left.\text{d}H_{\mathrm{pj}}/\text{d}t\right|_{\mathrm{Transf}}$ (red line, Equation \ref{eq:Ftransf_Ap}), $F_{\phi,\, Ap}$ (orange line, Equation \ref{eq:Fphi_Ap}), $\left.\text{d}H_{\mathrm{pj}}/\text{d}t\right|_{\mathrm{Bp, var}}$ (magenta line, Equation \ref{eq:Fvar_Ap}), $F_{\mathrm{Non-ideal,\, Ap}}$ (purple line, Equation \ref{eq:SNoId_Ap}), $\left.\text{d}H_{\mathrm{pj}}/\text{d}t\right|_{\mathrm{Diss}}$ (green line, Equation \ref{eq:Diss_Ap}). From left to right : the non-eruptive emergence simulation, the eruptive emergence and the simulation of the generation of a solar coronal jet. The yellow bands are the same as in Figure \ref{helicity}.}
\end{figure*}

\begin{figure*}
  \includegraphics[width=18cm]{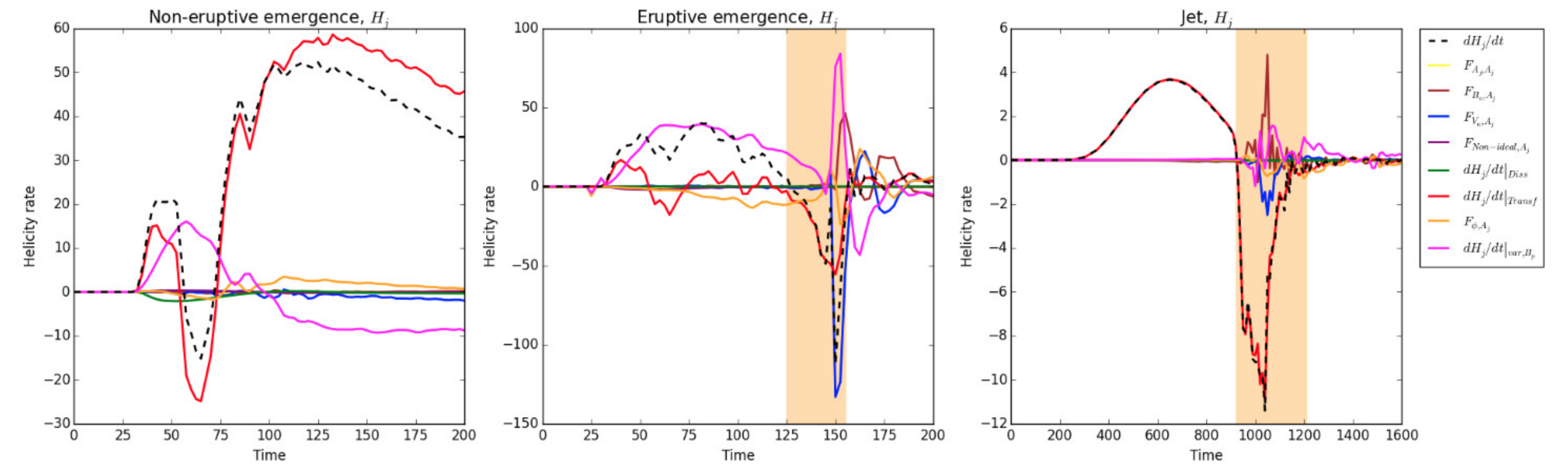}
   \caption{\label{flux_hj_bot} Same as Figure \ref{flux_hj} using Condition (\ref{eq:bot}) instead of Condition (\ref{eq:devore}).}
\end{figure*}

\begin{figure*}
  \includegraphics[width=18cm]{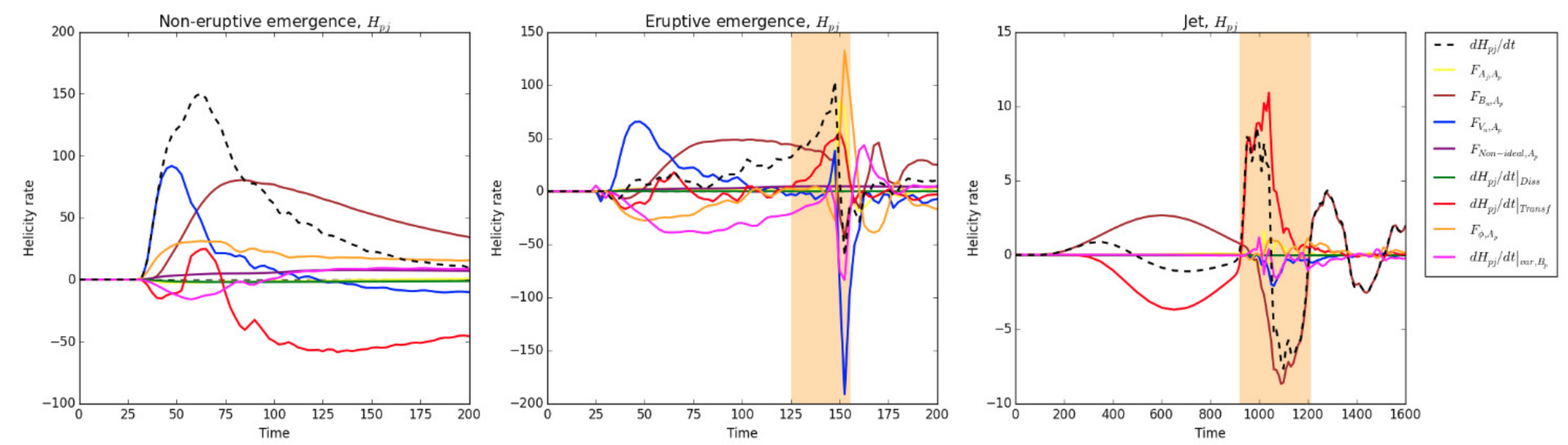}
  \caption{\label{flux_hpj_bot} Same as Figure \ref{flux_hpj} using Condition (\ref{eq:bot}) instead of Condition (\ref{eq:devore}).}
\end{figure*}

All the helicity evaluations performed in Section \ref{sec:num_test} are recomputed with this new gauge choice. We only present here the results in which the gauge choice has a significant impact. As expected, all the gauge independent quantities are not affected by the switch between condition (\ref{eq:devore}) and (\ref{eq:bot}). Only the figures presenting gauge dependent quantities are significantly affected, those representing the different terms in the helicity time variations of Equations (\ref{eq:dhjdt}) and (\ref{eq:dhpjdt}). These terms, computed with condition (\ref{eq:bot}) are presented in Figures \ref{flux_hj_bot} and \ref{flux_hpj_bot}. These figures should be compared with Figures \ref{flux_hj} and \ref{flux_hpj} (cf. Section \ref{sec:fluxes}). 

As expected, the curves in Figure \ref{flux_hj_bot} (resp. Figure \ref{flux_hpj_bot}) are markedly different from the curves in Figure \ref{flux_hj} (resp. Figure \ref{flux_hpj}). Indeed, the fluxes in Equations (\ref{eq:dhjdt}) and (\ref{eq:dhpjdt}) are gauge dependent (cf. Section \ref{sec:surr_envi}). Only the term $\left.\text{d}H_{\mathrm{x}}/\text{d}t\right|_{\mathrm{Transf}}$ is gauge invariant and is strictly identical. We also note that $\Delta H_{\mathrm{j}}/\Delta t$ and $\Delta H_{\mathrm{pj}}/\Delta t$ are almost identical, the marginal difference ($<4\%$) being due to the intrinsic precision on the estimation of the helicities due to the finite non-solenoidality of the datasets (cf. Sections \ref{sec:test}).

\begin{figure*}
  \subfigure{\label{invariant_noerupt-1} \includegraphics[width=9.4cm]{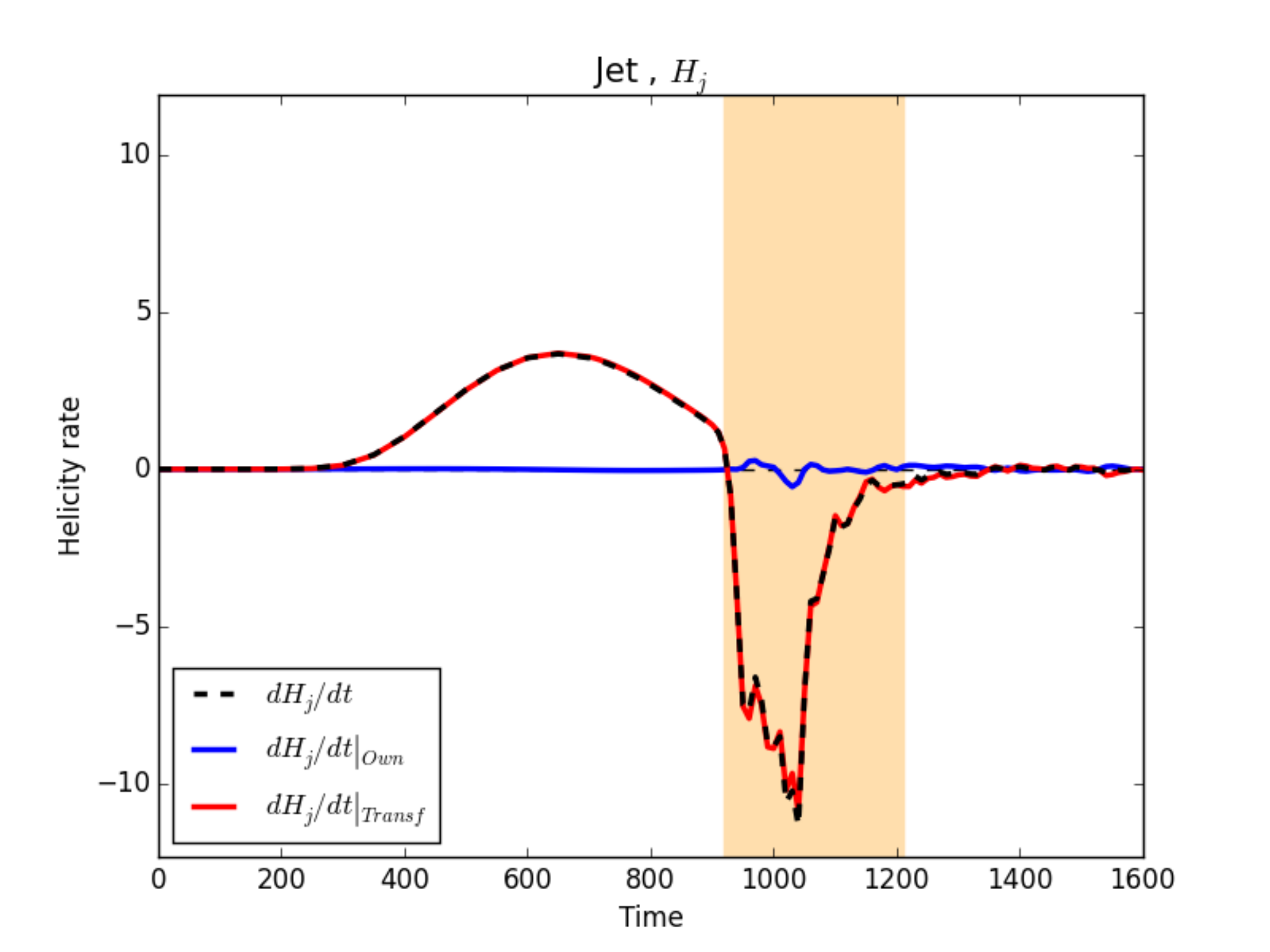}}
  \subfigure{\label{invariant_noerupt-2} \includegraphics[width=9.4cm]{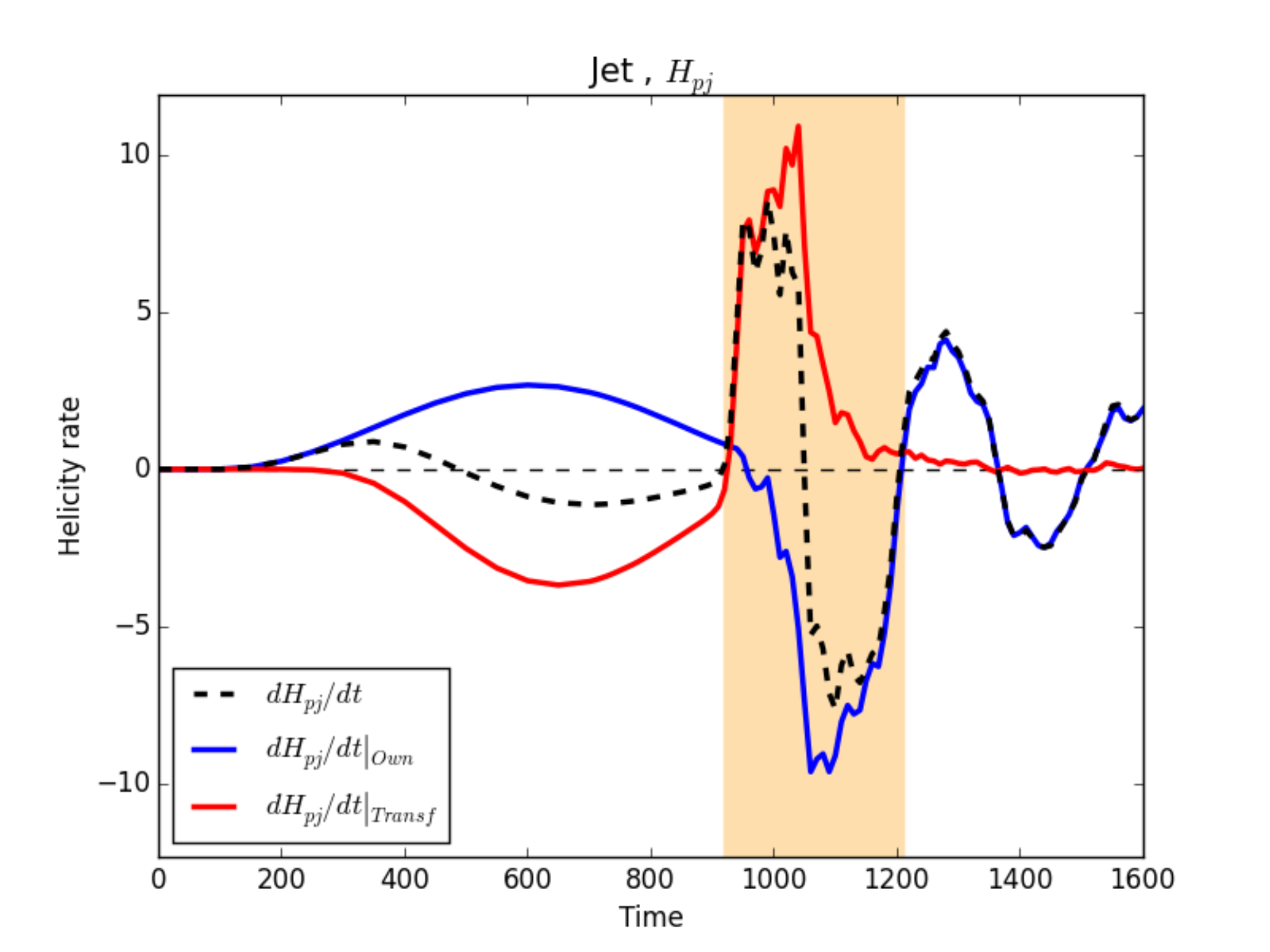}}
  \caption{Same as Figure \ref{invariant_noerupt} without the dissipation rates, $\left.\text{d}H_{\mathrm{j}}/\text{d}t\right|_{\mathrm{Diss}}$ (Equation \ref{eq:Diss_Aj}) and $\left.\text{d}H_{\mathrm{pj}}/\text{d}t\right|_{\mathrm{Diss}}$ (Equation \ref{eq:Diss_Ap}). For this simulation, these terms are assumed to null. The yellow bands correspond to the jet eruption phase.}
  \label{invariant_jet}
\end{figure*}

For the jet simulation (right panel of Figure \ref{flux_hj_bot}), the differences appear mainly during the non-ideal phase (after $t\sim900$). During that phase, a jet and a non-linear magnetic wave are passing through the top boundary. Hence the magnetic and the velocity fields vary significantly at the top boundary. Since with condition (\ref{eq:devore}), $F_{\mathrm{Bn,\, Aj}}$ and $F_{\mathrm{Vn,\, Aj}}$ are constantly null at the top boundary, these quantities presented very weak values in Figure \ref{flux_hj} (right panel) during the passage of the jet. The top constraint being lifted with condition (\ref{eq:bot}), $F_{\mathrm{Bn,\, Aj}}$ and $F_{\mathrm{Vn,\, Aj}}$  show significantly large values in Figure \ref{flux_hj_bot} around $t\sim1000$. The different boundary fluxes nonetheless cancel each other, and the gauge independent quantity $\left.\text{d}H_{\mathrm{j}}/\text{d}t\right|_{\mathrm{Own}}$ remains very low (as in Figure \ref{invariant_jet}), significantly smaller than $\left.\text{d}H_{\mathrm{j}}/\text{d}t\right|_{\mathrm{Transf}}$.
 
For the non-eruptive and eruptive emergence simulations, before $t\sim120$, with condition (\ref{eq:bot}) the contribution of $F_{\mathrm{Vn,\, Aj}}$ and $F_{\mathrm{Bn,\, Aj}}$ have completely disappeared (cf. Figure \ref{flux_hj_bot}). These quantities depend mostly on the change of the velocity and magnetic fields at the bottom boundary while the flux tube was emerging. The vector potential $\avecj$ being forced to be null at this boundary, the fluxes are null as well. The quantity $\text{d}H_{\mathrm{j}}/\text{d}t$ evolves almost only thanks to the volume terms; the transfer term and $\left.\text{d}H_{\mathrm{j}}/\text{d}t\right|_{\mathrm{Bp,\, var}}$. With this particular gauge choice $H_{\mathrm{j}}$ does not exchange with the outside during the pre-eruptive phase, both for the eruptive and the non-eruptive simulations.

During the eruptive phase of the eruptive emergence simulation (middle panels of Figures \ref{flux_hj_bot} and \ref{flux_hpj_bot}), we see that $F_{Vn,\, Ax}$, with $x$ being either $j$ or $pj$, is the main contribution of the helicity fluxes. The computation of $\text{D}_{n,x}$ (Equation \ref{eq:Dn}) and $\left.\text{d}H_{\mathrm{x}}/\text{d}t\right|_{\mathrm{Own}}$ (not shown here), inform us that this peak of $\text{F}_{Vn,Aj}$ is likely due to numerical errors: the time step is not sufficiently small to capture the sudden change of the velocity field at the top boundary. At that particular time the computation method and the specific choice of gauge can markedly influence the precision of the helicity flux estimation. This results had already been note by \citet{Pariat15b,Pariat17}.

\bibliographystyle{aasjournal}
   

\begin{thebibliography}{}
\expandafter\ifx\csname natexlab\endcsname\relax\def\natexlab#1{#1}\fi
\providecommand{\url}[1]{\href{#1}{#1}}

\bibitem[{Aly(2018)}]{Aly18}
Aly, J.-J. 2018, Fluid Dynamics Research, 50, 011408

\bibitem[{Antiochos(2013)}]{Antiochos13}
Antiochos, S.~K. 2013, arXiv.org, 772, 72

\bibitem[{Berger(1985)}]{Berger85}
Berger, M.~A. 1985, Astrophysical Journal Supplement Series (ISSN 0067-0049),
  59, 433

\bibitem[{Berger(2003)}]{Berger03}
---. 2003, Advances in Nonlinear Dynamos. Series: The Fluid Mechanics of
  Astrophysics and Geophysics, 20030424, 345

\bibitem[{Berger \& Field(1984)}]{BergerField84}
Berger, M.~A., \& Field, G.~B. 1984, Journal of Fluid Mechanics (ISSN
  0022-1120), 147, 133

\bibitem[{Brandenburg {et~al.}(2017)Brandenburg, Petrie, \&
  Singh}]{Brandenburg17}
Brandenburg, A., Petrie, G. J.~D., \& Singh, N.~K. 2017, The Astrophysical
  Journal, 836, 21

\bibitem[{Brandenburg \& Subramanian(2005)}]{Brandenburg05}
Brandenburg, A., \& Subramanian, K. 2005, Physics Reports, 417, 1

\bibitem[{Burlaga(1995)}]{Burlaga95}
Burlaga, L.~F. 1995, Interplanetary magnetohydrodynamics, 3

\bibitem[{Candelaresi(2012)}]{Candelaresi12}
Candelaresi, S. 2012, Ph.D. Thesis, 126

\bibitem[{Chae(2001)}]{Chae01}
Chae, J. 2001, The Astrophysical Journal, 560, L95

\bibitem[{Chae(2007)}]{Chae07}
---. 2007, Advances in Space Research, 39, 1700

\bibitem[{Dalmasse {et~al.}(2014)Dalmasse, Pariat, D{\'e}moulin, \&
  Aulanier}]{Dalmasse14}
Dalmasse, K., Pariat, E., D{\'e}moulin, P., \& Aulanier, G. 2014, Solar
  Physics, 289, 107

\bibitem[{Dalmasse {et~al.}(2013)Dalmasse, Pariat, Valori, D{\'e}moulin, \&
  Green}]{Dalmasse13}
Dalmasse, K., Pariat, E., Valori, G., D{\'e}moulin, P., \& Green, L.~M. 2013,
  Astronomy and Astrophysics, 555, L6

\bibitem[{Dalmasse {et~al.}(2018)Dalmasse, Pariat, Valori, Jing, \&
  D{\'e}moulin}]{Dalmasse18}
Dalmasse, K., Pariat, E., Valori, G., Jing, J., \& D{\'e}moulin, P. 2018, The
  Astrophysical Journal, 852, 141

\bibitem[{Dasso(2009)}]{Dasso09}
Dasso, S. 2009, Universal Heliophysical Processes, 257, 379

\bibitem[{Dasso {et~al.}(2003)Dasso, Mandrini, D{\'e}moulin, \&
  Farrugia}]{Dasso03}
Dasso, S., Mandrini, C.~H., D{\'e}moulin, P., \& Farrugia, C.~J. 2003, Journal
  of Geophysical Research, 108, 1362

\bibitem[{Dasso {et~al.}(2005)Dasso, Mandrini, D{\'e}moulin, Luoni, \&
  Gulisano}]{Dasso05a}
Dasso, S., Mandrini, C.~H., D{\'e}moulin, P., Luoni, M.~L., \& Gulisano, A.~M.
  2005, Advances in Space Research, 35, 711

\bibitem[{Del~Sordo {et~al.}(2010)Del~Sordo, Candelaresi, \&
  Brandenburg}]{DelSordo10}
Del~Sordo, F., Candelaresi, S., \& Brandenburg, A. 2010, Physical Review E, 81,
  036401

\bibitem[{D{\'e}moulin(2007)}]{Demoulin07}
D{\'e}moulin, P. 2007, Advances in Space Research, 39, 1674

\bibitem[{D{\'e}moulin(2008)}]{Demoulin08}
---. 2008, Annales Geophysicae, 26, 3113

\bibitem[{D{\'e}moulin {et~al.}(2016)D{\'e}moulin, Janvier, \&
  Dasso}]{Demoulin16}
D{\'e}moulin, P., Janvier, M., \& Dasso, S. 2016, Solar Physics, 291, 531

\bibitem[{D{\'e}moulin \& Pariat(2009)}]{Demoulin09}
D{\'e}moulin, P., \& Pariat, E. 2009, Advances in Space Research, 43, 1013

\bibitem[{D{\'e}moulin {et~al.}(2006)D{\'e}moulin, Pariat, \&
  Berger}]{Demoulin06}
D{\'e}moulin, P., Pariat, E., \& Berger, M.~A. 2006, Solar Physics, 233, 3

\bibitem[{DeVore(2000)}]{DeVore00}
DeVore, C.~R. 2000, The Astrophysical Journal, 539, 944

\bibitem[{Elsasser(1956)}]{Elsasser56}
Elsasser, W.~M. 1956, Reviews of Modern Physics, 28, 135

\bibitem[{Finn \& Antonsen(1985)}]{Finn85}
Finn, J.~H., \& Antonsen, T. M.~J. 1985, Comments on Plasma Physics and
  Controlled Fusion, 9, 111

\bibitem[{Georgoulis {et~al.}(2012)Georgoulis, Tziotziou, \&
  Raouafi}]{Georgoulis12}
Georgoulis, M.~K., Tziotziou, K., \& Raouafi, N.-E. 2012, The Astrophysical
  Journal, 759, 1

\bibitem[{Guennou {et~al.}(2017)Guennou, Pariat, Leake, \& Vilmer}]{Guennou17}
Guennou, C., Pariat, E., Leake, J.~E., \& Vilmer, N. 2017, Journal of Space
  Weather and Space Climate, 7, A17

\bibitem[{Guo {et~al.}(2013)Guo, Ding, Cheng, Zhao, \& Pariat}]{GuoY13b}
Guo, Y., Ding, M.-D., Cheng, X., Zhao, J., \& Pariat, E. 2013, The
  Astrophysical Journal, 779, 157

\bibitem[{Guo {et~al.}(2010)Guo, Ding, Schmieder, Li, T{\"o}r{\"o}k, \&
  Wiegelmann}]{GuoY10}
Guo, Y., Ding, M.-D., Schmieder, B., {et~al.} 2010, The Astrophysical Journal
  Letters, 725, L38

\bibitem[{Guo {et~al.}(2017)Guo, Pariat, Valori, Anfinogentov, Chen,
  Georgoulis, Liu, Moraitis, Thalmann, \& Yang}]{GuoY17}
Guo, Y., Pariat, E., Valori, G., {et~al.} 2017, The Astrophysical Journal, 840,
  40

\bibitem[{Hu {et~al.}(2014)Hu, Qiu, Dasgupta, Khare, \& Webb}]{Hu14}
Hu, Q., Qiu, J., Dasgupta, B., Khare, A., \& Webb, G.~M. 2014, The
  Astrophysical Journal, 793, 53

\bibitem[{James {et~al.}(2018)James, Valori, Green, Liu, Cheung, Guo, \& van
  Driel-Gesztelyi}]{James18}
James, A.~W., Valori, G., Green, L.~M., {et~al.} 2018, arXiv.org,
  arXiv:1802.07965

\bibitem[{Kazachenko {et~al.}(2010)Kazachenko, Canfield, Longcope, \&
  Qiu}]{Kazachenko10}
Kazachenko, M.~D., Canfield, R.~C., Longcope, D.~W., \& Qiu, J. 2010, The
  Astrophysical Journal, 722, 1539

\bibitem[{Kazachenko {et~al.}(2012)Kazachenko, Canfield, Longcope, \&
  Qiu}]{Kazachenko12}
---. 2012, Solar Physics, 277, 165

\bibitem[{Kazachenko {et~al.}(2009)Kazachenko, Canfield, Longcope, Qiu,
  Des~Jardins, \& Nightingale}]{Kazachenko09}
Kazachenko, M.~D., Canfield, R.~C., Longcope, D.~W., {et~al.} 2009, The
  Astrophysical Journal, 704, 1146

\bibitem[{Knizhnik {et~al.}(2015)Knizhnik, Antiochos, \& DeVore}]{Knizhnik15}
Knizhnik, K.~J., Antiochos, S.~K., \& DeVore, C.~R. 2015, The Astrophysical
  Journal, 809, 137

\bibitem[{Kusano {et~al.}(2004)Kusano, Maeshiro, Yokoyama, \&
  Sakurai}]{Kusano04}
Kusano, K., Maeshiro, T., Yokoyama, T., \& Sakurai, T. 2004, The Astrophysical
  Journal, 610, 537

\bibitem[{Leake {et~al.}(2014)Leake, Linton, \& Antiochos}]{Leake14a}
Leake, J.~E., Linton, M.~G., \& Antiochos, S.~K. 2014, The Astrophysical
  Journal, 787, 46

\bibitem[{Leake {et~al.}(2013)Leake, Linton, \& T{\"o}r{\"o}k}]{Leake13b}
Leake, J.~E., Linton, M.~G., \& T{\"o}r{\"o}k, T. 2013, The Astrophysical
  Journal, 778, 99

\bibitem[{Lim {et~al.}(2007)Lim, Jeong, Chae, \& Moon}]{Lim07}
Lim, E.-K., Jeong, H., Chae, J., \& Moon, Y.-J. 2007, The Astrophysical
  Journal, 656, 1167

\bibitem[{Linton \& Antiochos(2002)}]{Linton02}
Linton, M.~G., \& Antiochos, S.~K. 2002, The Astrophysical Journal, 581, 703

\bibitem[{Linton {et~al.}(2001)Linton, Dahlburg, \& Antiochos}]{Linton01}
Linton, M.~G., Dahlburg, R.~B., \& Antiochos, S.~K. 2001, The Astrophysical
  Journal, 553, 905

\bibitem[{Liu \& Schuck(2012)}]{LiuY12}
Liu, Y., \& Schuck, P.~W. 2012, The Astrophysical Journal, 761, 105

\bibitem[{Liu \& Schuck(2013)}]{LiuY13}
---. 2013, Solar Physics, 283, 283

\bibitem[{Longcope \& Beveridge(2007)}]{Longcope07b}
Longcope, D.~W., \& Beveridge, C. 2007, The Astrophysical Journal, 669, 621

\bibitem[{Longcope {et~al.}(2007)Longcope, Ravindra, \& Barnes}]{Longcope07a}
Longcope, D.~W., Ravindra, B., \& Barnes, G. 2007, The Astrophysical Journal,
  668, 571

\bibitem[{Low(1996)}]{Low96}
Low, B.~C. 1996, Solar Physics, 167, 217

\bibitem[{Low(2006)}]{Low06}
---. 2006, The Astrophysical Journal, 646, 1288

\bibitem[{Luoni {et~al.}(2005)Luoni, Mandrini, Dasso, van Driel-Gesztelyi, \&
  D{\'e}moulin}]{Luoni05}
Luoni, M.~L., Mandrini, C.~H., Dasso, S., van Driel-Gesztelyi, L.~L., \&
  D{\'e}moulin, P. 2005, Journal of Atmospheric and Terrestrial Physics, 67,
  1734

\bibitem[{Mandrini {et~al.}(2005)Mandrini, Pohjolainen, Dasso, Green,
  D{\'e}moulin, van Driel-Gesztelyi, Copperwheat, \& Foley}]{Mandrini05}
Mandrini, C.~H., Pohjolainen, S., Dasso, S., {et~al.} 2005, Astronomy and
  Astrophysics, 434, 725

\bibitem[{Miesch {et~al.}(2016)Miesch, Zhang, \& Augustson}]{Miesch16}
Miesch, M.~S., Zhang, M., \& Augustson, K.~C. 2016, The Astrophysical Journal
  Letters, 824, L15

\bibitem[{Moffatt(1969)}]{Moffatt69}
Moffatt, H.~K. 1969, Journal of Fluid Mechanics, 35, 117

\bibitem[{Moraitis {et~al.}(2018)Moraitis, Pariat, Savcheva, \&
  Valori}]{Moraitis18}
Moraitis, K., Pariat, E., Savcheva, A., \& Valori, G. 2018, submitted

\bibitem[{Moraitis {et~al.}(2014)Moraitis, Tziotziou, Georgoulis, \&
  Archontis}]{Moraitis14}
Moraitis, K., Tziotziou, K., Georgoulis, M.~K., \& Archontis, V. 2014, Solar
  Physics, 122

\bibitem[{Oberti \& Ricca(2018)}]{Oberti18}
Oberti, C., \& Ricca, R.~L. 2018, Fluid Dynamics Research, 50, 011413

\bibitem[{Pariat {et~al.}(2009)Pariat, Antiochos, \& DeVore}]{Pariat09a}
Pariat, E., Antiochos, S.~K., \& DeVore, C.~R. 2009, The Astrophysical Journal,
  691, 61

\bibitem[{Pariat {et~al.}(2016)Pariat, Dalmasse, DeVore, Antiochos, \&
  Karpen}]{Pariat16}
Pariat, E., Dalmasse, K., DeVore, C.~R., Antiochos, S.~K., \& Karpen, J.~T.
  2016, Astronomy and Astrophysics, 596, A36

\bibitem[{Pariat {et~al.}(2005)Pariat, D{\'e}moulin, \& Berger}]{Pariat05}
Pariat, E., D{\'e}moulin, P., \& Berger, M.~A. 2005, Astronomy and
  Astrophysics, 439, 1191

\bibitem[{Pariat {et~al.}(2017)Pariat, Leake, Valori, Linton, Zuccarello, \&
  Dalmasse}]{Pariat17}
Pariat, E., Leake, J.~E., Valori, G., {et~al.} 2017, Astronomy and
  Astrophysics, 601, A125

\bibitem[{Pariat {et~al.}(2015)Pariat, Valori, D{\'e}moulin, \&
  Dalmasse}]{Pariat15b}
Pariat, E., Valori, G., D{\'e}moulin, P., \& Dalmasse, K. 2015, Astronomy and
  Astrophysics, 580, A128

\bibitem[{Patsourakos \& Georgoulis(2017)}]{Patsourakos17}
Patsourakos, S., \& Georgoulis, M.~K. 2017, Solar Physics, 292, 89

\bibitem[{Patsourakos {et~al.}(2016)Patsourakos, Georgoulis, Vourlidas, Nindos,
  Sarris, Anagnostopoulos, Anastasiadis, Chintzoglou, Daglis, Gontikakis,
  Hatzigeorgiu, Iliopoulos, Katsavrias, Kouloumvakos, Moraitis,
  Nieves-Chinchilla, Pavlos, Sarafopoulos, Syntelis, Tsironis, Tziotziou,
  Vogiatzis, Balasis, Georgiou, {Karakatsanis, L. P.}, Malandraki,
  Papadimitriou, Odstrcil, Pavlos, Podlachikova, Sandberg, Turner, Xenakis,
  Sarris, Tsinganos, \& Vlahos}]{Patsourakos16}
Patsourakos, S., Georgoulis, M.~K., Vourlidas, A., {et~al.} 2016, The
  Astrophysical Journal, 817, 14

\bibitem[{Polito {et~al.}(2017)Polito, Del~Zanna, Valori, Pariat, Mason,
  Dud{\'\i}k, \& Janvier}]{Polito17}
Polito, V., Del~Zanna, G., Valori, G., {et~al.} 2017, Astronomy and
  Astrophysics, 601, A39

\bibitem[{Priest {et~al.}(2016)Priest, Longcope, \& Janvier}]{Priest16}
Priest, E.~R., Longcope, D.~W., \& Janvier, M. 2016, Solar Physics, 1

\bibitem[{Rudenko \& Myshyakov(2011)}]{Rudenko11}
Rudenko, G.~V., \& Myshyakov, I.~I. 2011, Solar Physics, 270, 165

\bibitem[{Russell {et~al.}(2015)Russell, Yeates, Hornig, \&
  Wilmot-Smith}]{Russell15}
Russell, A. J.~B., Yeates, A.~R., Hornig, G., \& Wilmot-Smith, A.~L. 2015,
  Physics of Plasmas, 22, 032106

\bibitem[{Rust(1994)}]{Rust94}
Rust, D.~M. 1994, Geophysical Research Letters, 21, 241

\bibitem[{Taylor(1974)}]{Taylor74}
Taylor, J.~B. 1974, Physical Review Letters, 33, 1139

\bibitem[{Temmer {et~al.}(2017)Temmer, Thalmann, Dissauer, Veronig, Tschernitz,
  Hinterreiter, \& Rodriguez}]{Temmer17}
Temmer, M., Thalmann, J.~K., Dissauer, K., {et~al.} 2017, Solar Physics, 292,
  93

\bibitem[{Thalmann {et~al.}(2011)Thalmann, Inhester, \&
  Wiegelmann}]{Thalmann11}
Thalmann, J.~K., Inhester, B., \& Wiegelmann, T. 2011, Solar Physics, 272, 243

\bibitem[{Valori {et~al.}(2012)Valori, D{\'e}moulin, \& Pariat}]{Valori12}
Valori, G., D{\'e}moulin, P., \& Pariat, E. 2012, Solar Physics, 278, 347

\bibitem[{Valori {et~al.}(2013)Valori, D{\'e}moulin, Pariat, \&
  Masson}]{Valori13}
Valori, G., D{\'e}moulin, P., Pariat, E., \& Masson, S. 2013, Astronomy and
  Astrophysics, 553, 38

\bibitem[{Valori {et~al.}(2016)Valori, Pariat, Anfinogentov, Chen, Georgoulis,
  Guo, Liu, Moraitis, Thalmann, \& Yang}]{Valori16}
Valori, G., Pariat, E., Anfinogentov, S., {et~al.} 2016, Space Science Reviews,
  201, 147

\bibitem[{Webb {et~al.}(2010)Webb, Hu, Dasgupta, \& Zank}]{Webb10}
Webb, G.~M., Hu, Q., Dasgupta, B., \& Zank, G.~P. 2010, Journal of Geophysical
  Research: Space Physics, 115

\bibitem[{Webb {et~al.}(2011)Webb, Hu, Dasgupta, \& Zank}]{Webb11}
---. 2011, Journal of Geophysical Research: Space Physics, 116

\bibitem[{Wiegelmann \& Sakurai(2012)}]{Wiegelmann12}
Wiegelmann, T., \& Sakurai, T. 2012, Living Reviews in Solar Physics, 9, 5

\bibitem[{Wiegelmann {et~al.}(2014)Wiegelmann, Thalmann, \&
  Solanki}]{Wiegelmann14}
Wiegelmann, T., Thalmann, J.~K., \& Solanki, S.~K. 2014, The Astronomy and
  Astrophysics Review, 22, 78

\bibitem[{Woltjer(1958)}]{Woltjer58}
Woltjer, L. 1958, in Proceedings of the National Academy of Sciences of the
  United States of America (National Acad Sciences), 489--491

\bibitem[{Yang {et~al.}(2013)Yang, B{\"u}chner, Santos, \& Zhang}]{YangS13}
Yang, S., B{\"u}chner, J., Santos, J.~C., \& Zhang, H.~Q. 2013, Solar Physics,
  283, 369

\bibitem[{Yeates \& Hornig(2013)}]{Yeates13}
Yeates, A.~R., \& Hornig, G. 2013, Physics of Plasmas, 20, 012102

\bibitem[{Yeates \& Hornig(2014)}]{Yeates14}
---. 2014, Journal of Physics: Conference Series, 544, 012002

\bibitem[{Zhao {et~al.}(2015)Zhao, DeVore, Antiochos, \& Zurbuchen}]{ZhaoL15}
Zhao, L., DeVore, C.~R., Antiochos, S.~K., \& Zurbuchen, T.~H. 2015, The
  Astrophysical Journal, 805, 61

\bibitem[{Zuccarello {et~al.}(2015)Zuccarello, Aulanier, \&
  Gilchrist}]{Zuccarello15}
Zuccarello, F.~P., Aulanier, G., \& Gilchrist, S.~A. 2015, The Astrophysical
  Journal, 814, 126

\bibitem[{Zuccarello {et~al.}(2018)Zuccarello, Pariat, Valori, \&
  Linan}]{Zuccarello18}
Zuccarello, F.~P., Pariat, E., Valori, G., \& Linan, L. 2018, in prep., mo

\end{thebibliography}

\IfFileExists{\jobname.bbl}{} {\typeout{}
\typeout{****************************************************}
\typeout{****************************************************}
\typeout{** Please run "bibtex \jobname" to obtain} \typeout{**
the bibliography and then re-run LaTeX} \typeout{** twice to fix
the references !}
\typeout{****************************************************}
\typeout{****************************************************}
\typeout{}}

%

\end{document}